\documentclass[journal=jacsat,manuscript=article]{achemso}

\usepackage[version=3]{mhchem} 
\usepackage{amssymb} 
\usepackage{slashed} 
\usepackage{tensor} 
\usepackage{dsfont} 
\usepackage{tikz-cd} 
\usepackage{mhchem} 
\usepackage{longtable} 
\usepackage{geometry} 

\author{Juan R. Chamorro}
\affiliation{Department of Chemistry, The Johns Hopkins University, Baltimore, MD 21218 USA}
\alsoaffiliation{Institute for Quantum Matter, Department of Physics and Astronomy, The Johns Hopkins University, Baltimore, MD 21218 USA}
\email{jchamor1@jhu.edu}

\author{Tyrel M. McQueen}
\affiliation{Department of Chemistry, The Johns Hopkins University, Baltimore, MD 21218 USA}
\alsoaffiliation{Institute for Quantum Matter, Department of Physics and Astronomy, The Johns Hopkins University, Baltimore, MD 21218 USA}
\alsoaffiliation{Department of Materials Science and Engineering, The Johns Hopkins University, Baltimore, MD 21218 USA}
\email{mcqueen@jhu.edu}

\author{Thao T. Tran}
\affiliation{Department of Chemistry, Clemson University, Clemson, SC 29634 USA}
\email{thao@clemson.edu}

\title[An \textsf{achemso} demo]
{The Chemistry of Quantum Spin Liquids}

\begin{document}

\begin{abstract}
  Quantum spin liquids are an exciting playground for exotic physical phenomena and emergent many-body quantum states. The realization and discovery of quantum spin liquid candidate materials and associated phenomena lie at the intersection of solid-state chemistry, condensed matter physics and materials science and engineering. In this review, we provide the current status of the crystal chemistry, synthetic techniques, physical properties, and research methods in the field of quantum spin liquids. We also highlight a number of specific quantum spin liquid candidate materials and their structure-property relationships, elucidating their fascinating behavior and connecting it to the intricacies of their structures. Furthermore, we share our thoughts on defects and their inevitable presence in materials, of which quantum spin liquids are no exception, which can complicate the interpretation of characterization of these materials, and urge the community to extend their attention to materials preparation and data analysis, cognizant of the impact of defects. This review was written with the intention of providing guidance on improving the materials design and growth of quantum spin liquids, and painting a picture of the beauty of the underlying chemistry of this exciting class of materials.
\end{abstract}
\begin{tableofcontents}

1. Introduction \par

2. Synthesis, Physical Properties, and Research Methods \par
    2.1 Experimental synthetic methods and phase analysis \par
        2.1.1 High-temperature solid state reactions\par
        2.1.2 Low temperature techniques\par
        2.1.3 Chemical vapor transport\par
        2.1.4 Flux growth\par
        2.1.5 Floating zone technique\par
        2.1.6 Phase analysis\par
        2.1.7 Thermodynamic stability calculations\par
    2.2 Measurement Techniques and Physical Properties\par
        2.2.1 Magnetization\par
        2.2.2 Heat Capacity\par
        2.2.3 Thermal conductivity\par
        2.2.4 Neutron scattering\par
        2.2.5 Electron spin resonance\par
        2.2.6 Nuclear magnetic resonance\par
        2.2.7 Muon spin spectroscopy\par
        2.2.8 High-angle annular dark-field (HAADF) imaging\par
3. Materials \par
    3.1 1D - Linear Chains\par
        3.1.1 \ce{GeCuO3}\par
        3.1.2 \ce{Sr2CuO3}\par
        3.1.3 \ce{CsNiCl3}\par
    3.2 2D \par
        3.2.1 Triangular Lattice\par
            3.2.1.1 \ce{NiGa2S4}\par
            3.2.1.2 \ce{YbMgGaO4}\par
            3.2.1.3 $6H$-\ce{Ba3CuSb2O9}\par
        3.2.2 Honeycomb Lattice\par
            3.2.2.1 $\alpha$-\ce{RuCl3}\par
            3.2.2.2 Iridates\par
        3.2.3 Kagome Lattice\par
            3.2.3.1 Herbertsmithite \ce{ZnCu3(OH)6Cl2}\par
            3.2.3.2 \ce{Ln3Sb3Mg2O14} and \ce{Ln3Sb3Zn2O14} materials\par
            3.2.3.3 \ce{LiZn2Mo3O8}\par
    3.3 3D\par
        3.3.1 Diamond Lattice\par
            3.3.1.1 \ce{FeSc2S4}\par
        3.3.2 Pyrochlores and Breathing Pyrochlores\par
    3.4 Quantum Spin Liquid Candidates\par
4. Defects and Problems with Measurements\par
    4.1 Current problems\par
    4.2 Defects\par
        4.2.1 Point defects\par
        4.2.2 Topological defects\par
5. Potential Applications\par
6. Summary\par
7. Author Biographies\par
8. Acknowledgments\par
9. References\par

\end{tableofcontents}
\section{\underline{Introduction}}

\noindent Chemistry is a diverse field that covers topics ranging from the fundamentals of manybody electron states to the design and development of synthetic tools and materials that improve everyday life. A, and perhaps the, unifying attribute of the chemical sciences is understanding the world in terms of electrons, and how they self-organize on an atomic scale to make (or not make) chemical bonds. These behaviors are inextricably linked to the spin of individual electrons, with the Pauli exclusion principle limiting the available electronic states. Consider a lattice of spins that can interact with each other. There are a fixed number of available microstates, $2J+1$, per site, giving a total of $(2J+1)^n$ spin configurations for a lattice with $n$ sites. In the case of an extended solid, $n$, and hence the number of spin configurations, is very large, of order 1 mol. In accordance with the third law of thermodynamics, where the entropy change must approach 0 as $T \rightarrow 0$ K, these spins will have the tendency to become part of a long range magnetically ordered state. In other words, these spins will go from pointing randomly in all directions, agitated by thermal fluctuations, to being locked into place, pointing in one direction over another, each a part of a long range ordered state obeying a subset of the underlying crystallographic symmetries. Magnetic orders are classified by the relative spin orientation between spins on lattice sites. The two extreme limits are antiferromagnetic, where adjacent spins point in opposite directions, or ferromagnetic, where spins point along the same direction. More formally, a magnetically ordered state is the lowest energy spin configuration (or subset of spin configurations). At temperatures above $T = 0$ K, thermal energy populates additional relative spin configurations/orientations, breaking the fixed relative orientations between adjacent sites; at sufficiently high temperature, a magnetic phase transition to a state with random orientations between adjacent sites, a paramagnetic state observable in many physical properties, occurs.

An inexact analogy can be drawn that nevertheless captures the essence of the interesting magnetic state we are about to discuss in this review. If we consider paramagnetism to be analogous to an ideal gas, where particles interact weakly with each other and are said to be in a highly entropic state; and magnetic order to be analogous to a solid, where particles form a static and orderly, regularly repeating lattice; then, a natural question arises: is there an analogous liquid of spins? Though the title of this review may seem like a spoiler, the answer is actually: ``kind of". Over the last twenty years or so, the field of spin liquids has emerged in condensed matter physics and has seen an enormous amount of attention no doubt primarily due to its possible relation to high-temperature superconductivity. It is optimistically believed that doped spin liquid materials may even be an avenue towards room temperature superconductors, materials which would undoubtedly change human civilization forever.\cite{Lee_2007}

Spin liquids are exotic materials where spins do not order at any finite temperature, and thus perhaps to the naked eye give the illusion that they are disordered paramagnets, but unlike paramagnets or ideal gasses, they have very strongly interacting spins that have energy scales comparable to ordered magnets, or greater. Normally, such strong interactions would result in a single, or small subset of, spin configurations having lower energy than the rest and drive magnetic order; however, if the interactions conspire to make all, or a large subset of, spin configurations equivalent in energy, then classical magnetic order is evaded. These spins are said to be frustrated. At the same time, something must happen to remove the entropy, because the third law of thermodynamics is a cruel mistress.  The quantum spin liquid is one such exciting possibility: a single state is selected as the ground state that consists of an admixture of many (or all) of the spin configurations. This state does not possess classic magnetic order, as when projected onto single spins each appears to be fluctuating, even at $T = 0$ K. At the same time, the spins are no longer randomly oriented relative to each other, and posses hidden correlations (the spins are entangled).  Where different classes of quantum spin liquid differ is in the details of precisely what these quantum ground states look like, and whether or not there is an energy gap to the first excited state.

But engineering or discovering such a state in real materials is exceptionally difficult: the Jahn-Teller theorem describes, very generally, how any non-linear electronic system with symmetry can lower its energy by removing some (or all) of the symmetry in the presence of degeneracy; spins, borne of electrons, are not immune to this effect, and the practical consequence of the removal of symmetry is the lowering of energy of some spin configurations relative to others, which favors the formation of classical magnetic orders. The situation is less dire if one is willing to accept materials which, eventually, do lose symmetry and magnetically order, but do so at a temperature much less than the interaction energy scale -- i.e., there is a wide range of temperatures over which spin liquid physics can dominate. This is the scenario found in well known 1-D chain magnets, where the $S = 1/2$ antiferromagnets ground state is a spin liquid, but where interchain interactions even $1/1000^{th}$ as strong as the intrachain interactions, eventually drives classical magnetic order. But the fact remains that engineering a quantum spin liquid is an exceptionally challenging task.

The key ingredient is to find ways to make a large number of spin configurations simultaneously have strong interactions between neighbors and have a large number of configurations with the same energy, i.e. to frustrate the spins. There are two ways to do so: either have it built in by the underlying geometric frustration of the lattice in which they sit, or by controlling the differences in the energies with which they interact with their nearest neighbors and their next nearest neighbors to force degeneracy when all interactions are included. 

The first case is the most common, and the basis for most of the compounds we will be discussing herein. A geometrically frustrated lattice of spins is built from one of two fundamental units: a triangle (for 2D) or a tetrahedron (for 3D) \cite{Cava2011}. \textbf{Figure 1} graphically demonstrates this intuitively. If we impose the so-called ``Ising" rule upon spins, constraining them to point along a single axis, and force the interactions between neighbors to be antiferromagnetic, then six of the eight possible spin configurations are forced to be equal in energy: any two spins can be aligned antiparallel, but then the total energy is the same whether the direction of the third is spin up or spin down. The remaining equal energy spin configurations are derived by applying the $C_3$ rotational symmetry, and time reversal symmetry (which flips the directions of all spins). In this toy model, only two eighths of the spin configurations are higher in energy then the rest -- the two ferromagnetic configurations (all spins up or all spins down). The local interactions combined with the lattice symmetry have forced a degeneracy of $3/4^{ths}$ of the spin configurations, i.e. have frustrated the spins. In the absence of considering the possible quantum mechanical fluctuations, this situation results in the theoretically well-studied classical spin liquid, the ``Ising" triangular lattice antiferromagnet.  The same happens in a tetrahedral arrangement of spins, as only two of the four can satisfactorily face opposite directions, yet the other two cannot, so none of them do at all. Since there are many real-world materials that have lattices based on triangular or tetrahedral units, it is natural to expect such frustrated behavior of spins in the real world.

\begin{center}                
    \includegraphics[width=0.7\textwidth]{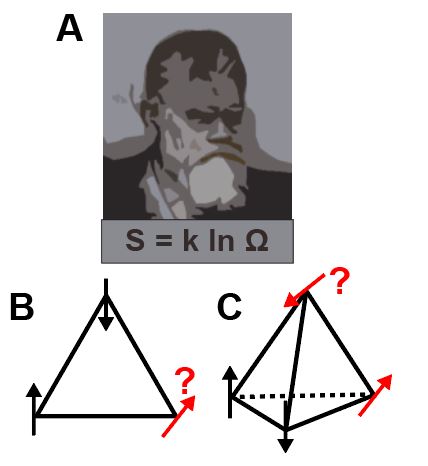} \par
    \small{\textbf{Figure 1.} 
    \textbf{A.} A portrait of Ludwig Boltzmann over his famous entropy equation, from which it can be inferred that given S $= 0$ at $T = 0$ K, the number of microstates must be $\Omega = 1$, implying a single ground state.
    \textbf{B.} A triangular arrangement of Ising spins cannot establish a single ground state, resulting in a seemingly impossible $\mathrm{S} = \mathrm{kln}(6)$ at $T = 0$ K.
    \textbf{C.}} A tetrahedral arrangement of atoms also showing frustration, with an even larger, impossible $T = 0$ K degeneracy of $\mathrm{S} = \mathrm{kln}(22)$.
\end{center}

A geometrically frustrated lattice is not enough to obtain a spin liquid, however. For one, certain kinds of magnetic order can emerge even in frustrated lattices, as shown in \textbf{Figure 2}. These kinds of order tend to emerge either due to lower energy scale perturbations or single ion effects, or when the size of each magnetic moment is large, i.e. usually greater than $S = 1$, as larger spins tend to show more classical behavior as opposed to quantum because the energy barrier between states scales as $S^2$, which strongly suppresses quantum fluctuations. In order to avoid the formation of magnetic order, therefore, the experimenter should opt to use ions with well-controlled single ion effects and with the smallest spin possible, $S = 1/2$. This also goes hand-in-hand with the second requirement for a spin liquid, which is entanglement. What separates a paramagnet from a spin liquid is the long range entanglement each spin experiences with the rest of the lattice. No one spin is an island entirely of itself; every spin is a piece of the continent. Facilitated by the large energetic degeneracies enabled by an underlying frustrated lattice, and with smaller magnetic moments that are much more susceptible to strong quantum fluctuations, long range entanglement can emerge and give rise to a quantum spin liquid ground state. \cite{Balents_2010,Broholmeaay0668}

\begin{center}                
    \includegraphics[width=1\textwidth]{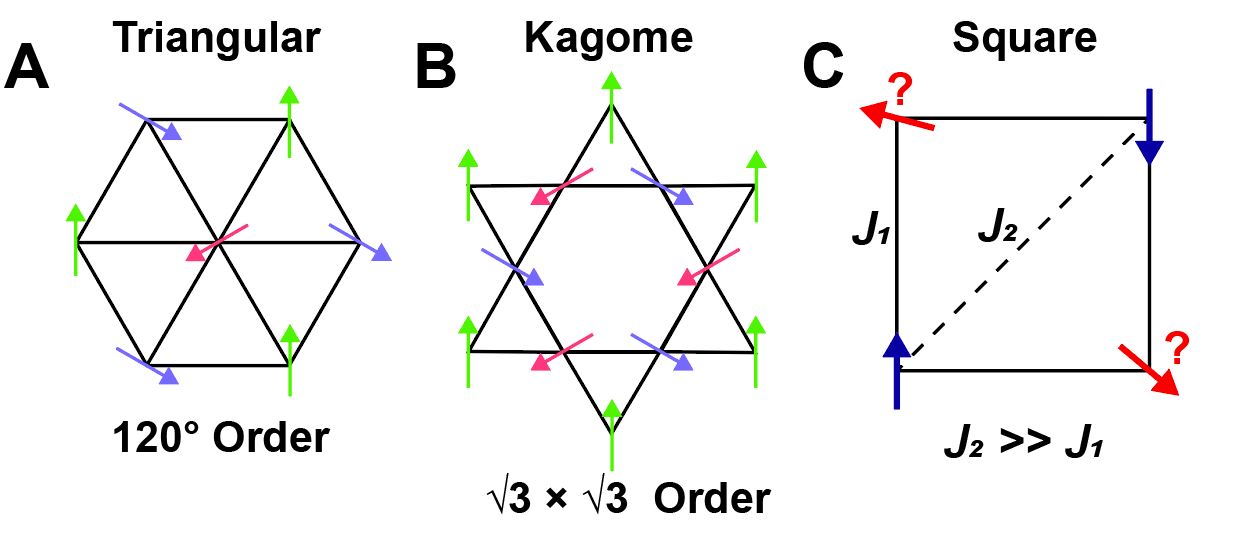} \par
    \small{\textbf{Figure 2.} 
    \textbf{A.} One possibility for magnetic order on a triangular lattice, the $120^{\circ}$ state.
    \textbf{B.} The $\sqrt{3} \times \sqrt{3}$ type of magnetic order on a kagome lattice.
    \textbf{C.} Theoretically in the presence of two exchange interactions $J_{1}$ and $J_{2}$, where the former is the nearest neighbor and the latter the next-nearest neighbor, frustration can be achieved if $J_{2} >> J_{1}$, a condition that can be met with low spins due to strong quantum fluctuations.
    }
\end{center}

Thus far, quantum spin liquid (QSL) behavior, or more precisely indirect indications of a QSL state (since measurement of entanglement in any form -- let alone of spins -- is exceptionally difficult experimentally), has only been convincingly observed in one dimensional materials and in materials with frustrated lattices. As aforementioned, these are real-world materials with real-world chemistry. In this work, we will review quantum spin liquids in terms of the synthesis of candidate compounds, the measurements routinely performed on them and what they can and cannot elucidate, and provide many examples. For these compounds, we seek to understand several things. We consider their structures and concomitant structure-property relationships, how the manner of their synthesis or their fundamental nature can give rise to defects that might affect quantum spin liquid behavior, and what physical properties they display and whether or not they are consistent with quantum spin liquid behavior. We will use the term "quantum spin liquid candidates" liberally, even applying it to systems that show magnetic order at some finite temperature, because some of these materials may have ranges in temperature where they exhibit signs of QSL behavior before they inevitably order at some lower temperature, or because they are so structurally similar to other, more genuine candidates that it is instructive to refer to them as such.

\section{\underline{Synthesis, Physical Properties, and Research Methods}}

\subsection{Experimental synthetic methods and phase analysis} \par
        
Synthesis is at the core of the chemistry of quantum spin liquids since it provides the materials for further studies and advanced characterizations. If no candidate materials are prepared, no quantum spin liquid phenomena can be realized. Thus, the development and application of a variety of synthetic techniques is key to producing quantum spin liquid materials with intriguing collections of electronic, magnetic and topological properties that may fulfill the needs of our society for advances in energy, memory, and computing-based applications. This section serves as a brief overview of the current status of synthetic techniques in the chemistry of quantum spin liquid materials, demonstrated by a number of select materials. These techniques are used for the synthesis of polycrystalline and single-crystal samples, the latter of which are naturally better for elucidating the properties of quantum spin liquid materials, as these tend to show complex and sensitive anisotropies that polycrystalline samples lack. We acknowledge that, however, it is nearly impossible to paint a perfectly complete picture in single review, given the wealth of existing synthetic protocols and their variations. As such, this section is rather a snapshot of current synthetic methodologies for both synthesis and analysis that may be a refection, to some degree, of the specific interests to the quantum spin liquid community. \par

\noindent \textbf{High-temperature solid state reactions} \par
        
Direct, high temperature solid-state reactions, also known as "shake-n-bake" reactions, are primarily driven by diffusion of particles, or rather, reactions that happen at the interfaces. Conventional solid-state synthesis requires high temperatures that are sufficient to overcome the activation energy barriers of reactions by providing particles with enough energy for proper diffusion. Thus, this synthetic technique primarily results in the thermodynamically most stable phases. The main advantage of high temperature solid-state synthesis is that it is straightforward, as all one needs is a well ground mixture of the reactants and a furnace. It is primarily used to obtain polycrystalline powders, either because they cannot be grown as crystals otherwise or as precursors for single-crystal growth. This technique enables the formation of a variety of quantum spin liquid candidate materials of varying frustrated lattices and physical properties, such as \ce{PbNi2V2O8},\cite{PhysRevLett.83.632,10.1143/PTPS.145.294} \ce{Ba4NbRu3O12},\cite{PhysRevB.93.180407} \ce{Na4Ir3O8},\cite{PhysRevLett.99.137207,PhysRevB.88.220413} \ce{RE3Sb3Mg2O14},\cite{doi:10.1002/pssb.201600256} and \ce{MnSc2S4}.\cite{PhysRevB.73.014413}. A major disadvantage of high-temperature solid state reactions is their inability to allow for the formation of potentially interesting metastable states, as they generally result in only the most thermodynamically stable product. Materials synthesized in this way are also prone to having more defects, especially vacancies and substitutional defects, as the higher the temperature the more volatile the reactants can be, and the more labile they are within the lattice. Achieving a more complete level of understanding of solid-state reactions in general, nevertheless, will open a route to better control over quantum spin liquid materials design and defect minimization \cite{doi:10.1021/acs.accounts.8b00382}. \par

\noindent \textbf{Low-temperature techniques} \par
        
Unsurprisingly, low-temperature techniques have the advantage to ability to generate materials with certain crystal structures that are inaccessible at high temperature, either because they are metastable or because they would decompose if heated to too high a temperature. Furthermore, in some cases, they enable the isolation of materials with less defects than the corresponding high-temperature synthesis, owing to the lower entropy involved. Unlike solid-state synthesis, the primary benefit in low-temperature reactions is that solid-solid diffusion rate is not the limiting step in the formation of desired products. This synthetic methodology enables the shift of conventional solid-state chemistry from thermodynamic control toward kinetic control where metastable phases can be controllably formed.\cite{doi:10.1021/acs.chemmater.6b04861}. One common low-temperature technique is hydrothermal synthesis, which involves performing chemical reactions in sealed vessels above ambient temperature and pressure, promoting the reactivity of reactants and control of interfaces and formation of unique phases\cite{doi:10.1002/anie.198510261,doi:10.1021/ja205456b,doi:10.1021/cm051791c}. Reactants that are usually insoluble in water under ambient conditions may dissolve in higher-than-100$^{\circ}$ conditions, for example, which can give rise to a vast variety of new materials. Some reactions can also proceed in water, or other solvents, under atmospheric conditions, further facilitating the synthesis of some materials. Low-temperature solution, solvothermal (non-aqueous) synthesis, and hydrothermal synthesis techniques have been utilized to produce a number of quantum spin liquid candidates, such as \ce{Cs2CuCl4},\cite{doi:10.1063/1.335093} \ce{KCuF3},\cite{RN17,PhysRevLett.88.106403,10.1143/PTPS.46.147} Zn$_{x}$Cu$_{4-x}$(OH)$_{6}$Cl$_{2}$,\cite{Shores2005,RN24,PhysRevLett.108.157202, Fu655,PhysRevB.83.100402}  and \ce{Cu3Zn(OH)6BrF}.\cite{PhysRevMaterials.2.044406,SMAHA2018123,RN25}. Synthetic Herbertsmithite \ce{ZnCu3(OH)6Cl2}, arguably the most popular mineral-based quantum spin liquid material (and discussed later in this text), is generally synthesized via hydrothermal methods\cite{Shores2005}.

\noindent \textbf{Chemical vapor transport} \par
        
The chemical vapor transport technique is extraordinarily useful in the preparation of pure and single-crystalline samples of materials. This technique involves a heterogeneous reaction where a solid, which usually has an insufficient pressure for vaporization, is volatilized in the presence of a volatile transport agent. This is usually done across a temperature gradient with a hot end and a cold end. Transport agent-assisted volatization occurs at the hot end (though not always) and products deposit as single crystals at the cold end.\cite{doi:10.1002/zaac.201300048}. Chemical vapor transport reactions have been used to produce a number of quantum spin liquid candidate crystals, including TiOBr,\cite{PhysRevLett.95.097203,doi:10.1002/zaac.19582950314} TiOCl,\cite{PhysRevB.67.020405, doi:10.1002/zaac.19582950314} $1T$-\ce{TaS2},\cite{Law6996, RN20} \ce{Li2IrO3},\cite{PhysRevLett.108.127203, PhysRevLett.114.077202,RN21, PhysRevLett.113.197201} and \ce{RuCl3}.\cite{Banerjee1055,RN23} Chemical vapor transport allows for the formation of crystals under constant thermodynamic conditions, which can improve crystal quality, however there is still much more to be done with and learned from this methodology. Furthermore, the preponderance of some transport agents over others, such as \ce{I2} vs. \ce{NH4Cl}, to incorporate into the final products provides an avenue towards defects, which may either help or hinder exotic quantum ground states. The controlled synthesis of new quantum spin liquid materials can benefit from a deeper understanding of the complex principles of chemical vapor transport, especially with regards to the thermodynamics of the involved solid and gas phases, and the kinetics associated with nucleation and growth.

\noindent \textbf{Flux growth} \par
        
The flux crystal growth technique involves two main components: (i) comprehensive knowledge of the chemical bonding and reactivity of the reagents involved, and (ii) multiple attempts to arrive at an appropriate flux and growth condition.\cite{Canfield_2019} This technique is an analogue of a low-temperature solvent-based reaction, wherein a solid-soon-to-be-molten-solvent (flux) is used to dissolve the reactants and allow for the crystallization of the product. The relevant reactants are placed in the same container as the flux and heated to high temperature. Once the flux melts, the reactants will dissolve in it, assuming they are soluble. As the reaction is cooled, the dissolved species will inevitably crash out and crystallize, as they are more supersaturated the closer the system is to the melting point of the flux. With the right flux and under the right heating and especially cooling conditions, large crystals can be obtained of desired products with high fidelity and quality. There are a variety of flux options, such as metal, salt and self-flux to choose from, depending on solubility of reactants in flux, reaction pathway and crystal nucleation.\cite{doi:10.1002/anie.201102676,B910749E,doi:10.1002/anie.200462170,Shoemaker10922,doi:10.1021/bk-2019-1333.ch006,doi:10.1021/cm2019873} The empirical nature of flux growth is, in part, due to the fact that the chemistry of this technique is far from equilibrium. \ce{Sm2CuO4},\cite{HUNDLEY1989102} \ce{CuGeO3},\cite{1999JCrGr.198..593R,PhysRevLett.70.3651} \ce{Ba3CuSb2O9},\cite{Katayama9305, PhysRevLett.106.147204,Man_2018} and \ce{Na2BaCo(PO4)2}\cite{Zhong14505} are a few examples of quantum spin liquid candidate crystals that were grown by flux technique. 

\noindent \textbf{Floating zone technique} \par
        
The optical floating zone technique has been used extensively to obtain high quality single crystals for quantum spin liquid candidate materials involving
\ce{Sr2CuO3},\cite{1999JCrGr.198..593R,PhysRevLett.76.3212,PhysRevB.62.R6108,RN15} \ce{SrNi2V2O8},\cite{PhysRevB.62.8921,PhysRevB.87.224423} \ce{FeSc2S4},\cite{MOREY2016128} \ce{Yb2Ti2O7},\cite{PhysRevB.95.094407,PhysRevX.1.021002, PhysRevLett.109.097205,PhysRevB.87.224423,RN26} and \ce{Er2Ti2O7}.\cite{Poole_2007,McClarty_2009} This technique involves taking two densely packed, counter-rotating rods of polycrystalline powder, often already of the same composition of the desired product, and using an extremely focused beam of light to melt one while holding the other in close contact. A molten zone forms in the middle that is held together by surface tension, giving the illusion of flotation and thus giving this technique its name. This technique is carried out at a range of temperatures depending on the melting point of the rods, which can range from a few hundred to a couple of thousand of degrees. The key for achieving a successful growth using the floating zone technique is to arrive at an appropriate combination of the principle growth parameters, such as feed rod quality, crystallization rate, growth atmosphere and gas pressure, temperature gradient profile and molten zone temperature and rotation rate.\cite{KOOHPAYEH2008121,doi:10.1146/annurev-matsci-070616-124006,Straker2020,doi:10.1021/acs.chemmater.0c01721} This technique is therefore highly complex and requires an unparalleled amount of attention in order to ensure the isolation of high quality crystals, and failure to address any of the aforementioned parameters may result in defect-rich, sub-optimal samples. 

Recent advances in this field included novel high pressure capabilities, supercritical fluid-assisted optical floating zone materials growth up to $P = 300$ bar, and laser-diode floating zone techniques.\cite{PHELAN2019705,zhang2019high,ITO2013264,Zhang2017} The invention and development of high pressure supercritical and laser-diode floating zone instruments is providing a unique route not just for the growth of known quantum spin liquid materials, but also for the discovery of new candidates, thereby expanding their potential synthesis space. \par


\noindent \textbf{Phase analysis} \par
        
Aside from synthesis itself, our primary understanding of phase analysis and structures of most solid-state materials is primarily achieved through the use of X-ray, synchrotron and neutron diffraction experiments, and quantum spin liquid materials are not an exception. Despite the usefulness of single crystals, powder X-ray, synchrotron and neutron diffraction measurements present several attractive features compared to single-crystal X-ray and neutron diffraction. For one, powder data collection is far more rapid, allowing refinements to be performed and completed in the time it may take to even obtain crystals. Thus experiments can be designed after a few trial-and-error runs, with the quickly attained powder data, that may then enable the growth of single crystals. Refinements of powder data are well suited not only for phase analysis, phase transition, temperature-dependent and pressure-dependent studies, but also for precise determination of unit-cell parameters and quantitative analyses of mixtures. Another advantage of power data acquisition is that crystal twinning is not an issue, though crystallites in powders are randomly oriented and can have non-negligible effects on diffraction patterns. Single-crystal X-ray and neutron diffraction, on the other hand, provide a complementary structural characterization, where the three-dimensional data in reciprocal space enables a more accurate examination and determination of atomic structural parameters, temperature factors and subtle structure disorders and defects. Other techniques such as various optical and electron microscopies can provide more macroscopic insights into samples, such as line defects and large-scale sample imperfections.  An appropriate combination of two or more of the aforementioned diffraction techniques allows us to have better understanding of the structure and quality of materials, thus providing elucidation how the underlying real-world chemistry impacts exotic quantum phenomena.

\noindent \textbf{Thermodynamic stability calculations} \par

The future of quantum spin liquid material design and discovery lies in dialogue between the experimentalist and the theorist. Solid-state chemists share their wisdom of thermodynamics, kinetics, chemical reactivities, periodic trends, and the underlying reaction mechanisms for preparing materials. Computational theorists in turn provide their insights into the theoretical prediction of new quantum spin liquid materials based on computationally predicted crystal structures, and concomitant interesting electronic and magnetic properties. Recently, new machine learning approaches have been developed for the theoretical prediction of new solid-state materials dictated by the stability of certain chemical bonding and periodic arrangements.\cite{doi:10.1021/cm100795d,Lu2020,doi:10.1146/annurev-matsci-110519-094700,doi:10.1021/acs.chemmater.0c01907,RN28,RN29} A collaborative effort in research that leverages quantum theory to study ensembles of spin liquid materials using first-principle calculations, along with an automated experimental feedback loop, is a potentially valuable avenue for the discovery and realization of new quantum spin liquid behavior and other novel related physical phenomena in materials. 

\subsection{Measurement Techniques and Physical Properties}

\subsubsection{Magnetization}
    
Magnetization measurements are perhaps the most basic measurement that can be performed on quantum spin liquid candidate materials. They can afford information about the magnitude of the magnetic interactions and whether or not long range magnetic order sets in at some finite temperature or under an applied magnetic field. Through the use of magnetic fields, they may also be used to track field-induced phase transitions to an ordered or disordered state\cite{PhysRevLett.87.206407}.
    
Most quantum spin liquid candidates display paramagnetic behavior across all measured temperature ranges. This does not apply to all spin liquid materials, however, as there is a subclass that is expected to manifest as a so-called "resonating valence bond" solid, whereby spins pair up into singlets that aren't static, but rather delocalized across the entire lattice, similar to delocalized electrons in a benzene molecule\cite{Anderson_1973}. Other measurements, such as nuclear magnetic resonance measurements (described below) can be used in conjunction with magnetization measurements to observe the expected downturn in susceptibility that in magnetization is clouded by the presence of ubiquitous magnetic impurities. Nevertheless, at high temperature, Curie-Weiss paramagnetic behavior is expected to be predominant. By fitting the data to the Curie-Weiss law, the magnitude of the magnetic moments can be extracted, as well as the strength of their interactions with each other. As aforementioned, magnetic moments close to the single unpaired electron value of $S = 1/2$ are highly desired as they are prone to much stronger quantum fluctuations which are necessary and conducive toward the stabilization of a quantum ground state. The Weiss constant $\Theta_W$, in kelvin, provides an estimate of the strength of magnetic interactions. In the materials section below, we highlight many quantum spin liquid candidates and give an overview of their interaction strengths as obtained from magnetization. The larger the Weiss constant, the greater the moment-moment interaction strength, and thus the higher the temperature  at which the material orders magnetically. 
    
Magnetization is also the primary measurement used in determining the popular "frustration index", invented by A. P. Ramirez in 1994  \cite{Ramirez_1994}. This value, which has no physical meaning but serves as a qualitative measure of magnetic frustration, is given by $f = \frac{\mid\Theta_W\mid}{T_N}$, where $T_N$ is the N\'eel ordering temperature for the system (assuming antiferromagnetism, though the Curie temperature $T_{C}$ for ferromagnets can be used just as well). In a system that does not show magnetic ordering at any temperature, $T_N$ is taken to be the lowest measured temperature. The larger the value of $f$, the more frustrated the system is said to be, and many reviews tend to rank or group quantum spin liquids based on their frustration indices. A true quantum spin liquid is expected to show $f \rightarrow \infty$, as they are expected to remain liquid down to absolute zero.
    
\subsubsection{Heat capacity}
    
Heat capacity is a powerful experimental tool routinely used to investigate spin liquid candidates and materials in general. Phonons, electrons, and magnetic excitations all heat, and heat capacity data can be disassembled to show the contributions of each. The magnetic contribution to the heat capacity in quantum spin liquid materials is usually obtained after subtraction of the phonon contribution, since most of the candidate materials are electrical insulators and thus have no electronic contribution, though it has been demonstrated that some disordered glassy materials can give rise to a linear heat capacity at low temperature \cite{elliott_drabold_1985}. Nevertheless, this magnetic contribution can then provide information valuable to determining what the underlying physics of the material is. The magnetic heat capacity can be integrated to obtain the entropy change as a function of temperature, via the equation $\Delta S = \int\frac{C_{v}}{T}dT$. Though it is the heat capacity at constant pressure $C_{p}$ that is measured for most materials, $C_{v} \approx C_{p}$ at low temperatures, and thus the magnetic entropy can be successfully derived from the data. Based on the chemical formula of and the size of the magnetic moments in the material being investigated, the change in magnetic entropy is expected to plateau at specific values, given by $\Delta \mathrm{S} = \mathrm{Rln}(2S+1)$, with spin $S$. Thus, a system containing a single magnetic atom per formula unit that possesses $S = 1/2$ spin is expected to show an entropy change that increases with temperature and plateaus at $\Delta \mathrm{S} = \mathrm{Rln}(2)$. Deviations from these expected plateau values, specifically plateaus that are under the expected values, are usually indicative of the absence of phase transitions into a magnetically ordered state that would serve to recover all of the expected entropy. Furthermore, in systems where the magnetic moments arise from heavier elements, such as $4d$, $5d$, and $4f$ elements, where spin-orbit coupling is strong, deviations from the expected value based solely on spin $S$ can be indicative of non-negligible orbital $L$ contributions to the magnetism, which can sometimes result in even more exotic phenomena. It is worth noting that phase transitions manifest as peaks in the heat capacity, and often serve to recover part or all of the expected entropy, as a transition into a magnetically ordered ground state serves to establish a macroscopic ground state. 
    
 Detection of QSL behavior requires extreme care, and rigorous exclusion of alternative possibilities, such as $T-linear$ contributions from an insulator or $T^{3/2}$ contribution from spin wave excitations. The temperature dependence of the magnetic contribution to the heat capacity is also very telling of the underlying physical behavior of the material. Within the class of materials known as quantum spin liquid materials, there are further subclasses depending on the symmetries and topologies of magnetic interactions. These different kinds of quantum spin liquid are expected to show power-law dependence depending on their type, such as the $U(1)$ spin liquid where the heat capacity follows a $T^{3/2}$ dependence \cite{PhysRevB.72.045105}, or the $\mathbb{Z}_2$ spin liquid which shows exponentially gapped behavior \cite{PhysRevLett.110.207208}. This arises because some spin liquids have itinerant excitations, which can carry heat, whereas others have localized ones that cannot. 

It is worthwhile to mention that such itinerant excitations, indirectly derived from heat capacity power laws, are known as spinons, and result from the fractionalization of electrons into their charge (holon) and spin (spinon) components. In some quantum spin liquids, the spinons can move around the lattice and form a Fermi surface, and can be detected in a number of measurement techniques. Electron fractionalization is a fundamental property of quantum spin liquids, along with long range entanglement that establishes them \cite{RN32}. 

\subsubsection{Thermal conductivity}
    
Thermal conductivity measurements are performed on materials to determine their ability to conduct heat. Heat is carried by itinerant excitations, such as phonon modes in the lattice and electrons in metals. As aforementioned, most spin liquid materials are electrically insulating and therefore, only the phonon contribution to the thermal conductivity is expected.

However, as previously discussed, electron fractionalization can occur in quantum spin liquids, resulting in a separation of the charge and spin components of electrons. While the holon component will remain localized, the spinon component is, in some cases, itinerant, and can therefore carry heat. The thermal conductivity will be modified and will contain a linear term consistent with the presence of heat-carrying fermions, and will therefore deviate from simple phonon behavior. 

This ability to measure the presence of spinons is an extremely powerful tool, and has, in some cases, even functioned as a "make or break" kind of measurement for some spin liquid materials. An example is \ce{YbMgGaO4}, which we will discuss more extensively below, where multiple other kinds of measurement such as magnetization, heat capacity, and neutron scattering all gave early indications of quantum spin liquid behavior (if one ignores the obvious chemical disorder issues!), specifically of the $U(1)$ subclass where spinons are expected to emerge and be itinerant, yet no magnetic thermal conductivity was observed, casting strong doubt on the proposed $U(1)$ ground state for this material \cite{PhysRevLett.117.267202}. On the hand, in two quantum spin liquid candidates with triangular lattices (discussed below), $\kappa$-\ce{(BEDT-TTF)2Cu2(CN)3} \cite{Yamashita2008} and \ce{EtMe3Sb[Pd(dmit)2]2}\cite{Yamashita2010}, magnetic thermal conductivity has been reported, which, if confirmed, would be indicative of highly mobile gapless excitations consistent with a quantum spin liquid ground state. Thermal conductivity measurements should not be used in isolation to demonstrate quantum spin liquid behavior, however, as there may be other sources of transport such as defects and impurities, but in conjunction with other measurement techniques, it is a very powerful tool for the elucidation of such interesting quantum phenomena.
    

\subsubsection{Neutron scattering} \par

Neutron scattering, both elastic and inelastic, is perhaps the most widespread technique used to measure the magnetic properties of quantum spin liquids. Elastic scattering functions in much the same way as X-ray scattering, with neutrons interacting with nuclei instead of electron clouds, allowing for diffraction experiments that can reveal the crystallographic structure of a material. Because neutrons have spin, unlike photons, they can also probe a material's magnetic structure, which in a diffraction pattern can appear as new peaks that are different from the crystallographic ones, or they can modify previously existing peaks (in the case of ferromagnetism, since the ferromagnetic structure will retain the same structure as the underlying crystal lattice). As such, neutron diffraction is very useful in the exploration of quantum spin liquid materials, as these materials are expected to show no signs of magnetic ordering at any temperature range and thus, no changes to the diffraction patterns are expected, barring temperature dependent changes to the underlying crystallographic structure. Neutron diffraction measurements are routinely carried out on spin liquid compounds at very low temperatures for this exact purpose.

One the other hand, inelastic neutron scattering can be used to study magnetic and crystal field excitations. Moderated, low energy neutrons, obtained as a byproduct from a nuclear reaction in a reactor, are passed through a monochromating chopper that sets their wavelength to some desired value. The neutrons, now of known initial energy and wavelength, are impinged upon a sample, and energy transfer occurs when the neutrons and the excitations they probe are of comparable energy scales. As a brief example of the power of neutron scattering to probe crystal field excitations, consider octahedral Ru$^{3+}$. As a $4d^5$ system, the general expectation is that, in the absence of spin-orbit coupling, this system will have a crystal field manifold of $t_{2g}^{5} e_{g}^{0}$, which is a Jahn-Teller active configuration, as shown in Figure 3. Ru$^{3+}$ has strong spin-orbit coupling, however, which lifts the degeneracy and instead splits the $t_{2g}$ manifold into a set of spin-orbital states, the bottommost of which is a $j = 1/2$ state. As a result, Ru$^{3+}$ in a frustrated lattice is a prime candidate for quantum spin liquid behavior as the $j = 1/2$ state is susceptible to the strong quantum fluctuations necessary for exotic behavior, which is the case in the honeycomb material $\alpha$-\ce{RuCl3}, which we discuss extensively below. The next state that is higher in energy is the $j = 3/2$ state, and these two are separated with a energy gap of magnitude $\frac{3\lambda}{2}$, where $\lambda$ is the spin-orbit coupling in meV. In one experiment on $\alpha$-\ce{RuCl3} \cite{Banerjee2016}, neutrons with initial energy of $E_{i} = 1.5$ eV were fired upon an aligned sample, and a $j = 1/2$ to $j = 3/2$ gap was measured of value $E = 195 \pm 11$ meV, corresponding to a spin-orbit coupling value of $\lambda = 130$ meV, which is very close to the free Ru$^{3+}$ ion expectation of $\lambda_{free} = 150$ meV \cite{Figgis1966}. This example illustrates the power of inelastic neutron scattering in the exploration of crystal field excitations, where effects such as Jahn-Teller distortions and spin-orbit coupling conspire to establish ground states necessary to the creation of quantum spin liquid and other exotic ground states.

\begin{center}                
    \includegraphics[width=180mm]{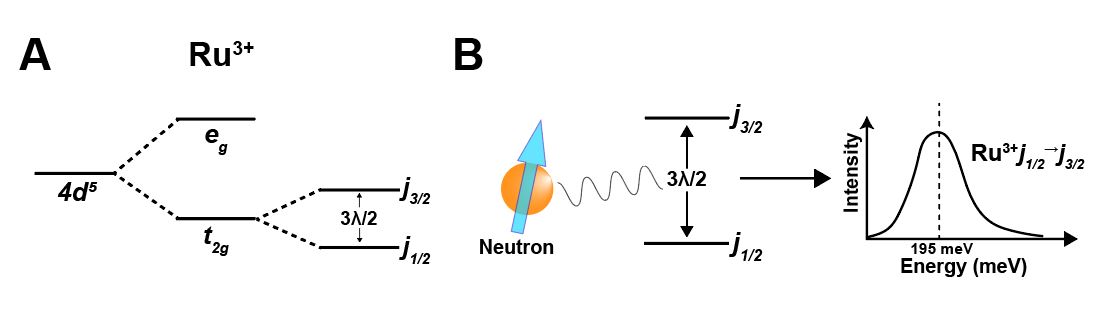} \par
    \small{\textbf{Figure 3.} Ru$^{3+}$ crystal field schematic.
    \textbf{A.} The crystal field diagram for Ru$^{3+}$, showing the impact of spin-orbit coupling, which reduces the manifold from the Jahn-Teller active $4d^{5}$ state to the $j = 1/2$ ground state.
    \textbf{B.} A neutron with energy greater than the $3\lambda/2$ gap can transfer some of its energy to an electron, which becomes excited. The energy of the outgoing neutron is then recorded, and the size of the gap can be determined. In the case of Ru$^{3+}$}
\end{center}

Magnetic excitations, unlike crystal field excitations, are instead on the order of meV, so much smaller neutron energy transfers (and thus in practice much lower incident energy due to limitations on resolving small transfers of energy) are required to probe them. As a simple example, imagine a linear chain of spins that are antiferromagnetically ordered. An excitation might be flipping a single spin so that it is now ferromagnetically interacting with its neighbors, which has a small energy cost on the order of meV. Neutrons with comparable energy can probe this excitation, and also measure its dispersion along the chain, as well as along the other two dimensional directions (which are always present in real-world materials). As such, neutrons can probe magnetic excitations and track them in reciprocal space, making them an extraordinarily powerful tool in the study of quantum spin liquid materials and all magnetic materials in general. 

\begin{center}                
    \includegraphics[width=0.8\textwidth]{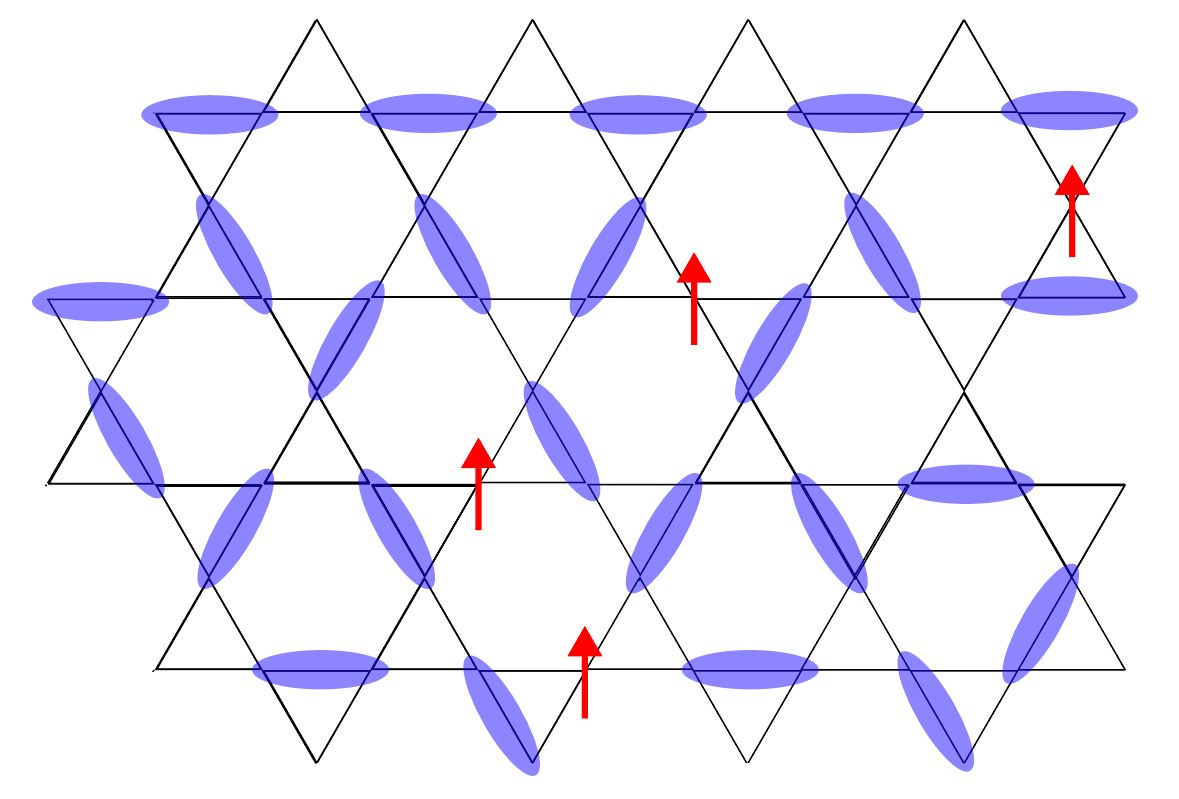} \par
    \small{\textbf{Figure 4.} Spinon excitations arise when valence bonds, shown as blue ovals, form between neighboring spins. Any unpaired electrons that are not participating in a singlet valence bond is a spinon, and if the valence bonds are resonating, these can travel around the lattice and are itinerant.}
\end{center}

In Figure 4 we schematically show what a spinon excitation looks like in the corner-sharing triangular kagome lattice of spins (a structure type we will discuss in more detail later). Borrowing the aforementioned Anderson interpretation of a quantum spin liquid as a resonating valence bond solid, where $S = 1/2$ spins form singlets with each other, any $S = 1/2$ moment that is not part of a singlet is a spinon excitation that is itinerant and can new singlets are continuously formed across the resonating lattice. These spinons can accept a wide range of energies, and thus fractionalized excitations in quantum spin liquids are not ubiquitously excited to the same higher energy state, but rather form a continuum of excitations with a large energy range, in all directions in reciprocal space. This is exactly what was observed in the quantum spin liquid candidate Herbertsmithite \cite{RN32}, aforementioned in the previous low-temperature synthetic techniques section and discussed more thoroughly in a later section, and thus provides direct evidence for fractionalized excitations in that material. This has also been observed in the triangular lattice material \ce{YbMgGaO4} \cite{RN35}, though in that system, as previously discussed, these exctitations may not necessarily be mobile.

This review is not a review of inelastic neutron scattering, but at least an introductory understanding to the technique is of great educational value to those who wish to venture into the synthetic field of quantum spin liquids, as this technique has proven invaluable in the study of their magnetic properties.  
    
\subsubsection{Electron Spin Resonance}

Electron spin resonance is the technique that uses the resonant absorption of electromagnetic radiation by unpaired electrons to probe the nature of defects in quantum spin liquids. Due to the Zeeman effect, magnetic moments will split in the presence of a magnetic field, in different ways. In the simple case of $S = 1/2$ states, the splitting will result in two spin states $M_s = +1/2$ and $-1/2$, where the magnetic moments align parallel and antiparallel to the applied field, respectively, and each state will be associated with a different Zeeman energy. Generally, spins pointing along the same direction as an applied field will be lower an energy than those pointing in the opposite direction. An unpaired electron can then transition between the two energy levels, either as the result of an absorption or emission of a photon of energy equal to the size of the splitting, such that the resonance requirement is satisfied. Electron spin resonance therefore provides a high-resolution and sensitive probe for the detailed examination of the characteristics of magnetic defects in quantum spin liquid materials and magnetic materials in general, by measuring unpaired electrons localized at defect sites. 

This technique can be used in conjunction with magnetization and nuclear magnetic resonance (discussed below) to isolate the contribution to the magnetic susceptibility from impurities, and to therefore establish the intrinsic magnetic behavior in quantum spin liquid materials. In addition, electron spin resonance enables the study of the coupling between defect spins and the intrinsic lattice spins, thus opening the possibility of the study of the role of defects in quantum spin liquid compounds. In the defects section later in this review, we explore examples in which electron spin resonance has been used to study imperfections in spin liquid materials.

\subsubsection{Nuclear Magnetic Resonance}

Nuclear magnetic resonance is a technique similar to electron spin resonance, but instead measures transitions between nuclear spin energy levels instead of electrons spin levels. The local nature of nuclear energy levels makes this technique a unique and specialized tool for the detailed inspection of defects, and whether or not they mask lattice spins, how spin fluctuations behave as a material reaches a critical point, and how the dynamics of lattice and spin degrees of freedom influence intrinsic magnetic ground states in quantum spin liquids. The line shapes of nuclear magnetic resonance data give us a sense of whether or not local spin susceptibilities give large and uniform contributions to the overall bulk susceptibility. A comparison of nuclear spin-lattice relaxation rates can provide for lattice freezing due to the orientational disorder of chemical bonds. Single or dual axis magic angle spinning is usually essential to eliminate crystal field broadening effects and enable detailed measurements. None-the-less, the nuclear magnetic resonance technique is a powerful tool used to study quantum spin liquid materials, especially when used in conjunction with other measurements such as magnetization and electron spin resonance. 
    
\subsubsection{Muon Spin Spectroscopy}

Muons are particles that are heavier than electrons and carry the same but opposite charge, i.e. $-1e$, and have the same spin $S = 1/2$. They are generated at large facilities by smashing high speed protons into target materials, which produces particles that quickly decay into muons. These muons are then fired at samples, where they stop at interstitial sites in the lattice and are, as a consequence of their spin, influenced by local magnetic fields around them. Though not always a gentle probe, their ability to detect small magnetic fields in a crystal environment makes them extremely useful in the study of both magnetic and electronic materials. 

At high temperatures, magnetic moments fluctuate at high speeds, thus only weakly depolarizing implanted muons before they decay. At low temperatures, however, the slower fluctuations of magnetic moments result in faster depolarizations. These depolarizations are measured in terms of muon decay asymmetry as a function of time on the order of microseconds, though they can happen at shorter time scales. In the case of magnetic order, oscillatory behavior is observed in the asymmetry data, and thus the absence of oscillatory behavior at all temperatures is already the first sign that a material does not order magnetically and can, under the right conditions, be a quantum spin liquid candidate. 

The depolarization (decay asymmetry) of muons can be fit to an exponential decay equation $A(t) = A(0)e^{-(\lambda t)\beta}$, where $\lambda$ is the depolarization rate  and $\beta$ is known as the stretching coefficient, which has a value of $\beta \approx 1/3$ in the case of systems with either glassy or magnetically ordered states\cite{PhysRevLett.72.1291, PhysRevB.32.7384}. Plateaus have been observed in the depolarization rate $\lambda$ for some spin liquid compounds \cite{PhysRevLett.98.077204, PhysRevLett.109.037208, PhysRevLett.117.097201, PhysRevB.93.214432}, and thus perhaps serve as an indicator of quantum spin liquid behavior, though an explanation as to why plateaus occur has not yet been fully provided. Time-field scaling of muon data can provide evidence for power laws that are indicative of exotic dynamics in magnetic materials, and unconventional power laws have been observed in some quantum spin liquid candidate materials, such as \ce{YbMgGaO4} \cite{PhysRevResearch.2.023191}.

The muon spin spectroscopy technique is therefore very useful in the study of quantum spin liquid materials and, through the various parameters obtainable such as the depolarization rate, stretching coefficient, muon implantation location, and time-field scaling parameters, conclusions can be drawn about the ground state of potential quantum spin liquid materials.

\subsubsection{High-angle annular dark-field imaging (HAADF)}

The HAADF technique is a unique tool for atomic-scale characterization of structural defects, especially topological defects such as dissociated superdislocations and anti-phase boundaries. Such characterization of imperfections becomes profoundly important in quantum spin liquids where subtle magnetic exchange interactions are possibly susceptible to minute perturbations, resulting in discrepancies in reported magnetic ground states of a material due primarily to differences in physical samples, due to varying purity levels of starting materials, rigor of synthetic techniques, and sometimes, luck. The presence of topological defects can induce significant strain, which in turn causes distortion of coordination environment of a magnetic cation and can have a cascading effect on the entire lattice, especially at extremely low temperatures. These defects potentially result in a disruption of long-range structural perfection of the material and lifting degeneracy of the spin states, thereby ultimately disturbing the delicate magnetic ground states. In a later section of this review we discuss the potential impact of defects in quantum spin liquid compounds, and the importance of characterizing all kinds of defects in order to conclusively determine magnetic ground states in materials.

\section{\underline{Materials}}

In this section, we discuss a number of specific quantum spin liquid candidate materials and discuss their crystallographic structures, some of their reported physical properties, and try to establish structure-property relationships that explain their fascinating behavior. We will divide the classification of such materials in terms of their dimensionality, starting with linear one-dimensional materials. For two-dimensional materials, we will discuss the triangular, honeycomb, and kagome lattice materials and give several examples of material instantiations of each. Lastly, for three-dimensional structures, we will discuss the diamond lattice and, more importantly, the pyrochlore lattice of corner-sharing tetrahedra, which along with the two-dimensional kagome lattice, has served as a shining beacon of quantum spin liquid phenomena.  

\subsection{1D - Linear Chains}
In the one-dimensional case, the only subclass of materials are linear chains. In real materials, these magnetic cations do not form a perfect one-dimensional chain but rather arrange in a quasi-one-dimensional arrangement, with ligands such as halides or O$^{2-}$ forming the magnetic exchange pathway between them. One-dimensional quantum magnets are a particularly exciting playground for quantum spin liquids, as the idealized theoretical model with an exactly known quantum ground state, i.e. a macroscopically entangled quantum spin liquid state, has been solved. Thus, under the right conditions, linear chains of $S = 1/2$ will always result in a quantum spin liquid. In this section we also briefly discuss Haldane chains of $S = 1$ cations, which aren't necessarily quantum spin liquid compounds, but show inter-neighbor singlet formation similar to a valence bond solid. We therefore find it of educational value to discuss this rather interesting class of material with interesting topological properties.

\subsubsection{Linear, $S = 1/2$ Chains}

\noindent \textbf{CuGeO$_{3}$} \par
            
\ce{CuGeO3} was one of the first realizations of an antiferromagnetic $S = 1/2$ Heisenberg chain. In Heisenberg magnetic materials, spins can point in any direction in three-dimensional space, as opposed to an Ising magnet, where they can only point in one of two directions, or an $xy$-magnet, where they can only point anywhere along a plane (usually defined as the $xy$ plane). It crystallizes in the orthorhombic space group $Pmma$ (no. 51), and its unit cell is shown in \textbf{Figure 5 A}. In this material, $S = 1/2$ edge-sharing \ce{CuO6} octahedra cations are arranged as chains along the \textit{c}-axis. The Cu$^{2+}$ cations are 2.94$\mathrm{\AA}$ apart from each other, and are just far apart enough so as to not establish metallic bonds, but close enough that the magnetic exchange is expected to be strong, as shown in \textbf{Figure 5 B}. The \ce{CuO6} are strongly Jahn-Teller distorted, with apical Cu-O bond lengths of 2.76 $\mathrm{\AA}$, and equatorial bond lengths of 1.94 $\mathrm{\AA}$, meaning the $S = 1/2$ unpaired electron sits in the Cu$^{2+}$ $3d_{x^{2}-y^{2}}$ orbital, pointing toward the O$^{2-}$ ligands. The equatorial Cu-O-Cu bond angles are 98.4$^{\circ}$ for all octahedra, which reduces the $d-p$ orbital overlap and weakens magnetic interactions, but results in net antiferromagnetic interactions as per the Goodenough-Kanamori rules. \cite{Kanamori1959, PhysRev.100.564,Goodenough1958}

\begin{center}
    \includegraphics[width=0.7\textwidth]{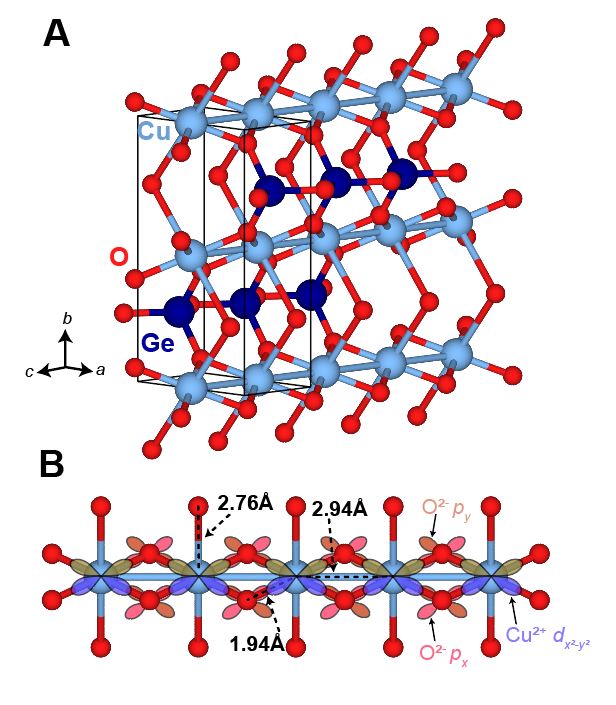} \par
    \small{\textbf{Figure 5.} \ce{GeCuO3}. 
    \textbf{A.} The unit cell of \ce{GeCuO3}}, demonstrating the Cu$^{2+}$ chains along the \textit{c}-axis, separated by \ce{GeO4} tetrahedra. 
    \textbf{B.} The \ce{CuO6} linear chain, with the Cu$^{2+}$ $3d_{x^{2}-y^{2}}$ orbitals shown on the Cu$^{2+}$ sites, which form bonding interactions with $p_{x}$ and $p_{y}$ orbitals from neighboring O$^{2-}$ anions.
\end{center}

Deep blue crystals of \ce{CuGeO3} were grown from floating zone technique with rods prepared from stoichiometric ratios of CuO and \ce{GeO2} under 1 atm \ce{O2} atmosphere \cite{1999JCrGr.198..593R,PhysRevB.54.469}. The chemistry of both CuO and \ce{GeO2} is not amenable to the formation of oxygen vacancies, implying that crystals grown under these conditions are likely to have a low concentration of such vacancies.

The magnetic interaction strength is approximately $J/k_{B} = 88$ K in \ce{GeCuO3}, where $J$ is the exchange interaction, a value that quantifies the strength of the magnetic interaction in eV.  however, it exhibits a magnetic phase transition to a long-range ordered spin-Peierls, non-magnetic state at $T_{SP} = 14$ K \cite{PhysRevLett.70.3651}, as demonstrated by a sudden and sharp drop in magnetization. Though this material, therefore, does not possess a quantum spin liquid ground state at the lowest temperatures, we nevertheless highlight it because it demonstrates an example of a trivial ground state in an $S = 1/2$ chain compound. The spin-Peierls distortion observed arises from a structural instability that is coupled to the magnetic interaction in this material, and is a common distortion in linear chain compounds \cite{PhysRevB.101.235107}. The symmetry of the unit cell is reduced to orthorhombic $Cmce$ (no. 64), and Cu$^{2+}$ forming linear chains are no longer equidistant from each other, but now are paired up into non-magnetic singlets. This distortion was confirmed by the inelastic neutron scattering \cite{PhysRevB.50.6508}, where an excitation gap consistent with a spin-Peierls distortion is seen on the order of $\Delta E = 1$ meV that decreases exponentially below $T_{SP}$. The magnetic excitation spectrum clearly reveals significant energy dispersion and the anisotropy of the magnetic exchange interactions.

\noindent \textbf{Sr$_{2}$CuO$_{3}$} \par
             
Another example of an antiferromagnetically interacting, $S = 1/2$ Heisenberg chain is \ce{Sr2CuO3}. It crystallizes in the orthorhombic $Immm$ (no. 71) space group, which has higher symmetry compared to \ce{GeCuO3}, and its structure is shown in \textbf{Figure 6 A}. In this material, Cu$^{2+}$ are fairly isolated from each other, as shown in \textbf{Figure 6 B}. In contrast to the \ce{CuO6} octahedra in \ce{GeCuO3}, the square-planar \ce{CuO4} units in \ce{Sr2CuO3} interact very strongly with each other with a magnetic interaction strength on the order of $J/k_{B} = 2200$ K. This is likely due to the direct overlap of the Cu$^{2+}$ $3d_{x^{2}-y^{2}}$ orbitals with the O$^{2-}$ $p_{x}$ orbitals, with a perfectly straight Cu-O-Cu bond angle of 180$^{\circ}$. As such, the antiferromagnetic interactions can be maximized and are extraordinarily strong. 

The material undergoes a phase transition to a long-range antiferromagnetically ordered state at $T = 5$ K, though above this temperature, there is believed to be a wide range of temperatures over which \ce{Sr2CuO3} displays physical properties consistent with a quantum spin liquid, including a sizable thermal conductivity attributed to spinons \cite{PhysRevB.62.R6108}, a drop in the magnetic susceptibility attributable to the formation of a valence bond non-magnetic ground state \cite {PhysRevLett.76.3212}, and the observation of spinon excitations in inelastic experiments \cite{RN15}. As such, \ce{Sr2CuO3} presents an example of a material which, despite a magnetically ordered ground state, exhibits spin liquid behavior over a range of temperatures instead. As a result, is an example of a linear $S = 1/2$ chain compound where the crystallographic structure enhances the magnetic interactions and pushes them into an interesting, quantum regime. 
             
\begin{center}                
    \includegraphics[width=0.7\textwidth]{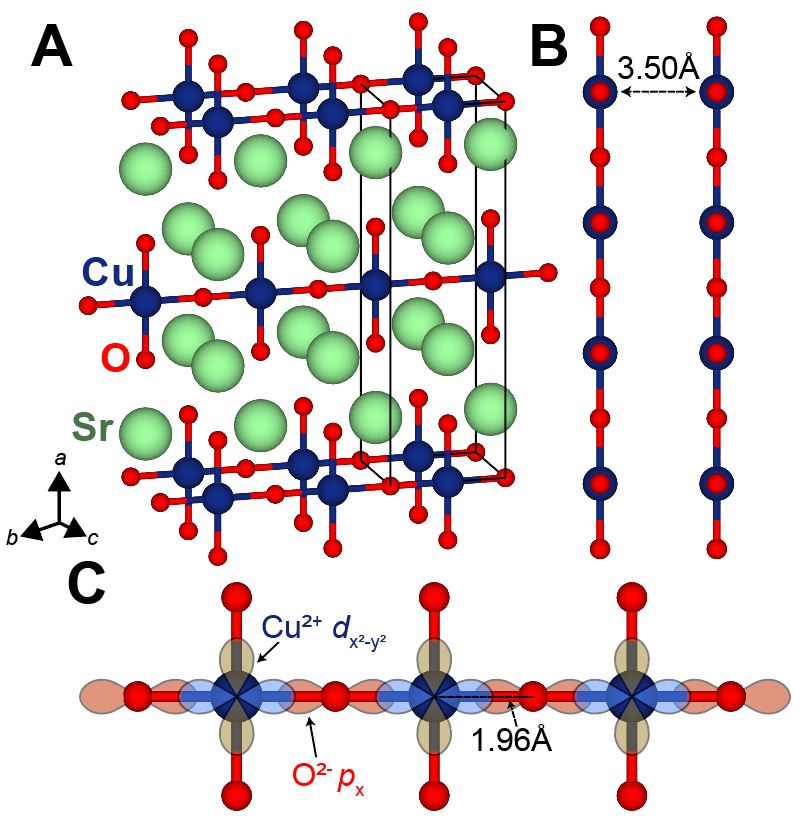} \par
    \small{\textbf{Figure 6.} \ce{Sr2CuO3} 
    \textbf{A.} The unit cell of \ce{Sr2CuO3}, demonstrating the linear arrangement of \ce{CuO4} square-planar units, along the \textit{b}-axis. 
    \textbf{B.} A side-ways view of the linear Cu$^{2+}$ chains, demonstrating their isolation from each other at 3.50 $\mathrm{\AA}$ apart. 
    \textbf{C.}} A close-up view of the Cu$^{2+}$ linear chain, highlighting the orbitals involved in bonding, i.e. the Cu$^{2+}$ $3d_{x^{2}-y^{2}}$ orbitals and the O$^{2-}$ $p_{x}$ orbitals.
\end{center}

 Other interesting $S = 1/2$ one-dimensional, antiferromagnetically interacting materials include KCuF$_{3}$\cite{PhysRevLett.70.4003,PhysRevB.44.12361, PhysRevB.52.13368}, \ce{Cs4CuSb2Cl12} \cite{PhysRevB.101.235107}, \ce{SrCuO2} \cite{Kim2006, PhysRevB.56.15589}, and \ce{Sr14Cu24O41} \cite{Blumberg2002}.

\subsubsection{Haldane, $S = 1$ Chains}

\noindent \textbf{\ce{CsNiCl3}} \par

A one-dimensional chain of integer spins, such as a chain of Ni$^{2+}$ $S = 1$ ions, can exhibit fascinating behavior in comparison to it's half-integer counterpart. The total, combined integer spin on each site can fractionalize, resulting in each site having an odd number of $S = 1/2$ spins that behave independently. In the $S = 1$ case, these independent spins can form valence bonds with their nearest neighbors, effectively creating singlets between magnetic moments. This ground state is therefore a non-magnetic, gapped one, as shown in \textbf{Figure 7 A}, and is known as an Affleck-Kennedy-Lieb-Tasaki (AKLT) chain \cite{Haldane1983, PhysRevLett.50.1153,PhysRevLett.59.799}. An excitation, such as a single spin flip, would have an energy cost associated with it on the order of $\Delta \sim J$, where $J$ is the nearest neighbor exchange interaction in eV. The first excited states are propagating bond triplets, which are very similar to the spinon excitations in quantum spin liquids, as both are a chargeless, spinful, itinerant excitation. 

\ce{CsNiCl3} was the first example evident for a Haldane gap in a $S=1$ Heisenberg antiferromagnetic chain \cite{Steiner1987}. Single crystals of this material were grown via the vertical Bridgman technique, which involves melting a stoichiometric amount of CsCl and \ce{NiCl2} and passing it through a hot zone slowly, so that only the hottest part can melt. By moving it slowly away from the hot zone, a crystal can form and continue to grow as more molten material is cooled and allowed to grow off it. This technique is an effective crystal growth technique, as molten zone refinement can occur as well, whereby impurities originally in the starting materials are pushed upward along the length of the grown crystal, and thus the material can be more or less purified during the process.

The crystal structure of \ce{CsNiCl3} is shown in \textbf{Figure 7 B}. It crystallizes in the hexagonal space group $P6_{3}/mmc$ (no.194). The linear chains consist of face-sharing, undistorted \ce{NiCl6} octahedra that extend along c, and a close-up of one is found in \textbf{Figure 7 C}, where the Ni$^{2+}$-Ni$^{2+}$ distance is $2.97 \mathrm{\AA}$. Specific heat measurements on crystals of this material, reveal a linear magnetic contribution in the low temperature region, which is an unexpected finding for any magnetic material \cite{Moses_1977}. The intrachain exchange interaction $J/k_B$ was deduced from magnetic susceptibility measurements to be $J/k_{B} = 14$ K \cite{deJongh1974}, a result confirmed by heat capacity. Inelastic neutron scattering found a gap in the excitations spectrum, as expected \cite{Steiner1987, Kakurai1991}. \ce{CsNiCl3} therefore presents an example of a material where spins form a non-magnetic ground state due to the strength of their interactions, with itinerant excitations that can propagate along the length of the chain. This is very similar to a quantum spin liquid, except for the lack of long range quantum entanglement that characterizes spin liquids. The singlets in an AKLT chain would have to be resonating and distributed across the chain in order to be classified as part of a resonating valence bond solid, i.e. a quantum spin liquid.
            
\begin{center}                
    \includegraphics[width=0.6\textwidth]{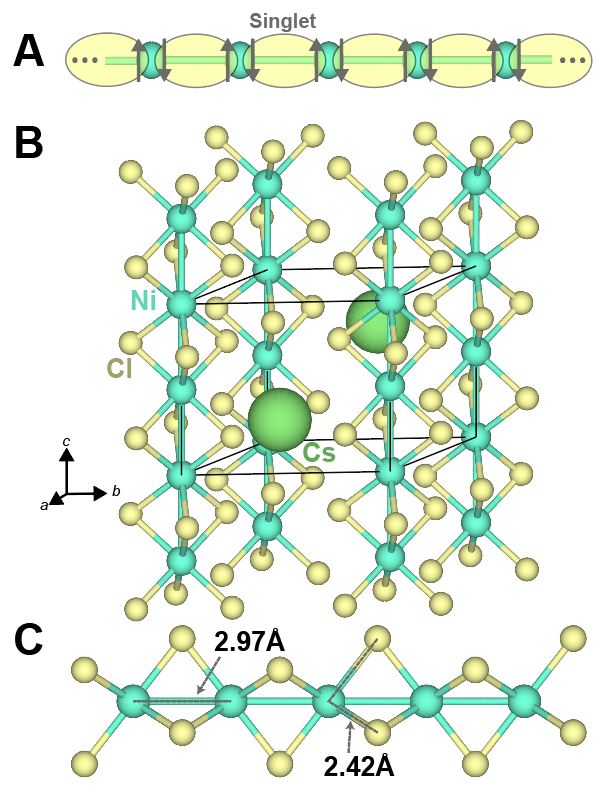} \par
    \small{\textbf{Figure 7.} \ce{CsNiCl3}. 
    \textbf{A.} An AKLT chain showing the formation of singlets between nearest neighboring spins.
    \textbf{B.} The structure of \ce{CsNiCl3}, demonstrating the \ce{NiCl6} octahedral chains along the \textit{c}-axis.
    \textbf{C.} A chain of face-sharing \ce{NiCl6} octahedra, with equivalent Ni-Cl bond lengths of $2.42 \mathrm{\AA}$ and Ni-Ni distances of $2.97 \mathrm{\AA}$}.
\end{center}

\subsection{2D}

There are many geometrically frustrated two-dimensional lattices that can assist in the generation of a quantum spin liquid ground state. In this review, we will discuss compounds with triangular, hexagonal honeycomb, and kagome sublattices of $S = 1/2$ or $S = 1$ magnetic ions, as these systems tend to have strong quantum fluctuations and thus may exhibit interesting quantum phenomena. It is worthwhile to note that these structures highly interrelated, and can derived from the triangle, the simplest frustrated two-dimensional unit, as shown in \textbf{Figure 8}. Note that triangles are not always necessary for frustrated interactions, as even a square lattice can exhibit frustration if next-nearest-neighbor interactions are comparable or stronger than nearest neighboring ones, as shown in \textbf{Figure 2 C}. \cite{Mustonen2018, ANDERSON1987}. 

\begin{center}                
    \includegraphics[width=1\textwidth]{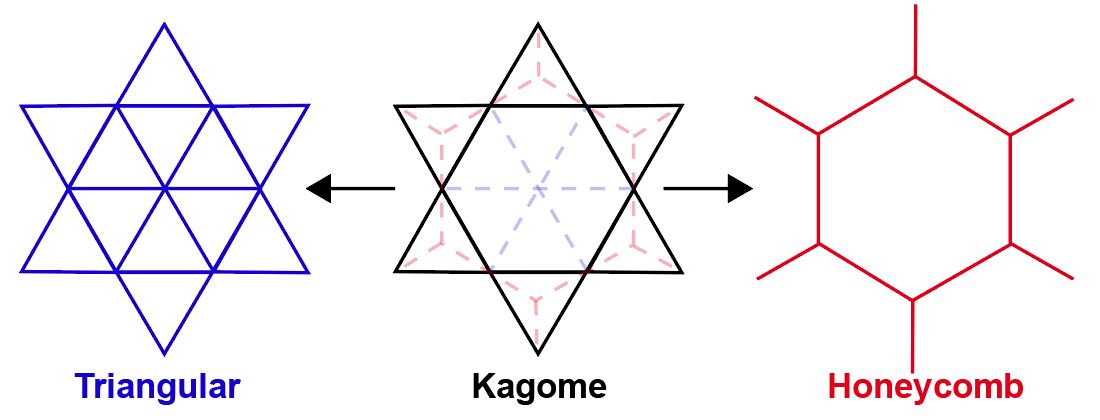} \par
    \small{\textbf{Figure 8.} The three two-dimensional lattice types discussed in this review are all related and can all be derived from each other.}
\end{center}
\subsubsection{Triangular Lattice}

The triangular lattice is the most basic of the two-dimensional frustrated lattices based on the triangular motif. It is build up of edge sharing triangles, and can be found in many real-world materials. In fact, the triangular lattice was the first lattice to be considered as a candidate quantum spin liquid host lattice by P. W. Anderson in 1973 \cite{Anderson_1973}, and has since been the subject of intense theoretical and experimental investigations. Several technologically important materials contain this sublattice, such as the battery materials \ce{LiCoO2} and \ce{NaCoO2} \cite{Kikkawa1986, Scrosati2000}

\noindent \textbf{\ce{NiGa2S4}} \par

\ce{NiGa2S4} is an interesting magnetic material that has well-separated triangular-lattice Ni$^{2+}$ $S = 1$ layers. It crystallizes in trigonal space group $P\bar{3}m1$ (no. 164), and has alternating layers of \ce{NiS6} octahedra and \ce{GaS4} tetrahedra, as shown in \textbf{Figure 9 A}. The \ce{NiS6} octahedra have Ni-S bond lengths of $2.42 \mathrm{\AA}$ and remain undistorted with cooling, as expected from the absence of a Jahn-Teller instability and weak spin-orbit coupling. \ce{GaS4} layers are separated from each other by a van der Waals gap, which further weakens interactions perpendicular to those layers, along \textit{c}. As shown in \textbf{Figure 9 B}, the Ni$^{2+}$ $S = 1$ cations sit on a triangular lattice and are $3.62 \mathrm{\AA}$ apart from each other. As such, \ce{NiGa2S4} presents an ideal two-dimensional triangular lattice material that given its low spin, could result in quantum spin liquid physics. However, due to the presence of a van der Waals gap, stacking faults, i.e. misalignment of the \ce{NiGa2S4}-\ce{NiGa2S4} layering can readily result in seemingly exotic intrinsic behavior, but care must be taken to take these non-negligible defects into account \cite{PhysRevB.77.054429}. 

\begin{center}                
    \includegraphics[width=0.8\textwidth]{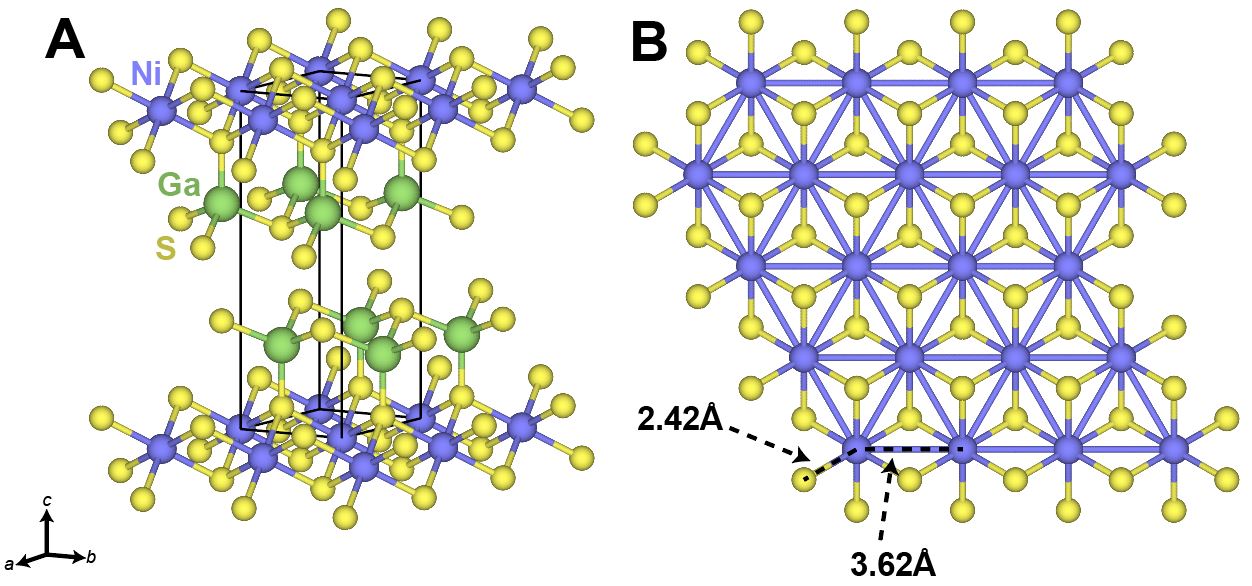} \par
    \small{\textbf{Figure 9.} \ce{NiGa2S4}.
    \textbf{A.}The unit cell of \ce{NiGa2S4}, showing the two-dimensional layers of \ce{NiS6} edge-sharing octahedra, separated from each other by a two \ce{GaS4} layers and a van der Waals gap.
    \textbf{B.} A top-down view of the Ni$^{2+}$ triangular sublattice.}
\end{center}

\ce{NiGa2S4} polycrystalline samples are usually prepared by through a high temperature solid state reaction of stoichiometric amounts of \ce{NiS} and $\alpha$-\ce{Ga2S3} at $1000^{\circ}$ C, and single crystals can be obtained via the chemical vapor transport technique using \ce{I2} as the transport agent \cite{Lutz1986}. While high temperatures can promote the vaporization of sulfur and thus introduce sulfur vacancies into the lattice, the use of the chemical vapor transport acts as a purification method that ensures that crystals that grow are of high quality and of correct composition. However, the impact of a layer-by-layer deposition, such as that likely in a chemical vapor transport growth, likely promotes stacking faults, and thus an alternative technique such as flux growth may prove useful in the mitigation of unwanted defects.

In terms of physical properties, \ce{NiGa2S4} does not shown signs of long range magnetic order down to at least $T = 0.35$ K in measurements of magnetization and heat capacity. \cite{Nakatsuji2005}
Anisotropic measurements of magnetization reveal Curie-Weiss paramagnetic behavior at high temperatures with Weiss temperatures of $\Theta_{W} = -77$ K along both the \textit{ab} and \textit{c} directions, consistent with moderately strong Heisenberg antiferromagnetic interactions. The Heisenberg interactions potentially stem from the fully octahedral, isotropic symmetry of the \ce{NiS6} units. However, a hump is seen at $T = 7$ K susceptibility measured along the \textit{ab} direction, potentially indicative of possible short-range magnetic interactions. Heat capacity measurements reveal a low temperature $C \propto T^{2}$ power law associated with gapless excitations \cite{Nakatsuji2007}. The low temperature heat capacity is also unaffected by the application of large magnetic fields, implying the presence of short-range magnetic interactions that give rise to spin clusters with unbroken crystallographic symmetry.

Neutron scattering experiments on \ce{NiGa2S4} have shown interesting results. Though $``120^{\circ}$ order" is expected for triangular lattices, where spins on each triangular vertex point $120^{\circ}$ away from each other, \ce{NiGa2S4} shows an incommensurate, i.e. violating the crystallographic symmetry, variation of this ordering whereby spins only locally interact with each other and form short range magnetic order at the nanoscale \cite{Nakatsuji2010}. This coherent behavior is indicative of strong quantum fluctuations in \ce{NiGa2S4}, and as such, \ce{NiGa2S4} displays quantum phenomena whereby interactions are very strong but magnetic order is destabilized by quantum fluctuations. As the first realization of Ni$^{2+}$ $S = 1$ on a triangular lattice, this compound sets an example and provides a platform upon which to continue the study of other integer spin triangular lattice materials

\noindent \textbf{YbMgGaO$_{4}$} \par

Another interesting triangular lattice compound is \ce{YbMgGaO4}, which is especially interesting because it is built up of rare-earth Yb$^{3+}$ $j = 1/2$ ions and is a rare example of a triangular lattice of lanthanide ions \cite{RN18}. Most triangular lattice sublattices in materials are composed of transition metal cations that are prone to strong single-ion anisotropies due to the anisotropies and high level of delocalization of $d$-orbitals. The magnetism in rare-earth compounds, however, arises from more localized lanthanide $f$-orbitals, and are therefore less susceptible to single-ion anisotropies. Instead, strong spin-orbit coupling, arising from their large atomic numbers, is usually the deciding factor in establishing what the ground state is for each rare-earth element. Lattice anisotropies, along with spin-orbit coupling, play a large role in establishing a ground state, quantum or otherwise. 

\ce{YbMgGaO4} was discovered fairly recently, and crystallizes in trigonal space group $R\bar{3}m$ (no. 166) \cite{RN18}. The unit cell for \ce{YbMgGaO4} is shown in \textbf{Figure 10 A}. Layers of \ce{YbO6} are separated from each other by layers of trigonal bipyramidal \ce{MgO5} and \ce{GaO5} units, where the Mg$^{2+}$ and Ga$^{3+}$ sites are completely disordered. Despite the disorder, the Yb$^{3+}$ sublattice remains intact, at least in measurements of global crystallographic symmetry, with constant Yb-Yb distances of $3.40 \mathrm{\AA}$. The average bond distance between Mg$^{2+}$ and Ga$^{3+}$ with O$^{2-}$ anions that are directly bonded to Yb$^{3+}$ cations is $1.89 \mathrm{\AA}$, as shown in \textbf{Figure 10 B} and \textbf{C}. In \textbf{Figure 10 D}, we see how this may have an influence on the magnetic \ce{YbO6} octahedra, as the bond distances between the octahedra and Mg$^{2+}$ or Ga$^{3+}$ units can change with magnitude $\Delta \sim 0.1 \mathrm{\AA}$. Note that though the disorder in this figure has been simplified by assuming an alternating Mg-Ga-Mg-Ga pattern, this is not the case in real life. Nevertheless, this figure highlights how structural site mixing of Ga and Mg sites can influence the underlying Yb-O bonding, which in turn may have an effect on the magnetism. Though bonding anisotropies play a smaller role in rare-earths such as Yb$^{3+}$, differences in their chemical bonding environments, such as the ones seen here, can potentially lead to enhancements of their relevance.

\begin{center}                
    \includegraphics[width=0.8\textwidth]{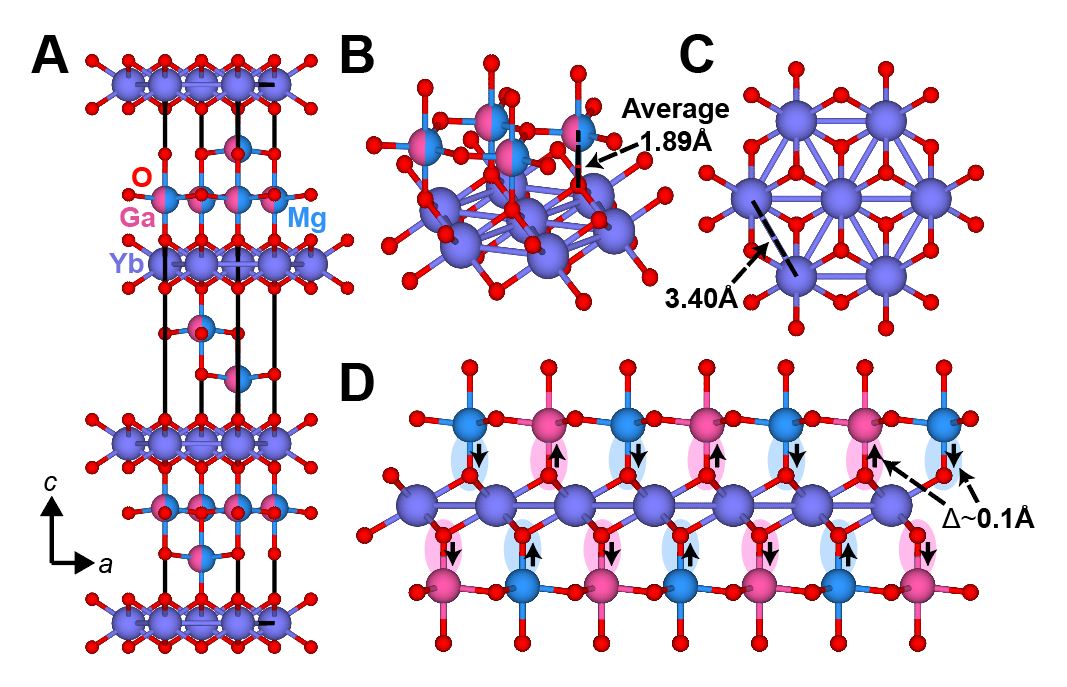} \par
    \small{\textbf{Figure 10.} \ce{YbMgGaO4}. 
    \textbf{A.} A sideways view of the unit cell of \ce{YbMgGaO4}, showing the Yb$^{3+}$ layers separated by \ce{GaO4} and \ce{MgO4} trigonal bipyramidal units.
    \textbf{B.} A schematic showing how the \ce{GaO4} and \ce{MgO4} units are bonded to the Yb$^{3+}$ layer, thus potentially impacting the triangular sublattice with their disorder.
    \textbf{C.} A top-down view of the unit cell, showing the triangular lattice of Yb$^{3+}$ rare-earth cations that are $3.40 \mathrm{\AA}$ away from each other.
    \textbf{D.} A schematic showing how the Mg$^{3+}$ and Ga$^{3+}$ site mixing might affect the \ce{YbO6} octahedra. Though this drawing shows a Mg-Ga-Mg-Ga type of ordering, real-world samples show complete disorder. Bond distances change with amplitude $0.1 \mathrm{\AA}$, potentially impacting the ground states of the Yb$^{3}$ units and the nature of their interactions with each other.}
\end{center}

Single crystals of \ce{YbMgGaO4} can be synthesized using the optical floating zone technique, through melting stoichiometric mixtures of \ce{Yb2O3}, \ce{MgO}, and \ce{Ga2O3}. The stability of these precursors as stoichiometric oxides ensures a complete stoichiometry of the product, and thus, the floating zone technique is the superior synthetic procedure for the generation of single crystals of this material. The physical properties of \ce{YbMgGaO4} are consistent with expectations for quantum spin liquids, yet subtle differences have stirred controversy. Large single crystals have been grown via the optical floating zone technique, by melting stoichiometric amounts of the metal oxides. Via measurements of heat capacity and magnetization, no magnetic order was observed down to $T = 0.06$ K \cite{RN18}. The interaction strength from the magnetization, via the Weiss temperature, is of order $\Theta_{W} = -4$ K, which implies very weak yet antiferromagnetic interactions. This is not unexpected in a rare-earth material with highly localized $f$-electron magnetism, and is unsurprising from a chemical perspective. The heat capacity shows a $C \propto T^{2/3}$ power law dependence, which is the expectation for a ``$U(1)$" quantum spin liquid, in which, fractionalization of electrons give rise to itinerant spinon excitations. These spinons have been shown to display a continuum of excitations in inelastic scattering, thus further demonstrating quantum spin liquid behavior \cite{RN35}.

Measurements of the thermal conductivity can provide evidence for the presence of itinerant spinons, as these are expected to carry heat. Such measurements were performed on \ce{YbMgGaO4} and, contrary to expectations based on aforementioned measurements, reveal the absence of any magnetic thermal conductivity, which strongly casts doubts on the proposed $U(1)$ quantum spin liquid ground state \cite{PhysRevLett.117.267202}. Some theoretical work has emerged claiming that the site mixing between Mg$^{2+}$ and Ga$^{3+}$ sites induces a structural disorder in the magnetic sublattice that can mimic spin liquid behavior, and thus result in physical properties such as those seen in magnetization, heat capacity, and neutrons scattering \cite{PhysRevLett.119.157201}. While the ground state of \ce{YbMgGaO4} remains a subject of active controversy and discussion, it has become clear that Mg$^{2+}$ and Ga$^{3+}$ site mixing plays a large role in the physics of this material and cannot thus be neglected.

\noindent \textbf{$6H$-\ce{Ba3CuSb2O9}}

Thus far we have discussed spin liquids as systems composed of large degeneracies of spin interactions. There exists, however, a subclass of spin liquids known as spin-orbital liquids, where both spin and orbital degeneracies give rise to an exotic, quantum ground state. From a chemical perspective, a spin-orbital liquid can emerge when Jahn-Teller-active cations form a frustrated lattice, yet do not undergo a Jahn-Teller distortion. In this review present two examples of such a phenomenon, $6H$-\ce{Ba3CuSb2O9} and \ce{FeSc2S4}, though we discuss the latter later in the context of frustrated three-dimensional diamond lattices.

\ce{Ba3CuSb2O9} is polymorphic and at high temperatures crystallizes in centrosymmetric, hexagonal space group $P6_{3}mmc$ (no. 194). In some samples, a low-temperature structure has been observed with orthorhombic, $Cmcm$ (no. 63) symmetry, where the \ce{CuO6} octahedra, which are undistorted in the hexagonal structure, have undergone a Jahn-Teller distortion and elongated along their local \textit{z}-axes \cite{Katayama9305}. When in the hexagonal structure, this material is referred to as $6H$-\ce{Ba3CuSb2O9}, and its unit cell is shown in \textbf{Figure 11 A}, which shows where the so-called Cu-Sb ``dumbbells" can be found. These dumbbells of Cu$^{2+}$ and Sb$^{5+}$ form a regular, stacked triangular lattice, as shown in \textbf{Figure 11 B}. Unfortunately, these dumbbells are prone to internal substitutional disorder that varies strongly among samples \cite{Khl1977}. 

One such dumbbell is shown in \textbf{Figure 11 C}. Though the high temperature structure retains three-fold Cu$^{2+}$ local symmetry and thus confirms the absence of a cooperative Jahn-Teller distortion affecting the entire lattice, the low temperature orthorhombic structure does not possess three-fold symmetry and displays differences in equatorial and apical Cu-O bond lengths consistent with a Jahn-Teller distortion.

\begin{center}              
    \includegraphics[width=1\textwidth]{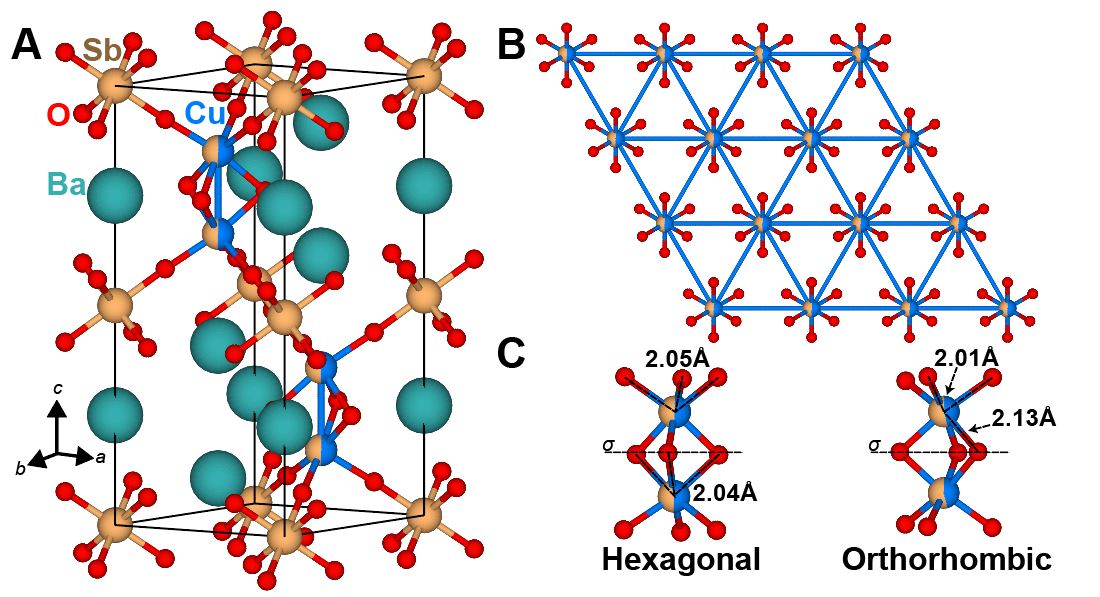} \par
    \small{\textbf{Figure 11.} $6H$-\ce{Ba3CuSb2O9}.
    \textbf{A.} The unit cell of $6H$-\ce{Ba3CuSb2O9}, showing the position of the Cu-Sb dumbbells and their separation from each other.
    \textbf{B.} A top-down view of the unit cell, highlighting the triangular lattice of dumbbell units.
    \textbf{C.} The two different dumbbell configurations for the hexagonal (left) and orthorhombic (right) structures. While the orthorhombic structure arises due to a cooperate Jahn-Teller distortion at the Cu$^{2+}$ sites, the Cu$^{2+}$ octahedra in the hexagonal structure are undistorted.}
\end{center}

It seems, however, that while some samples do undergo a structural phase transition from hexagonal $P6_{3}/mmc$ to orthorhombic $Cmcm$ upon cooling, some samples show no transition whatsoever down to at least $T = 20$ K \cite{Katayama9305}. Even in samples that undergo a transition, a modest volume fraction remains hexagonal. The origins of this are largely unexplored but are likely due to the subtle effects of Cu-Sb site mixing within dumbbells. The ionic radius difference between Cu$^{2+}$ and Sb$^{5+}$ is on the order of $0.1 \mathrm{\AA}$ according to the Shannon ionic radii \cite{Shannon1970},
and there is therefore little energetic cost to mixing between these sites.

Nevertheless, despite the disorder in this material, $6H$-\ce{Ba3CuSb2O9} possesses physical properties consistent with quantum spin liquid behavior. It has been showing via measurements of magnetization and heat capacity to not develop long range magnetic order down to $T = 0.2$ K, and has a Weiss temperature of $\Theta_{W} = -55$ K, indicative of strong, net antiferromagnetic interactions. The heat capacity at low temperature shows a large linear contribution, usually attributable to metallic, itinerant contributions, that potentially indicate the presence of heat-carrying spinons. The potentially unquenched orbital degree of freedom in undistorted Cu$^{2+}$ octahedra may then give rise to not just spin but also orbitally entangled states, and there have been some experimental results providing evidence for this proposal \cite{Man_2018}. Nevertheless, due to the extensiveness of defects in this material, further materials characterization and studies of the chemistry of this material are needed to truly establish what the ground state is.

Triangular lattice materials therefore offer an enticing avenue towards the realization of exotic quantum spin liquid phenomena. Other exciting and emerging compounds include the \ce{NaYbX2} family, with X = O, S, and Se \cite{PhysRevB.98.220409, 2003.09859, Sichelschmidt2019,PhysRevB.100.144432,PhysRevB.99.180401,2002.04772}. The materials contain perfect triangular lattices of Yb$^{3+}$, similar to YbMgGaO4, potentially with less structural disorder, though with greater chance for other types of defects (e.g. oxygen inclusion in the chalcogenides).
\subsubsection{Honeycomb Lattice}

The honeycomb lattice is not directly derived from the triangular motif, but has been shown to be a sublattice that leads to long range frustration and can give rise to very exotic quantum spin liquid states. The Kitaev model has become extremely popular in the field of quantum spin liquid materials, as it is an exactly solvable model for the honeycomb lattice that possesses the quantum spin liquid state as a possible ground state for a material with the right balance of exchange interactions \cite{PhysRevLett.102.017205, PhysRevLett.105.027204}. Honeycomb lattice materials, therefore, present a unique opportunity for the isolation of quantum spin liquid ground states, as they can be realized through the clever use of chemical principles to obtain just the right honeycomb structures necessary.

\noindent \textbf{$\alpha$-\ce{RuCl3}} \par

It is interesting that among the honeycomb quantum spin liquid candidates, the arguably most popular one is the binary compound ruthenium trichloride \ce{RuCl3}, which is probably the most widely used source of ruthenium in chemistry, and is readily commercially available from all chemical suppliers. \ce{RuCl3} is polymorphic and has two primary phases, $\alpha$-\ce{RuCl3} and $\beta$-\ce{RuCl3} \cite{Hyde1965}. $\beta$-\ce{RuCl3} can only be prepared at low temperatures and under very specific conditions, and irreversibly converts to $\alpha$-\ce{RuCl3} above approximately $500^{\circ}$ C. It therefore naturally follows that $\alpha$-\ce{RuCl3} single crystals are grown primarily using high temperature techniques, the most popular of which is the vapor transport technique, owing to the preponderance of this material (and other simple halides) to sublimation. While $\beta$-\ce{RuCl3} is diamagnetic, $\alpha$-\ce{RuCl3} is not, and shows rich physics owing to its honeycomb lattice of Ru$^{3+}$ ions.

$\alpha$-\ce{RuCl3} crystallizes in the non-centrosymmetric, trigonal space group $P3_{1}12$ (no. 151). A side view of the unit cell is shown in \textbf{Figure 12 A}. \ce{RuCl3} layers, built up of \ce{RuCl6} octahedra and arranged in a honeycomb lattice, are separated from each other by van der Waals gaps. The honeycomb arrangement of the Ru$^{3+}$ octahedra are shown in \textbf{Figure 12 B}, are separated from each other by a van der Waals gap that is approximately $3.5 \mathrm{\AA}$ in size. Each \ce{RuCl6} octahedron is nearly ideal, with six Ru-Cl bond lengths of $2.45 \mathrm{\AA}$. The Cl-Ru-Cl bond angles within an octahedron are very close to $90^{\circ}$, though small deviations are observed. There are two unique Ru sites and three unique Cl sites. Since there is a large gap between the layers, distortions in the \ce{RuCl6} octahedra are likely the result of single-ion anisotropies that are inherent to the Ru $4d$ energy levels, though they could also be the result of packing and steric effects. Due to strong spin-orbit coupling, arising from ruthenium's relatively high atomic number, Ru$^{3+}$ ions are effectively $j = 1/2$ units, as aforementioned in the inelastic neutron scattering experimental section of this review. The small $j = 1/2$ moment makes $\alpha$-\ce{RuCl3} more amenable to quantum behavior as smaller moments are more susceptible to quantum fluctuations and can more readily avoid long range order compared to their larger moment counterparts.

\begin{center}                           
    \includegraphics[width=1\textwidth]{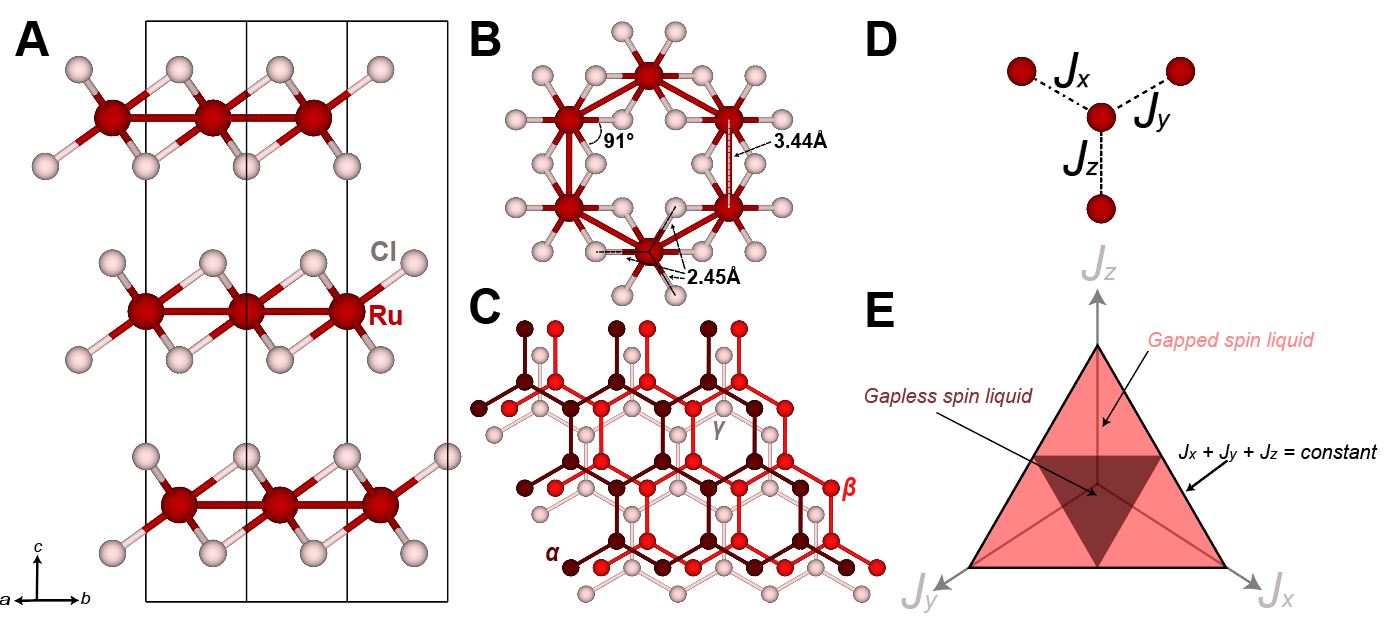} \par
    \small{\textbf{Figure 12.} $\alpha$-\ce{RuCl3}.
    \textbf{A.} A sideways view of the unit cell of $\alpha$-\ce{RuCl3}, demonstrating the van der Waals gap between the \ce{RuCl6} honeycomb layers.
    \textbf{B.} A top down view of a hexagon of \ce{RuCl6} octahedra, showing pertinent bond lengths and angles. The angles between Ru$^{3+}$ cations are exactly $120^{\circ}$.
    \textbf{C.} A top down view of the unit cell, showing the $\alpha-\beta-\gamma$ layering of the honeycomb sublattices that facilitate the presence of stacking faults.
    \textbf{D.} A schematic of four ions at a honeycomb vertex. The center ion has three exchange interactions with its nearest neighbors.
    \textbf{E.} The Kitaev model for the honeycomb lattice, showing two possible quantum spin liquid ground states based on the magnitudes of the three exchange interactions shown in \textbf{D}.}
\end{center}
    
 $\alpha$-\ce{RuCl3}, however, develops long range antiferromagnetic order at low temperature, as evidenced by neutron diffraction, magnetization, and heat capacity. \cite{Banerjee1055,PhysRevB.91.144420}. The ordering temperature varies from $T_{N} = 7$ K in bulk samples to $T_{N} = 14$ K in monolayer samples, that have been obtained by exfoliation, which is enabled by the extremely weak interlayer van der Waals bonding environment. The magnetic order in bulk samples has been shown to be of the zigzag type within layers, and the strength of magnetic interactions perpendicular to the layers, i.e. along the \textit{z}-direction, is weakened especially by the stacking fault disorder that is common in van der Waals layered materials. In order to access the potentially interesting quantum spin liquid behavior in this system, magnetic fields must be applied to excite electrons and melt away the low temperature magnetically ordered state. 
 
 The layers in $\alpha$-\ce{RuCl3} are staggered relative to each other, as shown in \textbf{Figure 12 C}. As shown in the figure, the stacking pattern is of $\alpha-\beta-\gamma$ type, and the weak interlayer bonding gives rise to stacking faults in this system, which have been shown to complicate interpretations of the ground state behavior of $\alpha$-\ce{RuCl3}. Sample-dependent variations in the behavior of $\alpha$-\ce{RuCl3} have been observed, potentially caused by variations in stacking faults across samples, though some properties have remained ubiquitous in all samples and have therefore become considered intrinsic properties. 

Nevertheless, $\alpha$-\ce{RuCl3} displays a wealth of fascinating, quantum phenomena in magnetization, heat capacity, neutron scattering, and thermal conductivity, to name a few.\cite{PhysRevLett.114.147201,Banerjee1055,PhysRevLett.120.217205,PhysRevB.95.241112,PhysRevB.94.020407}. The Kitaev model, as shown in \textbf{Figure 12 D} and \textbf{E}, is an exactly solvable model that considers the exchange interactions between cations on a honeycomb lattice, and has, as a solution, a Kitaev quantum spin liquid ground state \cite{PhysRevLett.102.017205, PhysRevLett.105.027204}. This kind of spin liquid, when the three $J$ exchange interactions are fine-tuned to be within specific limits of each other, can give rise to a gapless spin liquid ground state with Majorana fermions as the low-energy excitations. These particles are non-Abelian anyons, meaning that exchanging them not only changes their sign, but also results in a change in their state. As such, braiding Majoranas has been proposed as a basis for topological quantum computing, where exchanging Majorana particles with each other in specific ways is, in effect, a fault-tolerant quantum computation \cite{PhysRevX.6.031016}. Measurements of the thermal quantum Hall effect in $\alpha$-\ce{RuCl3} have demonstrated half-integer quantization, a key indicator of the presence of Majorana excitations in this system \cite{PhysRevLett.120.217205}. $\alpha$-\ce{RuCl3} therefore presents an example of a potentially technologically revolutionary material whose properties arise due to its quantum spin liquid behavior. 

\noindent \textbf{Iridates} \par

The honeycomb iridate compounds comprise a large and growing class of frustrated honeycomb lattice spin liquid candidate materials centered around Ir$^{4+}$, which is isoelectronic to Ru$^{3+}$ but with much stronger spin-orbit coupling, thus resulting in a much more energetically isolated $j = 1/2$ ground state. 

\begin{center} 
    \includegraphics[width=0.8\textwidth]{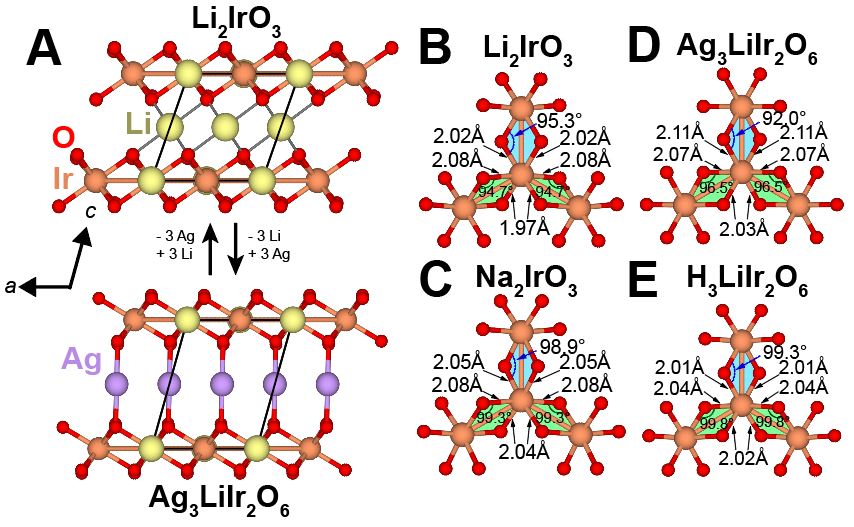} \par
    \small{\textbf{Figure 13.} Iridates.
    \textbf{A.} Sideways views of the unit cells \ce{Li2IrO3} and \ce{Ag3LiIr2O6}, highlighting the Ir$^{4+}$ honeycomb layers, separated by Li$^{1+}$ cations in the former and \ce{AgO2} linear units in the latter. These two structures are related to each other by the addition or removal of either three Ag or Li atoms.  
    \textbf{B - D.} Schematics of the local Kitaev units for four different iridates, \ce{Li2IrO3}, \ce{Na2IrO3}, \ce{Ag3LiIr2O6}, and \ce{H3LiIr2O6}. Due to the monoclinic symmetry of their structures, bond lengths and angles differ between Ir$^{4+}$ cations. The enclosing space between two Ir-O-Ir bonds have been highlighted for all Ir-Ir pairs, and are green for one group of parameters and light blue for another. These isosceles interactions are reflected in the variation in bond lengths and angles.}
\end{center}

\textbf{Figure 13 A} shows the general structure of the honeycomb iridates, with particular emphasis on the arrangement of the Ir$^{4+}$ honeycomb layers. The top part of the figure shows a side view of the unit cell of $\alpha$-\ce{Li2IrO3}, which is a prototypical honeycomb iridate compound. It has layers of \ce{IrO6} octahedra separated by Li$^{1+}$ cations. There are also Li$^{1+}$ cations within the magnetic layer, situated in the middle of each honeycomb hexagon. 

$\alpha$-\ce{Li2IrO3} crystallizes in the monoclinic space group $C2/m$ (no. 12), and it should be strongly emphasized that it is possible to have a honeycomb lattice even in the absence of global hexagonal or trigonal symmetry. In $\alpha$-\ce{Li2IrO3}, the Ir-Ir distances remain constant at $2.98 \mathrm{\AA}$ and bond angles are $120^{\circ}$ from each other $\pm 0.04^{\circ}$, implying a nearly perfect honeycomb lattice. In these materials, however, the lack of global hexagonal or trigonal symmetry necessitates a consideration of the Ir-O bond lengths and angles within each \ce{IrO6} octahedron, since the octahedra do not retain three-fold symmetry. As such, in the iridates, single-ion anistropies and Dzyaloshinskii–Moriya interactions, i.e. antisymmetric exchange interactions, may play a bigger role than they do in systems such as $\alpha$-\ce{RuCl3} \cite{PhysRevB.90.155126}. These anisotropies may then play a role in the stabilization of long range magnetically ordered ground states, and as such the further away bond angles deviate from the ideal $90^{\circ}$, the further the system may be from a Kitaev quantum spin liquid ground state. This acts both ways, however, as it applies a chemical consideration to the accessibility of exotic quantum ground states, and by tuning the crystal concentration of other elements, such as the size of the charge balancing cations in the structure, one can move between magnetic order and a Kitaev quantum spin liquid ground state \cite{PhysRevB.90.161110}. 

In \textbf{Figure 13 A}, we also show the reversible conversion of $\alpha$-\ce{Li2IrO3} to \ce{Ag3LiIr2O6}. $\alpha$-\ce{Li2IrO3} can be synthesized using a high temperature solid state reaction involving stoichiometric amounts of \ce{Li2CO3} and \ce{IrO2}, though the volatility of Li and the non-stoichiometry of \ce{IrO2} allow for non-stoichiometric variations of \ce{Li2IrO3} to form \cite{OMalley2008}. Nevertheless, the \ce{Li2IrO3} product formed, which is stable at low temperature in comparison to its three-dimensional $\beta$-\ce{Li2IrO3} and $\gamma$-\ce{Li2IrO3} polymorphic counterparts, can then be subjected to a low temperature, solution based deintercalation-intercalation reaction to form \ce{Ag3LiIr2O6}, which replaces the Li$^{1+}$ separation layers with linear \ce{AgO2} units. In this conversion, the overall global symmetry remains unchanged, and \ce{Ag3LiIr2O6} retains monoclinic $C2/m$ symmetry, but the interactions become more two-dimensional as the magnetic Ir$^{4+}$ layers move further apart, and the individual Ir-O bond lengths and angles within \ce{IrO6} octahedra change. Compounds such as \ce{Ag3LiIr2O6} and \ce{H3LiIr2O6}, that latter of which has hydrogen as the cation instead of silver, can only be synthesized using low temperature or high pressure techniques, such as hydrothermal techniques, as they are metastable materials that can only be kinetically trapped \cite{RN22}.

\textbf{Figure 13 B - D} show all of the pertinent bond lengths and angles for four different honeycomb iridate compounds. Bond angles are closest to $90^{\circ}$ in \ce{Li2IrO3} and \ce{Ag3LiIr2O6}, and furthest in \ce{Na2IrO3} and \ce{H3LiIr2O6}. However, the distribution in bond lengths is smallest in \ce{Na2IrO3} and \ce{H3LiIr2O6}, especially in the latter. Magnetization measurements indicate that \ce{Li2IrO3}, \ce{Na2IrO3}, and \ce{Ag3LiIr2O6} order antiferromagnetically at $T = 15$ K, $T = 15$ K, and $T = 12$ K respectively, though the magnetic ordering in the former are unconventional \cite{PhysRevB.93.195158}. \ce{H3LiIr2O6} does not show signs of a transition to a long range magnetically ordered state down to in magnetization and heat capacity down to $T = 0.05$ K, and NMR measurements give evidence for a strong sensitivity of the Kitaev quantum spin liquid ground state to defects and crystal imperfections \cite{PhysRevLett.122.047202, RN22, PhysRevB.101.201101}. The honeycomb iridate family of materials is therefore an extensive platform in which to explore structure-property relationships that may lead to Kitaev spin liquid behavior and other exotic, interesting quantum phenomena.

Recently, new proposals have emerged that predict that a Kitaev spin liquid can be stabilized by using high-spin Co$^{2+}$ $S = 3/2$ \cite{PhysRevB.97.014407}. A mechanism where spin-orbit coupling renders a $j = 1/2$ state, such as is the case for Ru$^{3+}$ and Ir$^{4+}$, can transform the $S = 3/2$ state in Co$^{2+}$ into a $j = 1/2$ state, which is primed for quantum spin liquid interactions. This is graphically shown in \textbf{Figure 14 C}. The specific requirements for this model to work are far more restrictive, however, as in order for cobalt honeycomb lattice materials to exhibit Kitaev spin liquid behavior, the Co-O-Co bond angles must be exactly $90^{\circ}$, or as close to it as possible, otherwise conventional order will be stabilized.

Many cobalt honeycomb lattice materials have been synthesized that have the potential to show interesting quantum magnetism. \textbf{Figure 14 A} shows a side view of the unit cell of \ce{BaCo2(AsO4)2}, a cobalt honeycomb material that crystallizes in the trigonal spacegroup $R\bar{3}$ (no. 148). This material has been demonstrated recently to exhibit a weak field-induced nonmagnetic state, potentially consistent with Kitaev behavior\cite{Zhong2020}. Other compounds include delafossite materials similar to the \ce{Ag3LiIr2O6} iridate, such as monoclinic, $C2/m$ (no. 12) \ce{Na3CoSb2O6}, whose unit cell is depicted in \textbf{Figure 14 B} \cite{PhysRevMaterials.3.074405}. This compound exhibits magnetic order, however, thus indicating that the bond angles surrounding each \ce{CoO6} octahedron may not be in the ideal regime necessary for Kitaev spin liquid behavior. Fortunately, chemical tunability may be possible in this family of materials, as various different kinds of charge-balancing cations can be used that could potentially modify the bond angles and realize a Kitaev state, such as in \ce{Li3CoSb2O6} \cite{2006.03724, Brown2019} and \ce{Ag3CoSb2O6} \cite{Zvereva2016}.

\begin{center} 
    \includegraphics[width=0.8\textwidth]{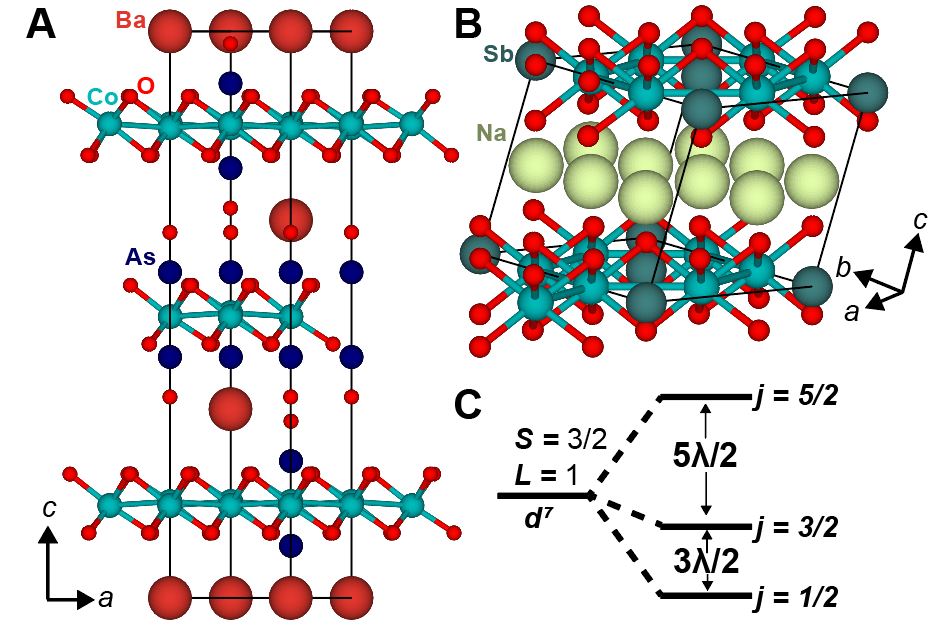} \par
    \small{\textbf{Figure 14.} Cobalt Honeycombs.
    \textbf{A.} A side view of the unit cell of \ce{BaCo2(AsO4)2}, showing the layering of the \ce{CoO6} octahedral honeycomb layers. 
    \textbf{B.} The unit cell of delafossite material \ce{Na3CoSb2O6}, with Co$^{2+}$ layers separated by a Na$^{1+}$ ionic layer.
    \textbf{C.} Under the influence of spin-orbit coupling, the $3d7$ crystal field of octahedral Co$^{2+}$ can be split and result in a $j = 1/2$ ground state, which is optimal for Kitaev spin liquid interactions.
    }
\end{center}      

\subsubsection{Kagome Lattice}

The two-dimensional Kagome lattice is named after a Japanese basket weave pattern and is built up of corner sharing triangles. Spins on a Kagome lattice have a higher geometric degeneracy in comparison to other frustrated lattices, and several Kagome lattice-containing materials are readily found in nature as minerals. This lattice, built up of interconnected Stars of David, is also directly derived from the triangular motif, and can be transformed into a pyrochlore lattice, as shown in\textbf{Figure 18}. By connecting the centers of each triangle to each other, a honeycomb lattice emerges, thus demonstrating the intricate interrelationships between all frustrated lattice types.

\noindent \textbf{Synthetic Herbertsmithite \ce{ZnCu3(OH)6Cl2}} \par

Synthetic Herbertsmithite is perhaps the most famous quantum spin liquid candidate material ever. As the name implies, it is a mineral named after mineralogist Herbert Smith, and has formula \ce{ZnCu3(OH)6Cl2} \cite{Braithwaite2004, RevModPhys.88.041002,Mendels2010}. It crystallizes in the trigonal space group $R\bar{3}m$ (no. 166). A side view of the unit cell is shown in \textbf{Figure 15 A}. Two-dimensional layers of strongly Jahn-Teller distorted \ce{Cu(OH)4Cl2} octahedra are separated from each other by \ce{Zn(OH)6} octahedra. The Cu$^{2+}$ $S = 1/2$ units form a perfect kagome lattice, as shown in \textbf{Figure 15 B}, where every Cu$^{2+}$ cation is $3.41 \mathrm{\AA}$ apart from one another.

\begin{center}                
    \includegraphics[width=1\textwidth]{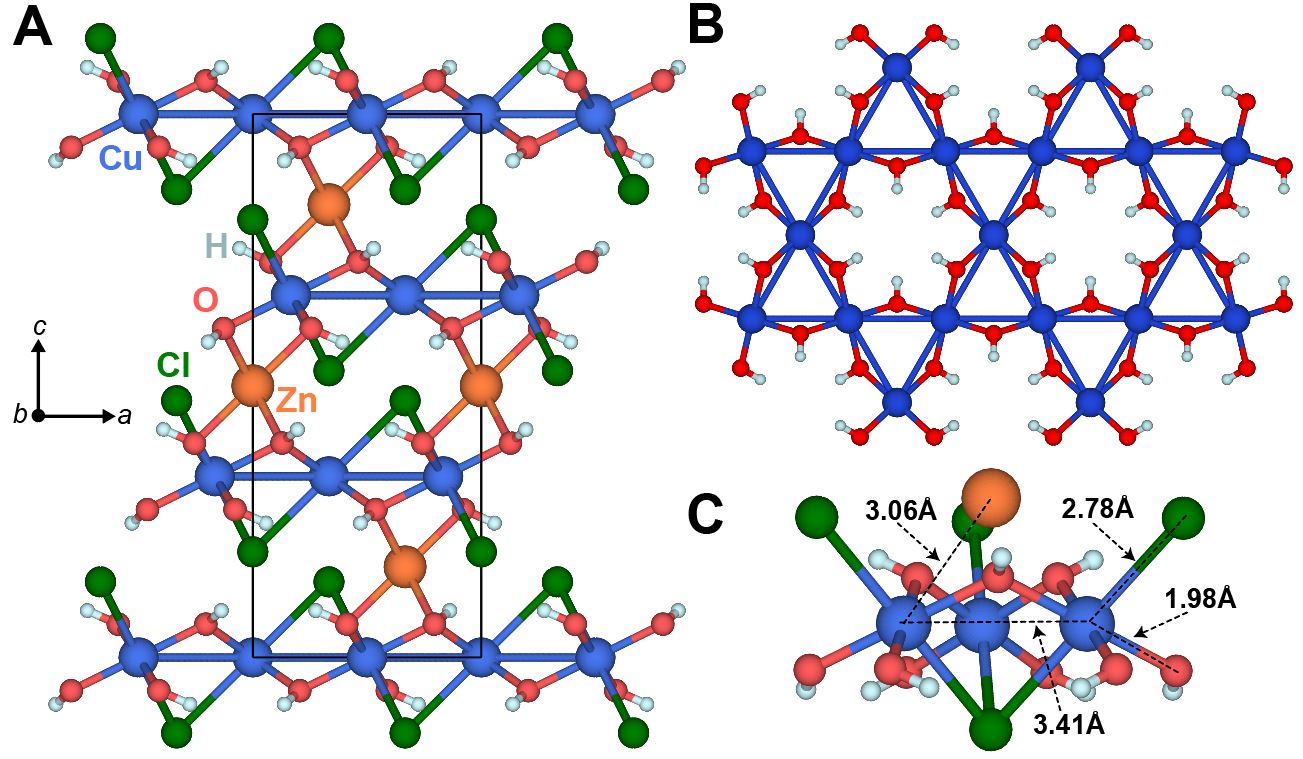} \par
    \small{\textbf{Figure 15.} Herbertsmithite.
    \textbf{A.} A side view of the unit cell of Herbertsmithite, highlighting the Cu$^{2+}$ kagome layering, separated by \ce{Zn(OH)6} octahedra.
    \textbf{B.} A top down view of the structurally perfect Cu$^{2+}$ kagome lattice.
    \textbf{C.} A close up of a triangular unit of \ce{Cu(OH)4Cl2} octahedral units, showing particularly relevant bond lengths.
    }
\end{center}

Polycrystalline samples of synthetic Herbertsmithite are usually easily synthesized via the hydrothermal method, where stoichiometric amounts of \ce{ZnCl2} and \ce{Cu2(OH)2CO3} are added to a hydrothermal bomb along with small amounts of \ce{H2O}, and heated to temperatures in excess of $200^{\circ}$ C for a day or so \cite{PhysRevX.6.041007}. Crystals can be grown in a manner similar to a chemical vapor transport growth, with a long, sealed reaction vessel that is subjected to a hot zone and a cold zone, but is filled with water, though reactions often need to run from weeks to months to allow for sizeable crystal growth \cite{Chu2011}. The conditions are still hydrothermal and in excess of the boiling point of water, which seemingly stabilizes the Herbertsmithite phase and allows for crystal growth. Though this technique is well-suited to the synthesis of this material, it is possible that the non-equilibrium conditions involved promote the development of defects in this material, specifically Cu-Zn site mixing.

Synthetic Herbertsmithite unfortunately suffers from a high degree of Cu$^{2+}$-Zn$^{2+}$ site mixing of up to an astronomical 15\% \cite{RN24}. This means that in the kagome lattice, 15\% of all Cu$^{2+}$ cations are replaced with non-magnetic Zn$^{2+}$ cations, and 15\% of the inter-kagome-layer Zn$^{2+}$ sites are replaced with magnetic Cu$^{2+}$, which can have a strong impact on the magnetic properties of this material and allow for inter-layer magnetic exchange pathways. We discuss these point defects in the defect section later in this review, but find it of great importance to mention that in Herbertsmithite, there are a sizeable amount of defects that should not and cannot be neglected.

Measurements of the physical properties of synthetic Herbertsmithite are quite extensive. Magnetization measurements down to at least $T = 0.1$ K reveal no signs of a magnetically ordered state at any temperature range \cite{Shores2005, PhysRevB.76.132411}. Inelastic neutron scattering measurements of the spin dynamics of Herbertsmithite also confirmed the absence of magnetic order down to $T = 0.05$ K, providing further proof for a possibly dynamic, fluctuating ground state in this material \cite{PhysRevLett.98.107204}. Muon spin spectroscopy measurements reveal a highly dynamical ground state, marked by the absence of any transition to antiferromagnetic order \cite{PhysRevLett.98.077204}. Electron spin resonance measurements shed light on the magnetic exchange interactions in this material, revealing strong anistropies arising from asymmetric exchange \cite{PhysRevLett.101.026405}. Heat capacity measurements have demonstrated behavior indicative of gapless (possibly spinon) excitations, though with unexpected, strong magnetic field dependence \cite{PhysRevLett.98.107204, PhysRevLett.100.157205}. This field dependence was later found to arise from the Cu$^{2+}$ cations sitting on the inter-layer Zn$^{2+}$, a finding that was confirmed by nuclear magnetic resonance measurements that showed a drop in the bulk susceptibility starting around $T = 50$ K, consistent with antiferromagnetically interacting kagome spins \cite{PhysRevLett.100.077203}. The Cu-Zn disorder further manifested in $^{18}$O nuclear magnetic resonance measurements, in the form of extra oxygen positions, possibly due to Jahn-Teller distortions of Cu$^{2+}$ at those sites. But perhaps the most telling measurements are of inelastic neutron scattering, which reveal the expected continuum of fractionalized excitations consistent with the presence of itinerant spinons \cite{RN32,PhysRevLett.104.147201}. 

All of these results, therefore, point to synthetic Herbertsmithite as being an optimal quantum spin liquid material where indications of defects can be observed in the data, and obfuscate interpretations of the intrinsic ground state. Herculean efforts were made to dope this material as theoretical predictions were made of exotic states \cite{Mazin2014}, though successful chemical doping resulted in even more insulating behavior, in part due to Anderson localization brought on by defects\cite{PhysRevX.6.041007, PhysRevLett.121.186402}. It is powerful that such an important and interesting material in the quantum spin liquid field is still susceptible to the same kinds of defects that permeate all materials, and thus goes to show that defects may play a very large role in quantum phenomena. It is part of a huge elucidation of many related mineral kagome-type phases (e.g. Volborthite \cite{Hiroi_Volborhite_2001} and Vesignieite \cite{200915843}), including development of new chemical methods for preparing minerals \cite{Winiarski_Volcanos2019} and fluorides such as \ce{(NH3CH3)2NaTi3F12} \cite{jiang_ramanathan_bacsa_la_pierre_2019}.
 
\noindent \textbf{\ce{Ln3Sb3Mg2O14} and \ce{Ln3Sb3Zn2O14} materials} \par

A particularly interesting family of materials is the double pyrochlore materials with formula \ce{Ln3Sb3Mg2O14} and \ce{Ln3Sb3Zn2O14}, with Ln$^{3+}$ ranging from La to Yb \cite{doi:10.1002/pssb.201600256, Sanders2016, PhysRevLett.116.157201}. They are particularly interesting because they are the only example of rare-earth cations on a kagome lattice. As shown in Figure 21 in the three-dimensional  pyrochlore section, the kagome lattice can be found within the pyrochlore lattice along many crystallographic orientations. A pyrochlore lattice is, in effect, two kagome lattices bound to each other by atoms above the corner-sharing triangles. Though we will discuss the pyrochlore structure in detail below, it is of use to explain some of its basic features here.

The basic pyrochlore stoichiometric formula is \ce{A2B2X7}, where both A and B cations sit on separate, interwoven corner-sharing tetrahedral pyrochlore sublattices, and X is the charge-balancing anion. If we were to explicitly write out the formula in a way to represent each atom in a tetrahedron on its own, the formula becomes \ce{A4B4X14}. If we then further label the tetrahedral capping atoms separately, we would arrive at A$_{3}$A'B$_{3}$B'O$_{14}$. The A' and B' sites can be thought of as being the sites that sit atop the corner-sharing kagome triangles, i.e. the sites that are responsible for transforming the two-dimensional kagome lattice into the three-dimensional pyrochlore. By substituting A' and B' with another element, it may be possible to isolate the kagome planes and synthesize a two-dimensional, kagome system. This is not a trivial feat, as large differences in size and charge are necessary between A and A', and B and B', in order to stabilize such a material.

This, however, is exactly what was achieved for the \ce{Ln3Sb3Mg2O14} and \ce{Ln3Sb3Zn2O14} family of materials, where A' and B' were both successfully set to be Mg$^{2+}$ and Zn$^{2+}$. The large difference in size and charge between rare-earth Ln$^{3+}$, Sb$^{5+}$, and Mg$^{2+}$ or Zn$^{2+}$ cations is enough to stabilize the formation of these interesting compounds. The general structure is shown in \textbf{Figure 16 A}, demonstrating the kagome lattice layering of the A-site cations (in this figure, Tb$^{3+}$) and B-site, Sb$^{5+}$ cations. The two perfect kagome lattices formed can be seen in \textbf{Figure 16 B}, where Tb-Tb and Sb-Sb distances are both $3.68 \mathrm{\AA}$. These materials crystallize in the trigonal $R\bar{3}m$ (no. 166) space group, a reduction from the cubic $Fd\bar{3}m$ seen in ordinary \ce{A2B2X7} pyrochlore materials.

Thus far it has been shown that this double pyrochlore structure type can only be stabilized with Sb$^{5+}$ on the B-site. When comparing Mg$^{2+}$ vs. Zn$^{2+}$ on the A'/B' sites, However, it has been found that Zn$^{2+}$ sites show positional disorder, in contrast to its Mg$^{2+}$ counterpart, which shows no disorder, as shown in \textbf{Figure 16 C}. In the \ce{Ln3Sb3Zn2O14} group of compounds, Zn$^{2+}$ can be found in any one of eight different positions around the site at the center of a kagome hexagonal void. Since this Zn cation does not sit on a special position, the \ce{Ln3Sb3Zn2O14} family also retains trigonal, $R\bar{3}m$ symmetry. However, the difference in bonding environments around the Zn may result in altered physical properties compared to the same Mg analogous compound. 

\begin{center}                
    \includegraphics[width=1\textwidth]{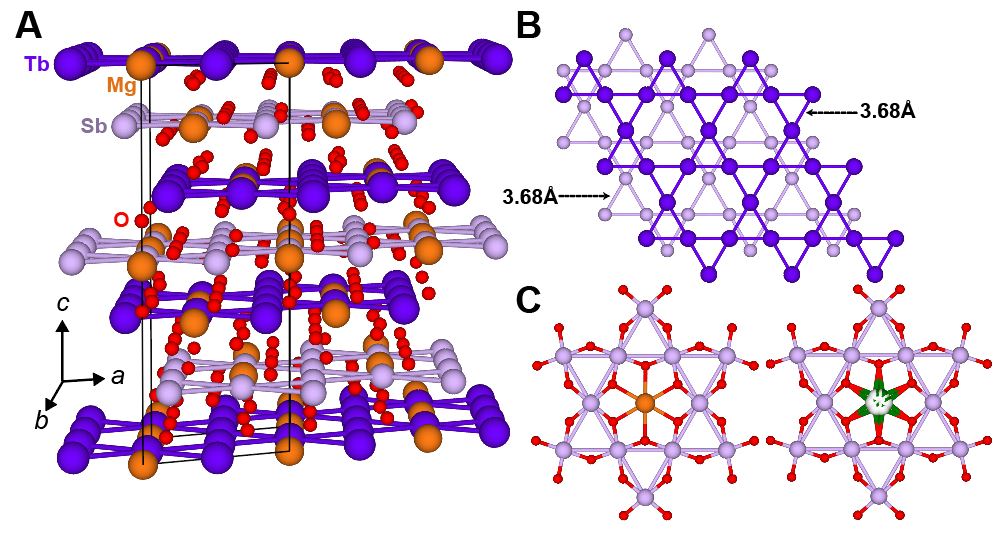} \par
    \small{\textbf{Figure 15.} \ce{Tb3Sb3Mg2O14}.
    \textbf{A.} A side view of the unit cell, showing the alternating Tb$^{3+}$ and Sb$^{5+}$ kagome sublattices.
    \textbf{B.} A top down view showing the Tb$^{3+}$ and Sb$^{5+}$ kagome lattices, which have equal Tb-Tb and Sb-Sb distances. The kagome lattices do not overlap directly over each other.
    \textbf{C.} A close up of the environments within kagome lattices, showing no disorder around the Mg$^{2+}$ as in the \ce{Ln3Sb3Mg2O14} family of materials, but occupational disorer in the Zn$^{2+}$ site as in \ce{Ln3Sb3Zn2O14}.
    }
    
\end{center}

Due to the diversity in phenomena exhibited by individual rare earth cations, the physical properties of \ce{Ln3Sb3Mg2O14} compounds are likely quite diverse, yet for the most part remain unexplored barring simple measurement techniques such as magnetization and heat capacity measurements. Interesting magnetic order has been observed in \ce{Dy3Sb3Mg2O14}, as evidenced by measurements of magnetization, heat capacity, and neutron scattering \cite{Paddison2016_2}, as well as in \ce{Nd3Sb3Mg2O14} \cite{PhysRevB.100.024414,PhysRevB.98.134401}. An extremely interesting quantum spin fragmentation state has been found in \ce{Ho3Sb3Mg2O14}, which shows quantum dynamics down to low temperatures \cite{1806.04081}. In summary, the physical properties of some \ce{Ln3Sb3Mg2O14} and \ce{Ln3Sb3Zn2O14} have indeed been explored in great depth, while others remain largely unexplored. These systems perhaps present an interesting opportunity in the field of quantum spin liquids and should be studied more closely with a wider range of techniques, so as to elucidate their properties.

\noindent \textbf{\ce{LiZn2Mo3O8}} \par

Most quantum spin liquids rely primarily on correlated spins that sit on atoms in a lattice. There exist, however, compounds where clusters act as the interacting spin centers. These clusters, which are in effect molecular units, can form strongly interacting emergent frustrated lattices. An example of this is the molybdenum-based cluster magnet \ce{LiZn2Mo3O8}.

The structure of \ce{LiZn2Mo3O8} is shown in \textbf{Figure 16 A}, shown from a sideways perspective to emphasize the layering and two-dimensionality of this material. It contains alternating \ce{LiZn2} and \ce{Mo3O8} layers and crystallizes in trigonal $R\bar{3}m$ (no. 166). The \ce{LiZn2} layers contain both \ce{LiO4} and \ce{ZnO4} tetrahedra, and \ce{LiO6} and \ce{ZnO6} octahedra where the Li-O and Zn-O bonds are shorter than $3 \mathrm{\AA}$. As such, covalent bonds with oxygen are possible for both cations, as well as ionic bonding. Li and Zn substitutional disorder can therefore potentially affect the oxygen atomic positions, which are shared with the \ce{Mo3O8} layer, and can therefore have a direct impact on the bonding around the magnetic molybdenum cations.

\begin{center}                
    \includegraphics[width=0.8\textwidth]{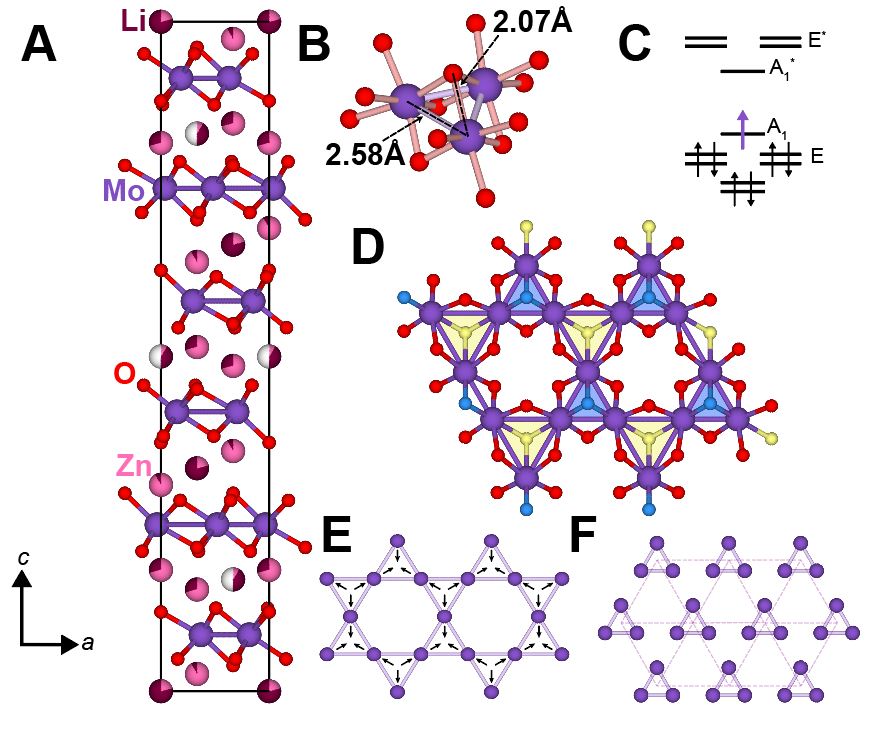} \par
    \small{\textbf{Figure 16.} \ce{LiZn2Mo3O8}.
    \textbf{A.} A sideways view of the unit cell, highlighting the layering of molydenum cluster-based triangular layers.
    \textbf{B.} An individual \ce{Mo3O13} cluster, with pertinent bond lengths.
    \textbf{C.} A molecular orbital diagram of the Mo clusters, as derived from calculations on isostructural and isoelectronic \ce{Mo3O13H15}.
    \textbf{D.} A top down view of the molybdenum layers, showing a breathing kagome lattice with small (blue) and large (yellow) triangles.
    \textbf{E.} The kagome lattice of molybdenum cations, with arrows indicating their movement with decreasing temperature.
    \textbf{F.} The resultant, emergent triangular lattice of \ce{Mo3} clusters, which arises after complete trimerization and molecular bonding of \ce{Mo3O13} cluster units.
    }   
\end{center}

The \ce{Mo3O8} layer appears at first to be a kagome lattice, with molybdenum occupying the corners of edge sharing triangles. However, this is not the case, as this material is a distorted kagome with alternating small and large triangles, also known as a breathing kagome lattice \cite{PhysRevB.97.035124,PhysRevB.95.054410}, where the smaller triangles have formed molecular trimers. The short bond distance between Mo cations of $2.58 \mathrm{\AA}$ small enough to result in direct Mo-Mo bonding, which is supported by the preponderance of molybdenum, as well as other refractory metals such as rhenium, to form direct metal-metal bonds \cite{Hughbanks1983,Schimek1994}. As such, the lattice distorts so as to form \ce{Mo3O13} molecular clusters, such as the one shown in \textbf{Figure 16 B}. These clusters are highly symmetric, and an analysis of the symmetries of symmetrically analogous compound \ce{Mo3O13H15} resulted in the molecular orbital diagram shown in \textbf{Figure 16 C}, which shows a single unpaired electron per cluster. As a result, each cluster acts effectively as a $S = 1/2$ unit. The kind of lattice distortion, as shown in \textbf{Figure 16 D}, can give rise to an emergent triangular lattice as in \textbf{Figure 16 E}. This particular kind of lattice distortion is not limited to cluster units, and has become of particular interest in $S = 1$ kagome lattice materials \cite{Kelly2019, PhysRevLett.124.167203}. An emergent lattice of \ce{Mo3O13} clusters acts as frustrated triangular lattice instead of a kagome one, and the trimers in this material contract upon cooling, further pushing this material more towards the triangular lattice limit. There have also been theoretical work indicating that, due to the nature of the exchange interactions, the emergent lattice of clusters may instead be honeycomb instead of triangular \cite{PhysRevLett.111.217201}.

Cluster magnetism is particularly interesting and a seemingly largely unexplored in the field of quantum spin liquids. The behavior of \ce{LiZn2Mo3O8} has been shown to behave as a valence bond solid, a similar ground state to the resonating one Anderson proposed for Ising $S = 1/2$ triangular lattices \cite{Anderson_1973}. Measurements of the magnetization of \ce{LiZn2Mo3O8} show a high temperature Curie-Weiss paramagnetic state with strong antiferromagnetic interactions on the order of $\Theta_{W} = -220$ K, and a sudden loss of magnetization upon cooling, at approximately $T_{VBS} = 96$ K consistent with the formation of valence bonds \cite{RN27}. The absence of a lambda transition in the heat capacity at this $T_{VBS}$ temperature, and the featureless behavior observed in the derived change in magnetic entropy both serve to exclude the possibility of a transition to a long range magnetically ordered state at this temperature.

Measurements of microwave and terahertz electron spin resonance, muon spin rotation, and $^7$Li nuclear magnetic resonance, all local magnetism measurements, confirm the lack of a transition to long range magnetic order down to $T = 0.07$ K, and indicate that the relevant $S = 1/2$ valence bonds are resonating \cite{PhysRevB.89.064407}. There is also evidence for the gapless spinon excitations attributable to a quantum spin liquid ground state, and molecular magnetism consistent with cluster physics \cite{PhysRevLett.112.027202}. As such, \ce{LiZn2Mo3O8} presents an example of a molecular cluster-based quantum spin liquid candidate. 

There has also been work exploring the stoichiometry of \ce{LiZn2Mo3O8} and its potential impact on both electronic and magnetic properties \cite{Sheckelton2015}. A recent study of \ce{LiZn2Mo3O8} suggests that alternative stoichiometries to those found in previous reports can be obtained through careful control of the reactants, specifically with regards to the Li and Zn content, which are rendered susceptible to off-stoichiometries due to their site mixing \cite{Sandvik2019}. While previous studies had stoichiometries of general formula Li$_{1+x}$Zn$_{2-y}$Mo$_{3}$O$_{8}$, a precise stoichiometry of Li$_{0.95}$Zn$_{1.92}$Mo$_{3}$O$_{8}$ can be obtained under careful reaction conditions. Samples measured with this stoichiometry also confirm the absence of a transition to a long range magnetically ordered state, and provide further support for an exotic ground state, though there is some variation in the physical properties as a result of stoichiometric changes.

Other magnetic molecular cluster compounds include \ce{Nb3Cl8}, which is also a trimerized kagome material with emergent triangular lattice layers that are separated by van der Waals gaps \cite{Sheckelton2017,Haraguchi2017,Pasco2019},
and $1T$-\ce{TaS2} \cite{Law6996,PhysRevB.96.081111,PhysRevB.96.195131}. In $1T$-\ce{TaS2}, thirteen Ta atoms form molecular clusters that, in turn, form an emergent triangular lattice that has been both theorized and shown, experimentally, to show behavior consistent with a quantum spin liquid. 

There exist a growing number of metallic materials with kagome lattices that possess interesting topological electronic phenomena such as \ce{Co3Sn2S2} \cite{Vaqueiro2009, Guin2019,PhysRevMaterials.4.024202}, \ce{Fe3Sn2}\cite{Caer1978, Kida2011, Ye2019}, and \ce{AV3Sb5}, where A = K, Rb, and Cs \cite{PhysRevMaterials.3.094407,1912.12288}. Though this review deals with insulating magnetic materials, the kagome lattices in those materials and related compounds remain largely unexplored in terms of their potential magnetism and is one of the most promising avenue for future exploration and may lead not only to interesting quantum magnetic ground states, but also new devices with practical applicability.

\subsection{3D}
In this section we will discuss compounds that contain diamond and pyrochlore three-dimensional sublattices. These sublattices, just like the two-dimensional ones, are interrelated, and can derived from the tetrahedron, though the the pyrochlore is best understood as the three-dimensional expansion of the kagome lattice, and the diamond lattice the expansion of the two-dimensional honeycomb. As the highest dimensionality lattices that can potentially give rise to quantum spin liquid materials, these lattice types are far more sensitive to anisotropies, including single-ion and lattice, and thus parameters must be fine tuned to give rise to exotic phenomena.
\subsubsection{Diamond lattice}
The diamond lattice of corner sharing tetrahedra is a popular three-dimensional frustrated lattice that has been shown to give rise to quantum spin liquid phenomena. Every atom is part of at least four tetrahedra, as opposed to the pyrochlore where they are only part of two, and thus frustration is caused by the tetrahedral motif. The name of this frustrated lattice is borrowed from the crystallographic diamond, which shares the same corner-sharing tetrahedral network \cite{Raman1944}. 

\noindent \textbf{\ce{FeSc2S4}}

\ce{FeSc2S4} is a particularly interesting material because it is a spin-orbital liquid candidate, just as $6H$-\ce{Ba3CuSb2O9}. In this system, however, it is tetrahedral Fe$^{2+}$ that shows a lack of a Jahn-Teller distortion instead of Cu$^{2+}$. In terms of stabilization energies, undistorted, octahedral Cu$^{2+}$ as equally energetically unfavorable as tetrahedral Fe$^{2+}$, and thus an undistorted Fe$^{2+}$ tetrahedron is highly unstable and rare \cite{Goodenough1964}. Tetrahedral and undistorted Fe$^{2+}$ exists, however, in \ce{FeSc2S4}, a spinel material. Spinels are a family of materials with general formula \ce{AB2X4}, where A-site cations form a diamond sublattice and B-site lattices form a pyrochlore sublattice, and X are the charge-balancing anions. The A-sites form \ce{AX4} tetrahedra, while the B-sites form \ce{BX6} octahedra. In \ce{FeSc2S4}, tetrahedral Fe$^{2+}$ is stabilized by the much higher octahedral crystal field stabilization energy of Sc$^{3+}$, and forms a three-dimensional diamond lattice, as shown in \textbf{Figure 17 A}. Relevant bond lengths and angles for individual and collective \ce{FeS4} tetrahedra are shown in \textbf{Figure 17 B}, along with the corresponding crystal field diagram for tetrahedral Fe$^{2+}$ is shown in \textbf{Figure 17 C}.

\begin{center}                
    \includegraphics[width=1\textwidth]{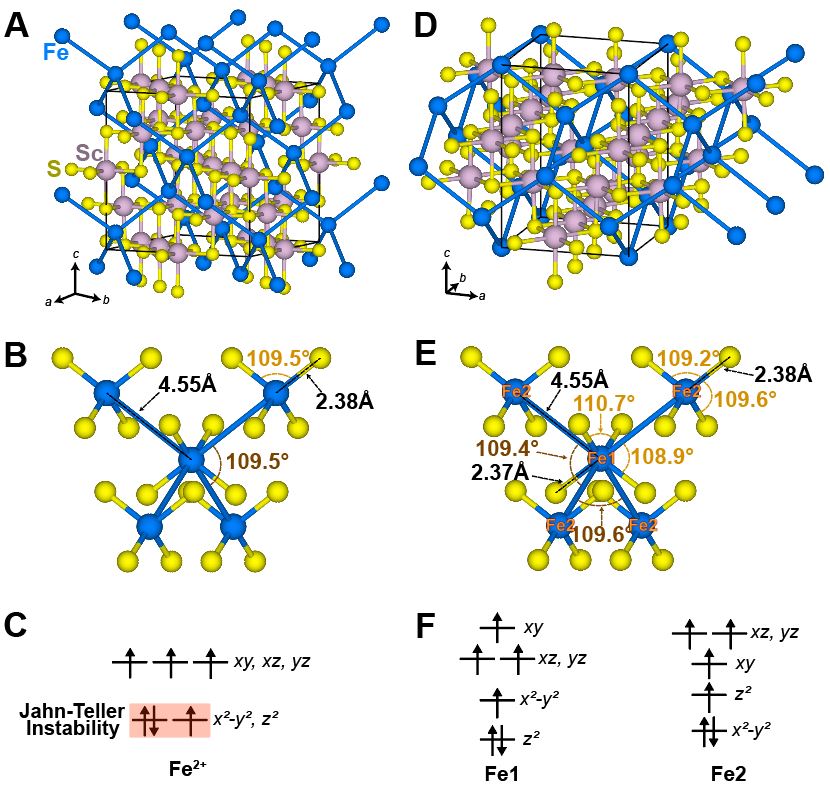} \par
    \small{\textbf{Figure 17.} \ce{FeSc2S4}
    \textbf{A.} The unit cell of cubic, $Fd\bar{3}m$ \ce{FeSc2S4}. The magnetic Fe$^{2+}$ diamond sublattice is highlighted.
    \textbf{B.} The distance between Fe-Fe cations is shown, along with the Fe-S bond lengths. The bond angles between Fe cations in the diamond lattice, and within \ce{FeS4} are both $109.5^{\circ}$, confirming that both the diamond lattice and tetrahedra possess perfect tetrahedral symmetry.
    \textbf{C.} The crystal field manifold for the tetrahedral Fe$^{2+}$ cations. The unstable orbital degeneracy can be seen in the lower energy $e_{g}$ manifold.
    \textbf{D.} The structure of low temperature, distorted \ce{FeSc2S4}, with tetragonal $I\bar{4}m2$ symmetry \cite{PhysRevX.6.041055}
    \textbf{E.} The local bonding environment of Fe$^{2+}$ cations in distorted \ce{FeSc2S4}, showing subtle deviations in both bond lengths and angles from ideal tetrahedral values.
    \textbf{F.} The predicted crystal field manifolds for both separate Fe$^{2+}$ sites in tetragonal \ce{FeSc2S4}.
    }
\end{center}

\ce{FeSc2S4} is synthesized using the high-temperature solid state technique, where pre-reacted FeS and \ce{Sc2S3} are ground and heated, under vacuum, to high temperature. Single crystals can be grown using a traveling solvent, Bridgman-like technique with FeS as the molten solvent \cite{MOREY2016128}, and the chemical vapor transport technique using \ce{I2} as a transport agent \cite{PhysRevB.96.054417}. Sulfide compounds tend to be prone to many kinds of defects, including oxygen interstitials, sulfur vacancies, and sulfide-sulfide bonds. Furthermore, variations in \ce{FeSc2S4} stoichiometry have been shown to result in difference in measureable physical properties across multiple samples, including the emergence of magnetic order. Thus, care must be taken in the synthesis so that the exact stoichiometry is realized. Moreover, the study of the impact of defects on the ground state of \ce{FeSc2S4} could potentially elucidate exotic behaviors that are otherwise unseen.

Aside from the lack of long range magnetic order at all temperatures, the Fe$^{2+}$ tetrahedra do not show a Jahn-Teller distortion at any temperature, implying the presence of an unquenched orbital degeneracy in the lower energy $e_{g}$ orbital manifold. As such, this material is expected to show behavior consistent with a spin-orbital liquid \cite{Kalvius2006,PhysRevLett.92.116401,PhysRevLett.102.096406}, where orbital moments, which usually order at far higher temperature scales compared to spins, do not order down to at least $T = 0.05$ K. Measurements of magnetization show a higher-than-spin $S = 2$ moment expected for Fe$^{2+}$, indicative of non-negligible orbital contributions to the total moment. Analyses of heat capacity measurements demonstrate an entropy that plateaus around $\Delta$S = Rln(5) + Rln(2), consistent with both a spin and orbital component \cite{PhysRevLett.92.116401}. 

A contrary report based on neutron scattering experiments of polycrystalline samples of \ce{FeSc2S4} revealed the subtle emergence of an antiferromagnetic ordering transition at $T_{N} = 11$ K, concomitant with a succinct structural phase transition that relieves the orbital degeneracy of the Fe$^{2+}$ sites \cite{PhysRevX.6.041055}. The reported low temperature structure is shown in \textbf{Figure 17 D}, and has tetragonal, $I\bar{4}m2$ (no. 119) symmetry. \ce{FeS4} tetrahedra alternate between contraction and expansion, similar to a breathing system, and pertinent bond lengths and angles are shown in \textbf{Figure 17 E}. The crystal field diagrams associated with these distortions are shown in \textbf{Figure 17 F}, where the orbital degeneracy has been relieved. Though this is a possible avenue towards tetrahedral Fe$^{2+}$ orbital frustration relief, counterclaims have arisen indicating that variations in sample stoichiometry are the origin of this behavior \cite{PhysRevB.96.054417}. Future studies on the nature and extent of defects in \ce{FeSc2S4} are necessary to determine what the true ground state of this material is, and whether or not it fulfills the requirements of a quantum spin liquid or spin-orbital liquid. 

Another interesting diamond lattice material is \ce{NiRh2O4}, which contains Ni$^{2+}$ on the diamond lattice \cite{PhysRevMaterials.2.034404}. Similar to \ce{FeSc2S4}, the large octahedral crystal field stabilization of the B-site cation, in this case Rh$^{3+}$ instead of Sc$^{3+}$, forces Ni$^{2+}$ into a tetrahedral arrangement, which is incredibly energetic, as it results in a Jahn-Teller instability in the higher energy $t_{2}$ manifold. Though motivated by the possibility of emergent topological phenomena in an $S = 1$ diamond lattice \cite{PhysRevB.91.195131}, and a potential spin-orbital liquid state arising from the $t_{2}$ orbital degeneracy, a high-temperature structural phase transition was observed at $T = 440$ K that relieves the orbital degeneracy and distorts the \ce{NiO4} tetrahedra. Nevertheless, this system does not show magnetic order down to $T = 0.1$ K, and magnetization, heat capacity, and neutron scattering data indicate unconventional behavior that is in contrast to many theoretical expectations \cite{PhysRevLett.120.057201,PhysRevB.100.140408,PhysRevB.96.020412}. Diamond lattice magnetic materials are therefore a potential source of interesting and exotic quantum behavior.

\subsubsection{Pyrochlores and Breathing Pyrochlores} 

\begin{center}                
    \includegraphics[width=1\textwidth]{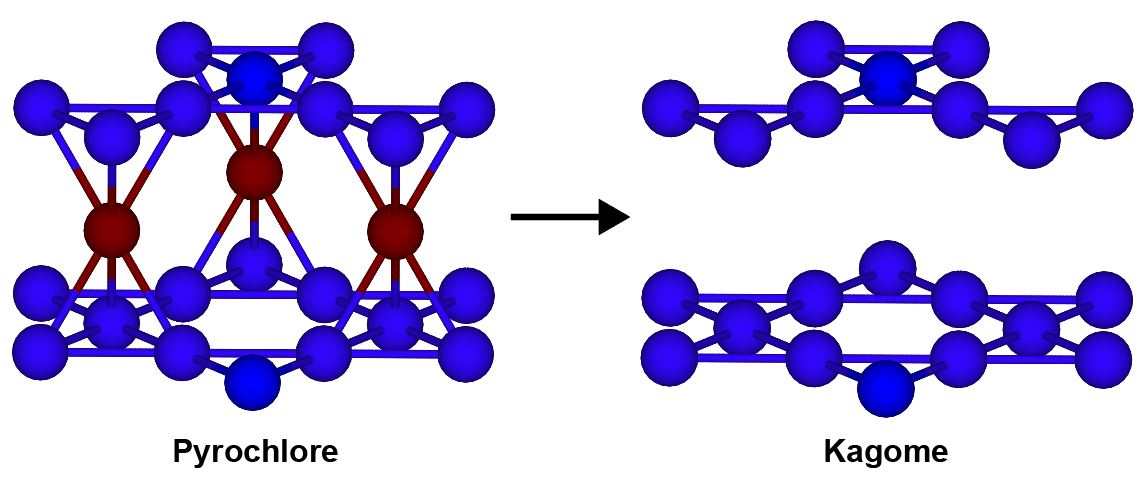}
    \small{\textbf{Figure 18.} The pyrochlore lattice can be transformed into a pyrochlore lattice by removal of the inter-layer atoms. These two frustrated structure types are intimately related and thus show how the two fundamental building blocks of frustration, the triangle and the tetrahedron, are homogenous. The transformation shown from pyrochlore to kagome is the same transformation applied to the \ce{Ln3Sb3Mg2O14} and \ce{Ln3Sb3Zn2O14} class of materials, discussed above. The atoms that bridge the two kagome lattices together form a triangular lattice, and thus the pyrochlore is the product of triangular and kagome lattices.}
\end{center}

The pyrochlore lattice is a lattice of corner sharing tetrahedra. As aforementioned, it is the three-dimensional expansion of the kagome and can also be described as an alternating stacking of kagome layers and triangular layers, as shown in Figure 18. The pyrochlore lattice is therefore the highest geometric degeneracy three-dimensional frustrated lattice. As aforementioned, its general pyrochlore formula is \ce{A2B2X7}, where both A and B form separate pyrochlore sublattices\cite{Subramanian1983}. There are many examples of pyrochlore materials with rare earth cations on the magnetic lattice, such as the rare-earth titanates, where non-magnetic Ti$^{4+}$ sits on the B-site. Rare earths cations sit on the A-site, resulting in compounds with extraordinary magnetic properties such as \ce{Dy2Ti2O7} \cite{Morris2009,Hiroi2003}, \ce{Ho2Ti2O7} \cite{Fennell2009,Matsuhira2000,Ehlers2004,GHASEMI201838}, and \ce{Yb2Ti2O7} \cite{PhysRevX.1.021002, Yasui2003, Hodges2001, Cao2009}.

\ce{Dy2Ti2O7} is a particularly interesting material. The material crystallizes in high symmetry cubic spacegroup $Fd\bar{3}m$ (no. 227) and its structure is shown in \textbf{Figure 19 A}. At low temperature, \ce{Dy2Ti2O7} forms an exotic state of matter known as spin ice, where there is residual entropy in the macroscopically degenerate ground state similar to water ice. The arrangement of the magnetic moments in spin ice, as the name implies, resembles the orientation of hydrogen atoms in water ice, i.e., two-in-two-out, as shown in \textbf{Figure 19 B} \cite{Pauling1935}. Magnetization measurements of \ce{Dy2Ti2O7} revealed strong magnetic anisotropy, which is unexpected in a cubic material \cite{PhysRevB.65.054410,PhysRevLett.90.207205}. Interestingly, applying a magnetic field along the [111] direction results in an entirely different value of the residual entropy compared to those of the other orientations such as [100] and [110]\cite{Matsuhira_2002,PhysRevB.66.144407}. This observation was also supported by low-temperature specific heat measurements down to $T = 100$ mK\cite{2004}. This low temperature spin state in \ce{Dy2Ti2O7} has been described as a classical, not quantum, spin liquid state, as it is thermal and not quantum fluctuations that are responsible for its emergence. The pyrochlore lattice can also establish long range magnetic order, such as the All-In-All-Out order shown in \textbf{Figure 19 C}. This, of course, has the tendency to occur in pyrochlore compounds where the rare-earth magnetic cations are of higher spin and thus tend to classical behavior, such as in \ce{Eu2Ir2O7} where both the Eu$^{3+}$ and Ir$^{4+}$ pyrochlore sublattices are magnetic \cite{PhysRevB.87.100403}. Furthermore, different classifications of magnetism have been observed across the rare earth pyrochlores, such as Ising in \ce{Dy2Ti2O7} \cite{PhysRevLett.83.1854}, Heisenberg in \ce{Gd2Ti2O7} \cite{PhysRevB.64.140407}, and $xy$-ordering in \ce{Er2Ti2O7} \cite{PhysRevLett.101.147205}.

Recently, the Ce$^{3+}$ $j = 1/2$ pyrochlore \ce{Ce2Sn2O7}, with a nonmagnetic Sn$^{4+}$ B-site sublattice, was proposed as a candidate quantum spin liquid \cite{PhysRevLett.115.097202}. Measurements of magnetization and muon spin spectroscopy indicate a lack of magnetic order down to $T = 0.02$ K, and a fluctuating dynamic ground state. Similar behavior has been observed in the compound \ce{Ce2Zr2O7} \cite{Gao2019}, where heat capacity, muon spin spectroscopy, and neutron scattering experiments were used to show a lack of magnetic order down to $T = 0.035$ K, and a continuum of excitations consistent with spinons. These materials were both synthesized via high-temperature solid state reactions involving stoichiometric amounts of \ce{CeO2} and ZrN in reducing environments and up to very high temperatures. The high proclivity of Ce$^{3+}$ towards oxidation in air renders this reaction to be very challenging, and care must be taken to ensure that final product stoichiometry is not oxygen rich. Crystals were then grown by the floating zone technique through the use of laser diodes, which offer a higher thermal gradient as opposed to the standard optical technique. 

A potential distortion of the pyrochlore lattice is similar to the trimerization seen in \ce{LiZn2Mo3O8} and $S = 1$ kagome magnets \cite{Kelly2019, PhysRevLett.124.167203}, where the corner-sharing tetrahedra alternate between big and small and thus give rise to a ``breathing" pyrochlore lattice. An example of such a system is \ce{Ba3Yb2Zn5O11}, whose unit cell is shown in \textbf{Figure 19 D} \cite{Haku2016}. This class of materials is fairly underexplored and could present an avenue towards new kinds of quantum spin liquid, as many theoretical predictions have emerged that predict interesting phenomena \cite{Benton2015, PhysRevB.94.075146}.

\begin{center}                
    \includegraphics[width=1\textwidth]{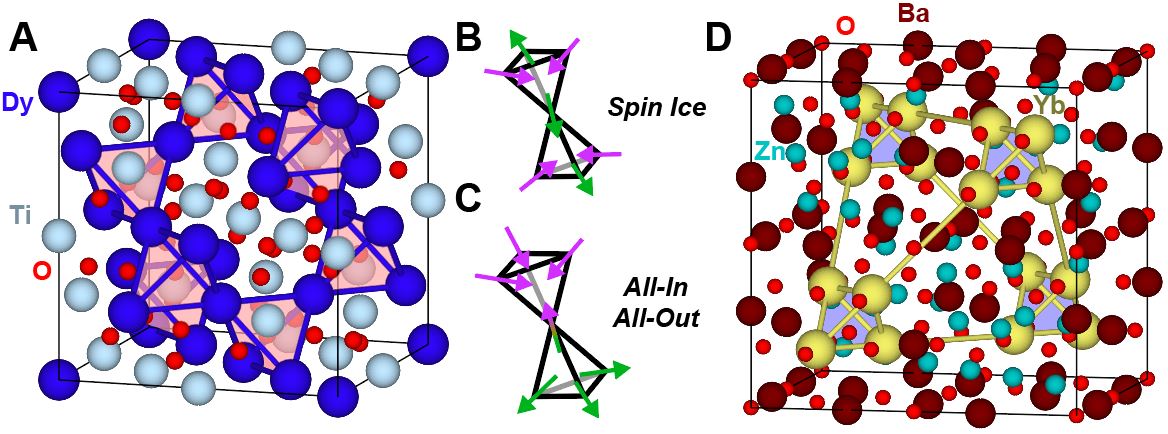}
    \small{\textbf{Figure 19.} The pyrochlore and the breathing pyrochlore.
    \textbf{A.} The representative unit cell of \ce{Dy2Ti2O7}, showing the pyrochlore sublattice of Dy$^{3+}$ cations.
    \textbf{B.} The spin-ice rule applied to two tetrahedra, with two-in-two-out order.
    \textbf{C.} The all-in-all-out order on the pyrochlore, an example of a kind of possible magnetic order.
    \textbf{D.} The unit cell of representative breathing lattice pyrochlore compound \ce{Ba3Yb2Zn5O11}.
    }
\end{center}

As previously mentioned, the spinel structure type with general formula \ce{AB2X4} contains a pyrochlore sublattice composed of the B-site cations. This pyrochlore lattice found in spinel materials is identical to that obtained in the pyrochlore family of materials. \textbf{Figure 20 A} shows the unit cell of \ce{ZnFe2O4}, which crystallizes in the cubic space group $Fd\bar{3}m$ (no. 227). This material contains a pyrochlore lattice of Fe$^{3+}$ octahedra, which are high spin $d^{5}$ $S = 5/2$ systems. As such, they are the furthest possible (for a transition metal) from a small $S = 1/2$ state. Nevertheless, \ce{ZnFe2O4} has been shown to exhibit quantum spin liquid-like behavior in measurements of diffuse neutron scattering \cite{PhysRevB.68.024412}, which has been proposed to be the result of unconventional exchange pathways \cite{Tomiyasu2011}. \ce{ZnFe2O4} serves as a prototypical spinel material with a magnetically active pyrochlore lattice, of which there are few compounds and thus offer a potentially profitable avenue of research that could yield new, interesting materials. 

Breathing pyrochlore lattices have also been observed in spinel materials, an example of which is \ce{LiInCr4O8}, whose unit cell is shown in \textbf{Figure 20 B}. The behavior of this material has been shown to exhibit a spin gap with a non-magnetic ground state, with potentially exotic origins\cite{PhysRevLett.110.097203}. The number of breathing pyrochlore materials is fairly low, and their existence within the spinel family offers an optimistic route towards the discovery of others that may display interesting quantum spin liquid phenomena. Chemical intuition and knowledge of chemical bonding and crystallographic transformations may therefore be of direct relevance to quantum spin liquid physics.

\begin{center}                
    \includegraphics[width=1\textwidth]{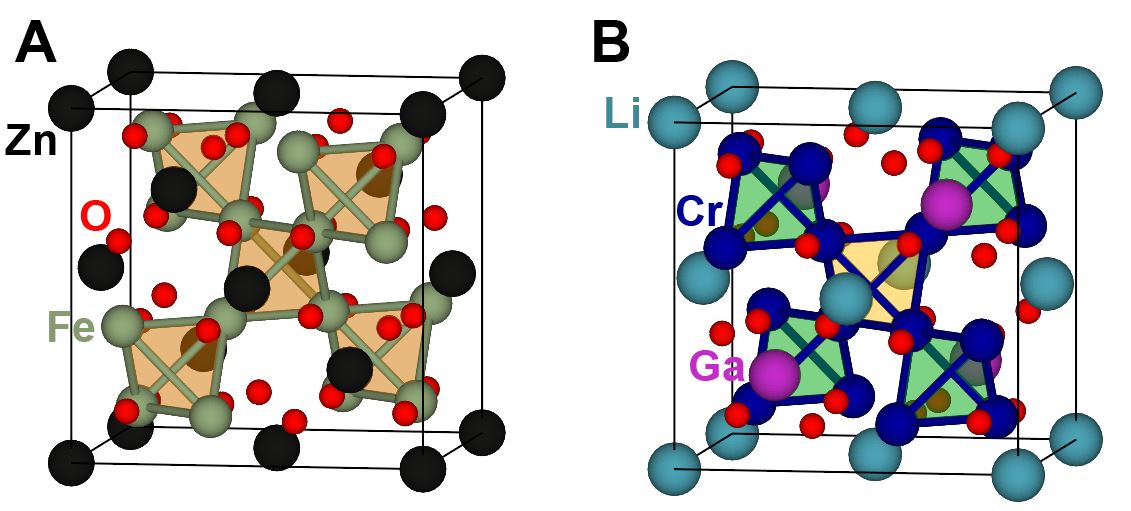}
    \small{\textbf{Figure 20.} Two representative spinel structures.
    \textbf{A.} The unit cell of prototypical spinel \ce{ZnFe2O4}, highlighting the Fe$^{3+}$ pyrochlore sublattice.
    \textbf{B.} The unit cell of breathing pyrochlore spinel material \ce{LiInCr4O8}, showing the large (green) and small (yellow) alternating pattern of corner-sharing tetrahedra.
    }
\end{center}

\subsection{Quantum Spin Liquid Candidates} \par
\begin{center}
\begin{longtable}{|p{0.01\textwidth}|p{0.25\textwidth}|p{0.1\textwidth}|l|l|l|l|}
\caption{A summary of spin liquid materials: Dimensionality; Material; Sub-lattice structure; Synthetic techniques involving solid-state (SS), flux, hydrothermal synthesis (HS), chemical vapor transport (CVT), floating zone (FZ), traveling solvent FZ (TSFZ), laser-diode FZ (LDFZ), Bridgeman and other; Exchange interaction $J/k_B$ (K) and Transition temperature $T_c$ (K)} \label{tab:long} \\
\hline \multicolumn{1}{c}{\textbf{Dimension}} & \multicolumn{1}{c}{\textbf{Material}} & \multicolumn{1}{c}{\textbf{Structure}} & \multicolumn{1}{c}{\textbf{Synthesis}} & \multicolumn{1}{c}{\textbf{$J/k_B$(K)}} & \multicolumn{1}{c}{\textbf{$T_c$(K)}} & \multicolumn{1}{c}{\textbf{Ref.}}\\ \hline 
\endfirsthead

\multicolumn{7}{c}%
{{\tablename\ \thetable{} -- continued from previous page}} \\
\hline \multicolumn{1}{c}{\textbf{Dimension}} & \multicolumn{1}{c}{\textbf{Material}} & \multicolumn{1}{c}{\textbf{Structure}} & \multicolumn{1}{c}{\textbf{Synthesis}} & \multicolumn{1}{c}{\textbf{$J/k_B$(K)}} & \multicolumn{1}{c}{\textbf{$T_c$(K)}} & \multicolumn{1}{c}{\textbf{Ref.}} \\ \hline 
\endhead

\hline \multicolumn{7}{|r|}{{Continued on next page}} \\ \hline

\endfoot

\hline \hline
\endlastfoot
1D   & \ce{Sr2CuO3} & Chain & TSFZ & 2200 & 5 & \cite{1999JCrGr.198..593R,PhysRevLett.76.3212,PhysRevB.62.R6108,RN15}   \\
     & \ce{Sr2Cu(PO4)2} &  & SS & 144 & 0.45 & \cite{BELIK2004883}  \\
     & \ce{Ba2Cu(PO4)2} &  & SS & 132 & 0.45 & \cite{BELIK2004883} \\
     & \ce{Cs4CuSb2Cl12} &  & HS & 186 & 0.7 & \cite{PhysRevB.101.235107}  \\
     & \ce{KCuMoO4(OH)} &  & HS & 238 & 1.5 & \cite{PhysRevB.96.104429,doi:10.1021/acs.inorgchem.5b00686}  \\
     & \ce{BaCuP2O7} &  & SS & 104 & 0.8 & \cite{doi:10.1021/ic050938z}  \\
     & \ce{Nd2CuO4} &  & Flux & 156 & 1.5 & \cite{HUNDLEY1989102} \\
     & \ce{BaCu2Ge2O7} & & FZ & 540 & 8.8 & \cite{PhysRevB.62.R6061}  \\
     & \ce{Bi6V3O16} &  & SS & 113 & 2 & \cite{PhysRevB.100.094431}  \\
     & \ce{Sm2CuO4} &  & Flux & 189 & 5.9 & \cite{HUNDLEY1989102}  \\
     & \ce{BaCu2Si2O7} &  & SS, FZ & 280 & 9.2 & \cite{PhysRevB.60.6601}  \\
     & \ce{VOSb2O4} &  & SS & 245 & 14 & \cite{RN16} \\
     & \ce{TiPO4} &  & SS & 965 & 74 & \cite{PhysRevB.83.180414,KINOMURA1982252}  \\
     & \ce{TiOBr} &  & CVT & 375 & 28 & \cite{PhysRevLett.95.097203,doi:10.1002/zaac.19582950314}  \\
     & \ce{TiOCl} &  & CVT & 660 & 67 & \cite{PhysRevB.67.020405, doi:10.1002/zaac.19582950314} \\
     & \ce{KCuF3} &  & HS & 395 & 39 & \cite{RN17,PhysRevLett.88.106403,10.1143/PTPS.46.147} \\
     & \ce{CuGeO3} & & TSFZ, Flux & 88 & 14 & \cite{1999JCrGr.198..593R,PhysRevLett.70.3651}  \\
     & \ce{CuSb2O6} &  & SS & 48 & 9 & \cite{KATO2000663}  \\
     & \ce{CuCl2} &  & Bridgeman & 90 & 24 & \cite{PhysRevB.80.024404}  \\
     & \ce{CuSiO3} &  & SS & 21 & 8 & \cite{PhysRevB.62.12201}  \\
     & \ce{Cs2CuCl4} &  & Low-temperature & 2 & 1 & \cite{doi:10.1063/1.335093} \\
     & \ce{CsNiCl3} &  &  Bridgeman  & 14  & 4.3 & \cite{Moses_1977,PhysRevLett.56.371,doi:10.1063/1.338595} \\
     & \ce{SrNi2V2O8} &  & TSFZ & 98  & 7 & \cite{PhysRevB.62.8921,PhysRevB.87.224423}  \\
     & \ce{PbNi2V2O8} &  & SS & 95 & & \cite{PhysRevLett.83.632,10.1143/PTPS.145.294}  \\
      & \ce{Pb(Ni1-xMgx)2V2O8} &  & SS & 95 & 3 & \cite{PhysRevLett.83.632, 10.1143/PTPS.145.294}  \\
   \hline
   2D   & \ce{NiGa2S4} & Triangular & SS, CVT & & 10 & \cite{Nambu_2011,PhysRevLett.99.157203}   \\
        & \ce{YbMgGaO4} &  & FZ & 1.5 & 0.06 & \cite{PhysRevLett.115.167203,RN18,RN19}    \\
        & \ce{Ba3CuSb2O9} &  & Flux & 32 & 0.2 & \cite{Katayama9305, PhysRevLett.106.147204,Man_2018}   \\
        & \ce{LiZn2Mo3O8} & & SS & & 0.07 & \cite{PhysRevB.89.064407,PhysRevLett.112.027202,PhysRevLett.111.217201,RN27}  \\
        & \ce{1T-TaS2} &  & CVT & & 0.07 & \cite{Law6996, RN20}  \\
        & \ce{Na2BaCo(PO4)2} &  & Flux & & 1 & \cite{Zhong14505}  \\
        & \ce{Ba4NbRu3O12} &  & SS & & 4 & \cite{PhysRevB.93.180407}  \\
        & \ce{Ba4NbIr3O12} &  & SS & & 1.8 & \cite{PhysRevMaterials.3.014412}  \\
        & \ce{(BEDT-TTF)2Cu2(CN)3} & & Electrochemical & 250 & 0.032 & \cite{PhysRevLett.91.107001}  \\
        & \ce{EtMe3Sb[Pd(dmit)2]2} & & Aerial oxidation & 220 - 250 & 1.37 & \cite{RN30,PhysRevB.77.104413}  \\
        & $\alpha$-\ce{Li2IrO3} & Honeycomb & CVT & & 15 & \cite{PhysRevLett.108.127203}   \\
        & $beta$-{\ce{Li2IrO3}} & & CVT & & 38 & \cite{PhysRevLett.114.077202,RN21}  \\ 
        & $\gamma$\ce{Li2IrO3} &  & CVT & & 40 & \cite{PhysRevLett.113.197201}    \\ 
        & \ce{Na2IrO3} &  & SS & & 13-18 & \cite{PhysRevB.82.064412,PhysRevLett.110.097204,PhysRevB.83.220403}   \\
        & \ce{Li3Co2SbO6} &  & SS & & 11 & \cite{vivanco2020competing}    \\
        & \ce{H3LiIr2O6} &  & HS & & & \cite{RN22}   \\
        & \ce{RuCl3} &  & CVT & & 8 & \cite{Banerjee1055,RN23}  \\
        & \ce{ZnCu3(OH)6Cl2} & Kagome & HS & 200 & & \cite{RN24,PhysRevLett.108.157202, Fu655}    \\
        & \ce{ZnxCu4-x(OH)6Cl2} & & HS & 200 & & \cite{PhysRevB.83.100402,Shores2005}    \\
        & \ce{Cu3Zn(OH)6BrF} &  & HS & 170 & & \cite{PhysRevMaterials.2.044406,SMAHA2018123,RN25}   \\ 
        & \ce{Cu4(OH)6BrF} &  & HS & 170  & 10-15 & \cite{PhysRevMaterials.2.044406,SMAHA2018123,RN25}   \\ 
        & \ce{Na4Ir3O8} &  & SS & 430 & & \cite{PhysRevLett.99.137207,PhysRevB.88.220413}   \\
        & \ce{PbCuTe2O6} &  & SS & 15 & 0.87 & \cite{PhysRevB.90.035141,PhysRevLett.116.107203}   \\
        & \ce{RE3Sb3Mg2O14} &  & SS &  & 2 & \cite{doi:10.1002/pssb.201600256}   \\
        & \ce{Nd3Sb3Mg2O14} &  & SS &  & 0.56 & \cite{PhysRevB.93.180407}   \\
        & \ce{Cu5V2O10(CsCl)} & & SS & & 24 & \cite{PhysRevB.98.054421}    \\
        & \ce{Ca10Cr7O28} & & TSFZ & & 0.3 & \cite{RN31}  \\
        & \ce{Cu3V2O7(OH)2(H2O)2} & & HS & & & \cite{Hiroi_Volborhite_2001,Kohama10686} \\
        & \ce{BaCu3V2O8(OH)2} & & HS & & 9 & \cite{C2JM32250A,200915843,C6TC02935C,PhysRevLett.121.107203} \\
    \hline
       3D   & \ce{FeSc2S4} & Diamond & TSFZ, SS & & 11 & \cite{MOREY2016128,SON2008e699,PhysRevLett.92.116401}  \\  
        & \ce{MnSc2S4} &  & SS & & 2.3 & \cite{PhysRevB.73.014413}   \\  
        & \ce{NiRh2O4} & & SS & & 0.1 & \cite{PhysRevMaterials.2.034404} \\
        & \ce{ZnFe2O4} & & Flux, Sol-gel, HS  & & 13 & \cite{PhysRevB.68.024412,KAMAZAWA19991261,doi:10.1021/cm0401802} \\
        & \ce{Yb2Ti2O7} & Pyrochlore & TSFZ, SS & & 0.27 & \cite{PhysRevB.95.094407,PhysRevX.1.021002, PhysRevLett.109.097205,PhysRevB.87.224423,RN26}   \\  
        & \ce{Er2Ti2O7} &  & TSFZ &  & 1.17 & \cite{Poole_2007,McClarty_2009} \\ 
        & \ce{Tb2Ti2O7} & & FZ, SS & & 0.05 & \cite{PhysRevLett.112.017203,PhysRevLett.109.017201,Takatsu_2011,Enjalran_2004,PhysRevLett.99.237202,PhysRevB.85.214420} \\
        & \ce{Dy2Ti2O7} & & FZ & & 1.2 & \cite{Matsuhira_2002,Kassner8549} \\
        & \ce{Ho2Ti2O7} & & TSFZ,Czochralski & 1 & 0.05 & \cite{GHASEMI201838,KANG2014104,PhysRevLett.79.2554} \\
        & \ce{Ho2Sn2O7} & & SS & & 0.75 & \cite{Matsuhira_2000}\\
         & \ce{Ce2Zr2O7} &  & LDFZ, SS  &  &  & \cite{PhysRevLett.122.187201,RN34} \\
        & \ce{NaSrMn2F7} &  & FZ & & 2 & \cite{Sanders_2016}    \\ 
        & \ce{NaCaFe2F7} &  & FZ & & 2 & \cite{Sanders_2016}    \\ 
        & \ce{NaSrFe2F7} &  & FZ & & 2 & \cite{Sanders_2016}    \\
\end{longtable}
\end{center}

\section{\underline{Defects and Problems with Measurements}}

Quantum spin liquids are materials that, in theory, never undergo a phase transition into a long-range magnetically ordered state. These materials are no exception to defects, and oftentimes contain more of them than otherwise expected. Here we consider some of these defects and the problems they may pose when studying their interesting behaviors, and some potential solutions to these problems. We appeal to all those experimentally (and theoretically) studying spin liquid materials to further consider and expand the role of defects in these materials, so as to completely understand these fascinating materials in the context of the real, entropic world. Thus far, large amounts of quantifiable point defects have been observed in most spin liquid candidates, such as in Herbertsmithite,\cite{Shores2005,RevModPhys.88.041002} Barlowite,\cite{PhysRevLett.113.227203,PhysRevMaterials.2.044406}, \ce{FeSc2S4},\cite{PhysRevB.96.054417,MOREY2016128} and \ce{YbMgGaO4},\cite{RN35,PhysRevLett.119.157201}, to name a few. A list of various spin liquid candidates, and possible or experimentally observed sources of disorder is shown in \textbf{Table 2}. 

\begin{table}
    \centering
    \caption{Several popular spin liquid candidates are shown, along with the origins of their experimentally observed structural disorders}
    \begin{tabular}{|p{0.4\textwidth}|p{0.4\textwidth}|p{0.2\textwidth}|}
    \hline
     Material & Source of disorder & Ref. \\
    \hline
    & Point defects, Substitutional &  \\ 
      Herbertsmithite & Cu$^{2+}$ on Zn$^{2+}$ interlayer site &\cite{Shores2005,RevModPhys.88.041002,RN24,PhysRevLett.104.147201,PhysRevB.82.144412,PhysRevLett.103.237201,Nilsen_2013,PhysRevLett.118.017202} \\
     \ce{ZnCu3(OH)6Cl2} & Possible configurational disorder of (OH) groups & \\
    \hline
    & Point defects, Substitutional & \\
    Zn-doped Barlowite & Cu$^{2+}$ and Zn$^{2+}$ disorder in interlayer site & \cite{PhysRevLett.113.227203,PhysRevMaterials.2.044406,Feng_2017,PhysRevB.98.155127,PhysRevB.92.220102,SMAHA2018123} \\
    Cu$_{4-x}$Zn$_{x}$(OH)$_{6}$Cl$_{2}$ & Possible configurational disorder of (OH) groups & \\
    \hline
       & Point defects, Atomic disorder & \\
       Vesigniete & \ce{VO4} tetrahedra show strong V and O positional disorder & \cite{C2JM32250A,200915843,C6TC02935C,PhysRevLett.121.107203}  \\
       \ce{BaCu3V2O8(OH)2} & Possible configurational disorder of (OH) groups & \\
    \hline
    & Point defects, Atomic site mixing & \\
    \ce{YbMgGaO4} & \ce{Mg^2+} and \ce{Ga^3+} are disordered and randomly occupy each other’s sites & \cite{RN35,PhysRevLett.119.157201,PhysRevLett.115.167203,PhysRevLett.118.107202,PhysRevX.8.031028} \\
    \hline
    & Atomic site mixing & \\
    \ce{$6H$-\ce{Ba3CuSb2O9}} & Cu$^{2+}$ and Sb$^{5+}$ are disordered and can occupy each other’s sites & \cite{Katayama9305,PhysRevLett.115.147202,PhysRevB.93.184425} \\
    \hline
    & Point defects, Off-stoichiometry & \\
    \ce{FeSc2S4} & Off-stoichiometries, especially involving \ce{Fe^2+} & \cite{PhysRevB.96.054417,MOREY2016128,FICHTL20052793} \\
    \hline
    & Inter-layer stacking faults  & \\
    \ce{RuCl3} & Off-stoichiometries, especially involving other halides and oxygen & \cite{Ziatdinov2016,Banerjee1055,PhysRevB.93.134423} \\
    \hline
    & Point defects, Off-stoichiometry & \\
    \ce{Pr2Zr2O7} & Excess \ce{Pr^3+}, such that the new formula becomes Pr$_{2+x}$Zr$_{2-x}$O$_{7-x/2}$ & \cite{PhysRevLett.118.107206,KOOHPAYEH2014291,PhysRevX.7.041028,RN36}  \\
    \hline
    & Point defects, Off-stoichiometry & \\
    \ce{Yb2Ti2O7} & ``stuffing", anti-site disorders & \cite{PhysRevB.86.174424,PhysRevB.95.094431,RN33,RN37} \\
    & substitution Yb on Ti B-site & \\
    & oxygen vacancies & \\
    & Topological defects, Superdislocations & \\
    \hline
    & Molecular distortions & \\
    \ce{(BEDT-TTF)2Cu2(CN)3} & Random distortions of the BEDT-TTF molecules & \cite{RN30} \\
    \hline
    & Vacancies, Off-stoichiometry & \\
    \ce{Na2IrO3} & Na$^{1+}$ and O$^{2-}$ deficiencies & \cite{PhysRevB.85.180403, Wallace2015} \\
    \hline
    \end{tabular}
    
    \label{tab:my_label}
\end{table}

In this section, we briefly discuss the possible impact of defects on spin liquid physical properties measurements such as magnetization, heat capacity, inelastic scattering, and resonance techniques. Furthermore, we discuss several key techniques, such as heat capacity, and how problems arise when trying to interpret the data that require immediate solutions to ensure reliability and reproducibility in the spin liquid field. We discuss how some data manipulation techniques, such as determining the phonon contribution to the heat capacity, can askew interpretations of the data, and propose some ways to try and consolidate these techniques in a manner conducive toward an honest description of intrinsic behavior in this interesting class of materials.

\subsubsection{Current Problems}

Heat capacity is a particularly useful tool to study a material’s physical properties and is universally employed to study spin liquid candidates. It provides information about the entropy in a system, which arises from a combination of phonon, electronic, magnetic, and other degrees of freedom, as aforementioned in the methods section above. The entropy derived from the magnetic contribution to the heat capacity provides information about magnetism in a system, and can give indications as to exotic, residual entropies at low temperatures, entropy changes across phase transitions, and unexpected spin-orbital contributions \cite{doi:10.1021/ja01315a102}. The phonon contribution to the specific heat arises from the vibrational degrees of freedom of atoms in a lattice and cannot be avoided. In order to determine the change in entropy as a function of temperature for the magnetic degrees of freedom in an insulating magnetic compound, the phonon contribution must be subtracted from the total heat capacity. However, determining the phonon contribution to the heat capacity of a material is not always an easy task. Two common ways include synthesizing and measuring the heat capacity of a closely related (in atomic mass) diamagnetic analog; and calculating the phonon dispersion curves and concomitantly extracting the theoretical phonon heat capacity using density functional theory. These two ways of estimating the phonon contribution to the specific heat, however, make an unlikely assumption: that the material in question is completely free of defects. This is obviously not possible. Estimating the phonon contribution based on the Debye $T^3$ phonon dependence law, is also possible, but is only applicable under $T$ = 10 K, and is strongly hindered by the presence of any other contributions from other effects in that temperature scale, such as ubiquitous Schottky defects.

To illustrate complications in phonon heat capacity analysis, we turn to Herbertsmithite, with formula \ce{ZnCu3(OH)6Cl2} \cite{Shores2005}. This material contains Cu$^{2+}$ cations sitting in kagome planes that are separated by diamagnetic Zn$^{2+}$ cations.  It has been shown that even in the highest quality single crystals of this material, there exists Cu$^{2+}$-Zn$^{2+}$ site mixing, of up to 15\%, which modifies the chemical formula of Herbersmithite to (Zn$_{0.85}$Cu$_{0.15}$)(Cu$_{2.85}$Zn$_{0.15}$)(OH)$_{6}$Cl$_{2}$\cite{RN24}. A representative image of the impact 15\% site mixing has on a kagome lattice can be found in Figure 21. In the ideal case, the Zn$^{2+}$ cations sit in \ce{Zn(OH)6} octahedra with equal axial and equatorial bond lengths. A Cu$^{2+}$ cation sitting on this same site, however, would be strongly susceptible to a first-order Jahn-Teller distortion. Consequently, strong electron-phonon coupling is expected, which should result in an altered phonon dispersion compared to the truly ideal case. This, in turn, should have an impact on the phonon contribution to the heat capacity, and thus renders values estimated using a diamagnetic analog or through density functional theory calculations inadequate.

\begin{center}                
    \includegraphics[width=1\textwidth]{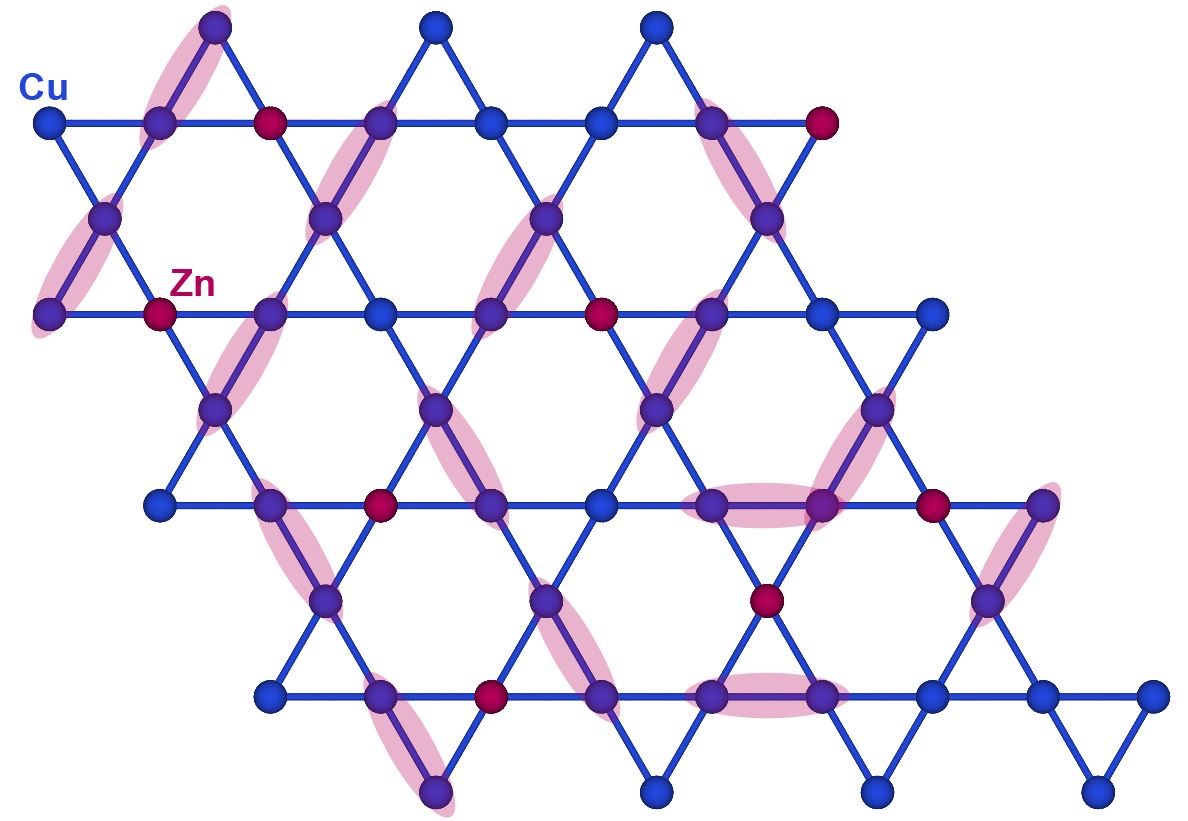} \par
    \small{\textbf{Figure 21.} The consequences of a replacement of 15\% of kagome lattice Cu$^{2+}$ sites with Zn$^{2+}$ in pure Herbertsmithite. The ovals represent Cu$^{2+}$ cations whose nearest neighbor interactions are affected as a result of the substitutional disorder.}
\end{center}

The electronic contribution to the heat capacity can also be strongly affected by defects. While most spin liquid candidates are insulators, substitution of atoms on sites may result in local metallicity or increased metallic behavior which manifests in the heat capacity as a non-zero, linear electronic term. The introduction of defect spins may also result in glassy or glass-like behavior, which has been known to give rise to linear terms in low temperature heat capacity \cite{elliott1983physics,tari2003specific}. It is important to note that the magnetic contribution to the heat capacity obtained by subtracting the phonon term can then be offset by the presence of an electronic term, which can have a non-negligible effect on any determinations of power law dependencies.

The magnetic contribution to the specific heat can also be affected by defects. Defects that have a strong effect on the phonon dispersions in a material can have a non-negligible effect on the magnon dispersions, due to magnon-phonon coupling, which results in an altered magnetic contribution to the heat capacity and altered power laws. At sufficiently high concentrations, defect spins can add to the heat capacity and alter it, making it difficult to distinguish the contributions from the magnetic moments of interest from those of defect spins.

There are, however, other possible intrinsic contributions to the heat capacity that may obscure quantum spin liquid signatures in heat capacity. Schottky anomalies are regularly observed and are often expected. They are the manifestation of an energy gap, in either the electronic or nuclear energy levels, and are often observed at low temperatures as characteristic broad peaks. These peaks can be fit to models, though often not without difficulty. Furthermore, defect spins can generate Schottky anomalies of their own that scale relative to their concentration.

Fortunately, magnetization, nuclear magnetic resonance, and electron spin resonance measurements can provide information about the intrinsic and defect-driven behaviors in a material. Magnetization as a function of magnetic field can yield information about the kinds of interactions in a system. A purely linear relationship is expected for a non-interacting paramagnet, and any deviations are indicative of interactions. In a spin liquid candidate, one would expect the magnetic moments sitting on a frustrated lattice to remain paramagnetic but strongly interacting, which has an overall net effect of deviating from a fully linear behavior in the magnetization. Defect spins, however, if truly non-interacting and paramagnetic, should contribute a linear dependence to the magnetization that scales to their concentration. Electron spin resonance measurements can yield information about unpaired electrons localized at defect sites, and nuclear magnetic resonance measurements can detect nuclear contributions from defects. Measurements of the temperature dependent magnetization of magnetic materials, or measurements of the temperature dependent magnetic susceptibility, are often marked by an upturn in the values at low temperatures. This upturn (Curie tail) is directly caused by the presence of defect spins in a material, and can be fit to models to extract the impurity concentration. Subtracting this contribution from the total magnetization or magnetic susceptibility, in conjunction with results from electron spin and nuclear magnetic resonances, can yield the contribution exclusively from the relevant, frustrated moments. 

In short, the inevitable presence of defects in real-world materials can pose many problems in the analysis of measurements of spin liquid candidate materials, and must be explicitly accounted for to ensure high fidelity.

\subsection{Defects} \par
As aforementioned, the uncertainties over the nature of defects has hindered the extraction and understanding of intrinsic magnetic ground states in quantum spin liquids. In addition, lattice imperfections are important aspects for materials as they affect physical properties such as diffusion pathways, material plasticity and strength, and device fabrication. Thus, it is critical to make efforts to untangle the effects of defects in local or real-space probe techniques. Though it costs enthalpy to form a defect, there is a significant entropic driving force that quickly becomes the dominant energy scale. Now let us look at a few types of defects observed in quantum spin liquid candidate materials and how they have been characterized.

\noindent \textbf{Point Defects} \par 
The most common point imperfections in crystals are lattice vacancies, extra, interstitial atoms not in regular positions, and chemical impurities. Point defects in Herbertsmithite \ce{ZnCu3(OH)6Cl2} have been observed to exhibit weak coupling with the kagome spins and have dominant contribution to thermodynamic signatures at low energy and low temperature.\cite{RN32,PhysRevB.94.060409,PhysRevLett.103.237201,PhysRevLett.100.087202,PhysRevB.84.020411,PhysRevLett.100.077203} In an attempt to try and understand the deep impact of defects in this material, a two-dimensional nuclear magnetic resonance investigation of Herbertsmithite was performed to estimate the weakly coupling defect moments arising from Cu$^{2+}$ in Zn$^{2+}$ sites\cite{PhysRevB.84.020411}. The ability to identify the source of enhancement of magnetic susceptibility in Herbertsmithite enables more accurate and faithful analysis of the bulk, intrinsic spin susceptibility data. In addition to this technique, the high resolution and sensitivity of local probe electron spin resonance measurements enable detailed inspection of different contributions to magnetic susceptibility. Electron spin resonance characterizations for Herbertsmithite resulted in the discovery of two types of intrinsic magnetic defects that possess considerably different interactions with the kagome spins\cite{PhysRevLett.118.017202}, as shown in \textbf{Figure 22}. Strongly coupled defects are associated with broader resonance responses, while the weakly exchange-coupled imperfections are characterized by their narrowness. 

\begin{center}                
    \includegraphics[width=1\textwidth]{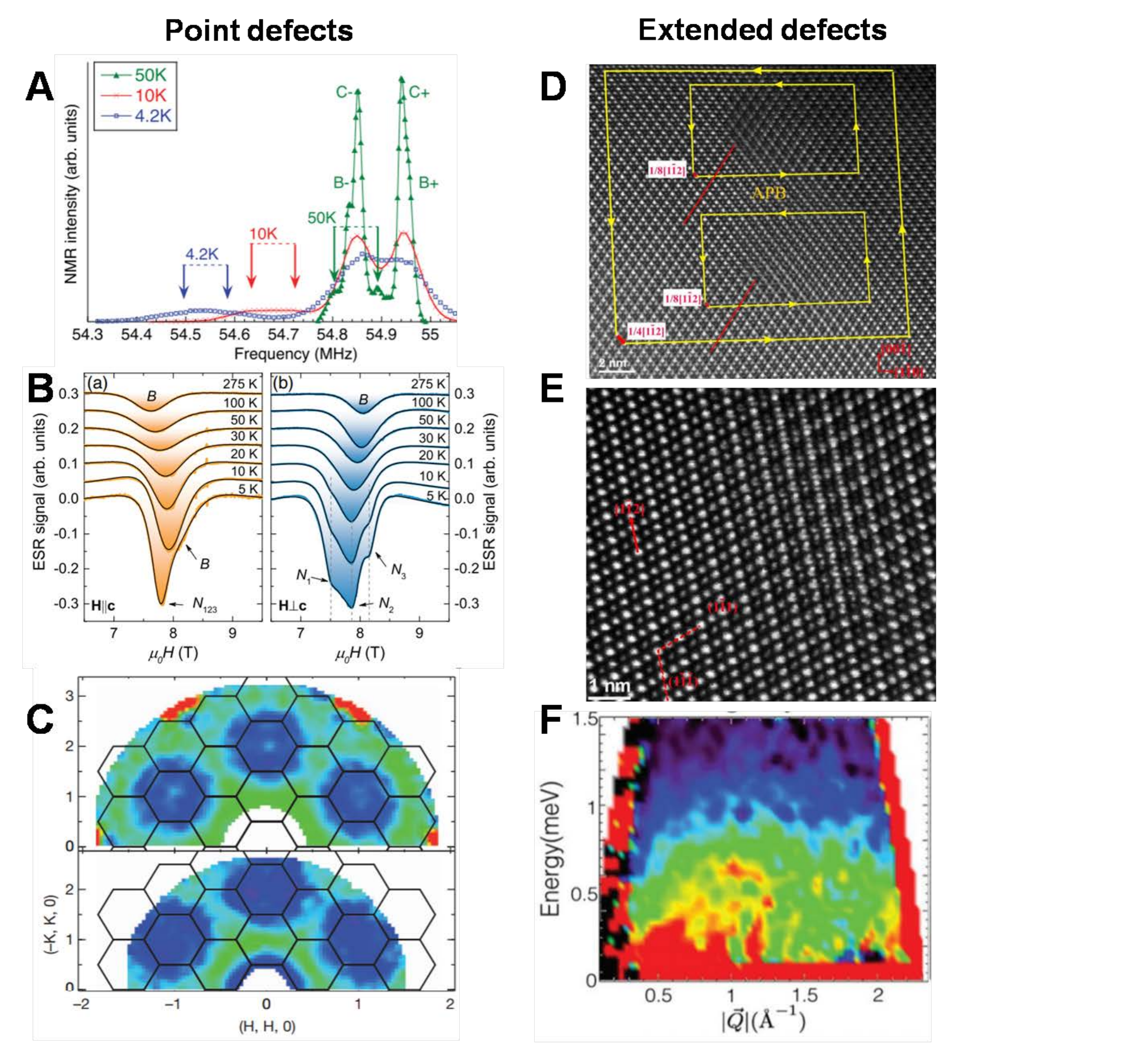}
    \small{\textbf{Figure 22.} Defects in quantum spin liquid materials.
    \textbf{A.} $^2$D nuclear magnetic resonance data on Herbetsmithite \ce{ZnCu3(OD)6Cl2}, revealing weakly coupled defect moments arising from Cu$^{2+}$ in Zn sites, adapted in part with permission from \cite{PhysRevB.84.020411} Copyright 2011 American Physical Society;
    \textbf{B.} Electron spin resonance spectra for \ce{ZnCu3(OH)6Cl2} indicating two types of intrinsic magnetic defects associated with broad and narrow ESR components, adapted in part with permission from \cite{PhysRevLett.118.017202} Copyright 2017 American Physical Society;
    \textbf{C.} Inelastic neutron scattering data on Herbertsmithite, collected at $T = 1.6$ K, showing the spin excitations plotted in reciprocal space, adapted in part with permission from \cite{RN32} Copyright 2012 Springer Nature;
    \textbf{D. and E.} Atomic resolution high-angle-annular-dark-field scanning transmission electron microscopy (HAADF-STEM) images of \ce{Yb2Ti2O7} crystals grown from floating-zone technique show line and planar lattice imperfections \cite{RN33}; Reproduced without changes from ref. \cite{RN33} under CC-BY-4.0;
    \textbf{F.} Inelastic neutron scattering for single crystal \ce{Yb2Ti2O7} at $T = 0.1$ K, showing the possible effects of crystal imperfections on the data, adapted in part with permission from \cite{PhysRevB.93.064406} Copyright 2016 American Physical Society.
    }
\end{center}

\noindent \textbf{Extended defects} \par
Another class of lattice imperfections in crystals include extended defects such as dislocations and grain boundaries. These involves shear displacement of two or more planes of atoms, stacking faults (one part of a crystal slides as a unit across an adjacent part), edge, planar and screw dislocations. For the characterization of line or planar lattice imperfections, atomic resolution high-angle-annular-dark-field scanning transmission electron microscopy (HAADF-STEM) imaging and electron energy loss spectroscopy (EELS) are powerful tools. \ce{Yb2Ti2O7} crystals grown from floating-zone technique, for example, exhibit line and planar dislocations, which are dissociated superdislocations and anti-phase boundaries (APBs) as shown in \textbf{Figure 22 D and E}\cite{RN33}. In contrast to the floating-zone crystal of \ce{Yb2Ti2O7}, significantly less defects were observed in the stoichiometrically prepared polycrystalline samples obtained from solid-state synthesis and essentially no lattice imperfections were found in the single crystals grown from traveling-solvent floating-zone method \cite{RN33}. The results from these characterizations not just uncover the nature of existing imperfections in quantum spin liquid candidate materials, but also highlight the important dependence of crystal quality on synthetic methods and growth conditions. This strongly urges the materials synthesis and crystal growth community to extend their extra cares to the sample creation and preparation of these delicate, quantum materials.

Defects can obfuscate interpretations of data related to spin liquid behavior, but there is hope. Though it may be impossible to fully characterize every defect in a sample and determine what their effects will be on the overall physical properties, a general understanding may be enough to greatly improve our understanding of spin liquid behaviors. We propose some ways to do this, to improve some of the aforementioned problems in the interpretation of data, such as heat capacity data.

Inelastic neutron scattering is a particularly useful technique to study spin liquid materials as it can give information about magnetic excitations, dynamics, and other exotic behavior. It can also be used to probe phonons. Since neutrons interact inelastically with atomic nuclei, a nearly complete understanding of the phonon dispersions is possible. This experimental dataset can be fit to models, from which the phonon heat capacity can be calculated. A key difference between this method and any of the aforementioned ways of estimating the phonon contribution to the heat capacity is that this is experimentally determined. The derived values are obtained by direct fitting of actual data, and thus requires no assumptions such as perfect crystallinity. To further strengthen the understanding of the phonons, Raman scattering can then also be used, as it may provide insight where neutrons are limited. There is also the possibility of cross-coupling heat capacity and magnetization measurements to gain a more complete understanding of the effects defects can have on the physical properties. By fitting the Curie tail in the magnetization to known models, the concentration of defect spins can be determined, and this information can be used in the heat capacity to improve fits to the magnetic contribution, Schottky anomalies, and electronic contribution anomalies. Measuring the magnetization of a spin liquid candidate up to very high magnetic fields, on the orders of tens of Tesla, may also yield information about the concentration of defect spins. At sufficiently high fields, the magnetization is expected to saturate for frustrated magnetic moments of the lattice. Any defect spins will not tend toward saturation, and a sufficiently high magnetic field would enable one to separate the contributions of the frustrated moments from those of the defect spins. The concentration of defect spins can therefore be readily determined. Furthermore, information about the kinds of interactions between defect spins (such as whether or not they are truly non-interacting) can then be obtained by fitting the defect spin magnetization to either Brillouin or Langevin functions. High field magnetization measurements can therefore be invaluable in understanding the role defect spins play in spin liquid candidate materials.

Systematic studies of the kinds and concentrations of point defects can also help in establishing the validity of a spin liquid observation claim. Taking atomic vacancies as an example, these can reduce the overall volume of a crystal and may, in some cases, even act as a source of negative chemical pressure. This may then completely alter the behavior of a material, renormalizing the electronic band structure and the phonon dispersions. The concentration of vacancies in a material can be systematically determined through careful dilatometry measurements, wherein the macroscopic volume of a sample is carefully measured across a temperature range and compared to the lattice parameter change across the same range. Monte Carlo simulations can then be employed, for example, to understand how the physical properties are modified as a function of vacancies in a material. 
We have focused primarily on point defects because these have been already observed, and reported on, in all spin liquid candidates. The effects of extended defects, however, are currently understudied. To provide an example, many materials contain magnetic ions arranged in the pyrochlore lattice of corner-sharing tetrahedra, a structure which is particularly susceptible to both structurally fixed and labile extended defects. In the case of the pyrochlore lattice, the arguably most common structurally fixed extended defect is ‘stuffing’, whereby extra A-site cations are incorporated into the structure \cite{LAU20063126,PhysRevB.86.174424,LAU200845,RN39,RN26}.Though extreme care is required so as to obtain correct stoichiometries when synthesizing pyrochlores, potential stuffing is often overlooked. In contrast, structurally labile extended defects have been recently observed in pyrochlore materials, in the form of dynamic structural fluctuations smaller than a nanometer, that are sufficient to increase the structural complexity of these materials and have a measurable effect on the physical properties.\cite{RN33}
There are many other kinds of extended defects, however, such as anti-phase domain boundaries, shear planes, topological defects, and others. These are present in all solids, though their effects on the physical properties of spin liquid materials remain, for the most part, unexplored. Greater efforts must be made to understand their impact on spin liquids, as they may present yet another subtle route towards the stabilization of a ground state and subsequent release of entropy. 

We would like to point out that the goal of this discussion is not to point out what role defects play in stabilizing or destabilizing spin liquid states. Arguments have been made either way for whether defects help, as is the case in the possible Coulomb quantum spin liquid pyrochlore candidate \ce{Pr2Zr2O7}, whose quantum properties are actually enabled by disorder \cite{PhysRevLett.118.107206,Kimura2013,PhysRevX.7.041028}, or hinder spin liquid physics. We merely point out the importance of characterizing the effect of defects on the overall physical properties, and point towards some ways of incorporating defect considerations into final analyzes of data. Materials as interesting and exotic as quantum spin liquids, with their entangled macroscopic ground states subject to electron fractionalization and potential for quantum computing, should have their physical properties scrutinized to the highest possible standard.

\section{\underline{Potential Applications}}
Richard Feynman envisioned a future where innovative leaps in quantum systems served as the fundamental paradigm shift for information processing. Quantum computers can store, transmit, compute and process information with remarkably reduced computational resource, exponentially expanding the size and complexity of computation systems. The field of quantum materials and quantum information science has emerged and the cooperative effort between interdisciplinary science and engineering has continued to contribute to the advances of quantum computation. One of the avenues to push the field forward is to make a substantial effort leveraging materials chemistry approaches combined with quantum information science. Quantum spin liquids featuring inherently quantum mechanical nature hold the potential to revolutionize the way quantum information processing is developed. Many of the requirements for quantum information science, including a manifold of available microstates, long coherence time, high scalability, sufficient error tolerance and faithful transmission of information, are governed by the underlying chemistry and elucidation of exchange interactions, chemical bonding, lattice geometry and spin and orbital degrees of freedom of quantum spin liquids.

Potential, real-world applications of quantum spin liquids arise from their ability to give rise to exotic quasiparticle excitations, such as Majorana fermions, which are non-Abelian, braidable anyons, as aforementioned in the $\alpha$-\ce{RuCl3} section. Majorana fermions are thought to be ideal for quantum computing for the two following reasons: (i) the particles can exchange positions resulting in a change in underlying quantum states and (ii) they have long range entanglement that can be exploited to result in fault-tolerant quantum computations \cite{RN23,doi:10.1146/annurev-conmatphys-030212-184337}. This highlights the capability of performing complex mathematical operations by employing quantum spin liquids in quantum information science.

Though at present most considerations of the potential use of quantum spin liquids is limited to quantum computing, the highly entangled, quantum nature of their ground state should be the exact kind of physics required for other technologies, including novel detectors and sensors, and technologies that have no comparable counterpart in the world today.

\section{\underline{Summary}}
The review of the chemistry of quantum spin liquids may leave one with a couple of queries, such as “Does the quantum spin liquid ground state actually exist in real, defectuous materials, and do they have any relevance to the real world?” The answer is probably yes, for both. Even though quantum entanglement of spins has not (at present) been directly observed in quantum spin liquids, it has been experimentally convincingly proven to exist in one dimensional quantum spin liquids, and rigorously, theoretically proposed in real materials. The underlying chemistry of quantum spin liquid resonates with the rigorous requirements that must be fulfilled in order to sufficiently prepare them for quantum computing use. Discoveries in this field have significant implications for a list of societal branches including national security, information technology, clean energy, health care, communications, and many more. By sharing this review, we hope to provide a perspective to the materials community, demonstrating how far the community has come, and, at the same time, encouraging more attention and resources to quantum spin liquid materials design, preparation, characterization and technological implementation that would result in not just the development of new chemistry routes, but also in understanding genuine quantum spin liquid behavior. In doing so, it is reasonable to feel optimistic that we, as a group of scientists sharing the same set of interests and beliefs, will someday make the technological implementation of quantum spin liquids a reality. 

\section{Author Biographies}

\noindent Juan R. Chamorro is a PhD candidate at The Johns Hopkins University in Baltimore, Maryland. He received his B. A. in chemistry in 2017 from The Johns Hopkins University. He is interested in using chemical principles to study quantum materials, specifically focusing on topological, superconducting, and exotic magnetic materials.
 
\noindent Tyrel M. McQueen is a professor of chemistry, physics and astronomy, and materials science and engineering at the Johns Hopkins University, and director of the PARADIM Bulk Materials Discovery Facility. McQueen is the recipient of numerous awards including the Packard Science and Engineering and Sloan Research Fellowships. His research focuses on the synthesis, discovery, and analysis of new quantum materials, with a vision of their current and future utility.

\noindent Thao T. Tran is an assistant professor of chemistry at Clemson University. Her research focus on the synthesis, discovery and characterization of new materials with exotic physical properties. She received her B.S. in chemistry from Vietnam National University - Ho Chi Minh, and her Ph.D. in materials chemistry in 2015 from University of Houston. She conducted postdoctoral research (2016-2019) at Johns Hopkins University working on photovoltaic and quantum materials.

\begin{acknowledgement}
This work was supported as part of the Institute for Quantum Matter, an Energy Frontier Research Center funded by the U.S. Department of Energy, Office of Science, Office of Basic Energy Sciences, under Award No. DE-SC0019331. T.T.T acknowledges the support from Clemson University, College of Science, Department of Chemistry. 

\end{acknowledgement}

\bibliography{main_biblio}

\newpage
\begin{center}                
    \includegraphics[width=1\textwidth]{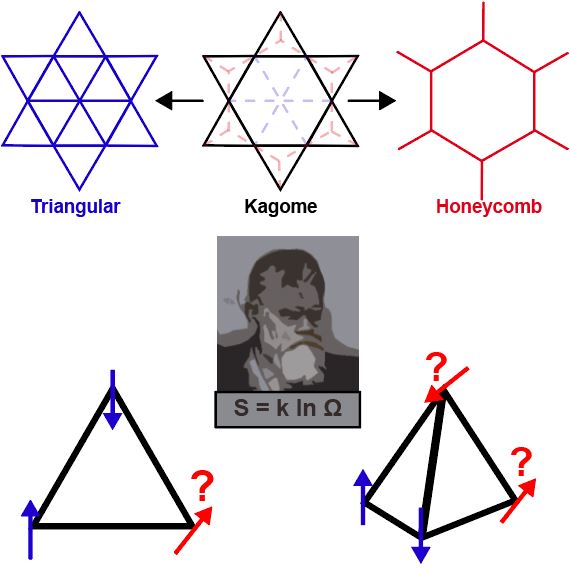}
    \small{\textbf{Table of Contents Figure}}
\end{center}

\end{document}